\documentclass[12pt]{article}

\usepackage{epsfig}

\def\be{\mbox{\bf e}}

\def\bphi{\mbox{\boldmath $\phi$}}

\def\balpha{\mbox{\boldmath $\alpha$}}
\def\bbeta{\mbox{\boldmath $\beta$}}

\def\bpsi{\mbox{\boldmath $\psi$}}

\def\bgamma{\mbox{\boldmath $\gamma$}}
\def\bdelta{\mbox{\boldmath $\delta$}}

\def\beeta{\mbox{\boldmath $\eta$}}
\def\bzeta{\mbox{\boldmath $\zeta$}}

\def\bPhi{\mbox{\boldmath $\Phi$}}

 \def\tfrac#1#2{{\textstyle{{#1}\over{#2}}}}

\def\half{\tfrac{1}{2}}
\def\third{\tfrac{1}{3}}
\def\quart{\tfrac{1}{4}}
\def\sixth{\tfrac{1}{6}}
\def\twothirds{\tfrac{2}{3}}

\def\phivac{\bPhi_{\mathrm{\scriptscriptstyle{VAC}}}}

\def\tr{\mathop{\rm tr}\nolimits}
\def\Tr{\mathop{\rm Tr}\nolimits}

\usepackage{amssymb}
\def\bbc{{\mathbb C}}

\def\bbr{{\mathbb R}}

\def\bbz{{\mathbb Z}}

\begin{document}

\begin{titlepage}

\baselineskip 24pt

\begin{center}

{\Large {\bf A Closer Study of the Framed Standard Model Yielding Testable New
Physics plus a Hidden Sector with Dark Matter Candidates}}

\vspace{.5cm}

\baselineskip 14pt

{\large Jos\'e BORDES \footnote{Work supported in part by Spanish
    MINECO under grant FPA2017-84543-P,
Severo Ochoa Excellence Program under grant SEV-2014-0398, and
Generalitat Valenciana under grant GVPROMETEOII 2014-049. }}\\
jose.m.bordes\,@\,uv.es \\
{\it Departament Fisica Teorica and IFIC, Centro Mixto CSIC, Universitat de 
Valencia, Calle Dr. Moliner 50, E-46100 Burjassot (Valencia), 
Spain}\\
\vspace{.2cm}
{\large CHAN Hong-Mo}\\
hong-mo.chan\,@\,stfc.ac.uk \\
{\it Rutherford Appleton Laboratory,\\
  Chilton, Didcot, Oxon, OX11 0QX, United Kingdom}\\
\vspace{.2cm}
{\large TSOU Sheung Tsun}\\
tsou\,@\,maths.ox.ac.uk\\
{\it Mathematical Institute, University of Oxford,\\
Radcliffe Observatory Quarter, Woodstock Road, \\
Oxford, OX2 6GG, United Kingdom}

\end{center}

\vspace{.3cm}

\begin{abstract}

This closer study of the FSM
\begin{itemize}
\item [I] retains the earlier results of \cite{tfsm} in offering explanation 
for the existence of three fermion generations, as well as the hierarchical 
mass and mixing patterns of leptons and quarks;
\item [II] predicts a vector boson $G$ with mass of order TeV which mixes with 
$\gamma$ and $Z$ of the standard model.  The subsequent deviations from the
standard mixing scheme are calculable in terms of the $G$ mass.  While these 
deviations for (i) $m_Z - m_W$, (ii) $\Gamma(Z \rightarrow \ell^+ \ell^-)$, and 
(iii) $\Gamma(Z \rightarrow {\rm hadrons})$ are all within present experimental 
errors so long as $m_G > 1$ TeV, they should soon be detectable if the 
$G$ mass is not too much bigger;
\item [III] suggests that in parallel to the standard sector familiar to us,
there is another where the roles of flavour and colour are interchanged. Though
quite as copiously populated and as vibrant in self-interactions as our own, 
it communicates but little with the standard sector except via mixing through 
a couple of known portals, one of which is the $\gamma-Z-G$ complex noted in 
[II] above, and the other is a scalar complex which includes the standard model
Higgs.  As a result, the new sector appears hidden to us as we appear hidden to them, 
and so its lowest members with masses of order 10 MeV, being electrically 
neutral and seemingly stable, but abundant, may make eligible
candidates as constituents of dark matter.
\end{itemize}
\noindent A more detailed summary of these results together with some remarks 
on the model's special theoretical features can be found in the last
section of the text.

\end{abstract}

\end{titlepage}

\clearpage

\section{Introduction}

The framed standard model (FSM) \cite{tfsm} is constructed from the standard 
model (SM) by adding to the usual gauge boson and matter fermion fields the frame 
vectors in internal space as dynamical variables (framons), thus making the 
particle theory more similar in spirit to the vierbein formulation \cite{EKS}
of gravitation.  In particle physics itself, the following then immediately 
result:

\begin{itemize}

\item {\bf (i)} The standard Higgs boson, as framon in the electroweak sector.

\item {\bf (ii)} A global $\widetilde{su}(3)$ symmetry, ``dual'' to the local 
colour symmetry, to act as fermion generations.

\item {\bf (iii)} A fermion mass matrix of the form:
\begin{equation}
m = m_T \balpha \balpha^\dagger,
\label{mfact}
\end{equation}
(with $\balpha$ universal and only $m_T$ depending on the fermion species). 
At tree level, $\balpha$ is constant so that only the top generation has a mass
and there is no mixing, but under renormalization by framon loops, the vector 
$\balpha$ rotates with changing scale $\mu$, leading to a hierarchical fermion 
mass spectrum and mixing between up and down fermion states, namely the CKM 
matrix for quarks and neutrino oscillations for leptons.  Indeed, a fit with 
the renormalization group equation (RGE)  so derived to 1-loop level
\cite{tfsm}, with 7 real parameters, gives 
already the close agreement with experiment shown in Table \ref{tfsmfit}, 
effectively replacing, to this accuracy, 17 independent parameters of the 
standard model by just 7.

\end{itemize}
The FSM seems thus to have given a geometric meaning to the Higgs field, and
offered a solution to the fermion generation problem as well as an explanation
for the bewildering mass and mixing pattern of quarks and leptons.  It gives
in addition a new solution to the strong CP problem translating it via the 
rotation of $\balpha$ to the CP-violating phase in the CKM matrix.  Even if
taken only as just a parametrization of the data, there is in the literature, 
as far as we know, no other fitting the same wide range of data to 
comparable accuracy with so few adjustable parameters. 

\begin{table}
\centering
\begin{tabular}{|l|l|l|l|l|}
\hline
& Expt (June 2014) & FSM Calc & Agree to & Control Calc\\
\hline
&&&& \\
{\sl INPUT} &&&&\\
$m_c$ & $1.275 \pm 0.025$ GeV & $1.275$ GeV & $< 1 \sigma$&$1.2755$ GeV\\
$m_\mu$ & $0.10566$ GeV & $0.1054$ GeV & $0.2 \%$ & $0.1056$ GeV\\
$m_e$ & $0.511$ MeV &$0.513$ MeV & $0.4 \%$ &$0.518$ MeV\\
$|V_{us}|$ & $0.22534 \pm  0.00065$ & $0.22493$ & $< 1 \sigma$ &$0.22468$\\
$|V_{ub}|$ & $0.00351^{+0.00015}_{-0.00014}$& $0.00346$ & $< 1 \sigma$&$0.00346$ \\
$\sin^2 2\theta_{13}$ & $0.095 \pm 0.010$ & $0.101$ &$< 1 \sigma$ &$0.102$\\
\hline
&&&& \\
{\sl OUTPUT} &&&&\\
$m_s$ & $0.095 \pm 0.005$ GeV & $0.169$ GeV & QCD &$0.170$ GeV \\
& (at 2 GeV) &(at $m_s$) &running& \\
$m_u/m_d$ & $0.38$---$0.58$ & $0.56$ &  $< 1 \sigma$&$0.56$ \\
$|V_{ud}|$ &$0.97427 \pm 0.00015$ & $0.97437$ & $< 1 \sigma$&$0.97443$ \\
$|V_{cs}|$ &$0.97344\pm0.00016$ & $0.97350$ & $< 1 \sigma$&$0.97356$ \\
$|V_{tb}|$ &$0.999146^{+0.000021}_{-0.000046}$ & $0.99907$ &$1.65
\sigma$&$0.999075$ \\
$|V_{cd}|$ &$0.22520 \pm 0.00065$ & $0.22462$ & $< 1 \sigma$ &$0.22437$\\
$|V_{cb}|$ & $0.0412^{+0.0011}_{-0.0005}$ & $0.0429$ & $1.55 \sigma$&
$0.0429$ \\
$|V_{ts}|$ & $0.0404^{+0.0011}_{-0.0004}$ & $0.0413$ &$< 1 \sigma$& 
$0.0412$\\  
$|V_{td}|$ & $0.00867^{+0.00029}_{-0.00031}$ & $0.01223$ & 41 \% & $0.01221$\\
$|J|$ & $\left(2.96^{+0,20}_{-0.16} \right) \times 10^{-5}$ & $2.35
\times 10^{-5}$ & 20 \% &$2.34\times 10^{-5}$ \\
$\sin^2 2\theta_{12}$ & $0.857 \pm 0.024$ & $0.841$ &  $< 1 \sigma$& $0.840$\\ 
$\sin^2 2\theta_{23}$ & $>0.95$ & $0.89$ & $> 6 \%$ &$0.89$\\
\hline 
\end{tabular}
\caption{Calculated fermion masses and mixing parameters compared with
experiment, reproduced from \cite{tfsm}} 
\label{tfsmfit}
\end{table}

Despite these attractive features, however, the FSM begs a question which needs
an urgent answer.  The framons represent altogether 11 complex degrees of 
freedom, of which two correspond to the standard electroweak Higgs, which is 
already seen in experiment.  But this still leaves 9, corresponding to the 
framons in the colour sector.  Then:
\begin{itemize}
\item {\bf Q} Why have we not been aware of colour framons in experiment?
\end{itemize}

The question {\bf Q}, however, is not immediately answerable because these
framons are colour triplets, and since colour is confined, they cannot propagate
as particles in free space, but can manifest themselves, at best, as confined 
constituents of colourless bound states.  A colour framon can combine with 
its conjugate in the s-wave to form scalar bosons (which we shall label 
generically as $H$), or in the p-wave (hence involving the gluon via the 
covariant derivative) to form vector bosons (which we shall call generically 
$G$), or else it can combine with a coloured fermion to form fermionic bound 
states (which we shall call generically $F$).  The question {\bf Q} can 
thus be rephrased as:

\begin{itemize}

\item {\bf Q'} Why have we not seen these $H$, $G$ and $F$ in experiment?

\end{itemize}
to answer which, we shall need first to know something about the properties of
these particles.

The particles $H$, $G$, and $F$  are in the colour theory the exact parallels 
of respectively the Higgs boson $h_W$, the vector bosons $W^\pm, \gamma-Z^0$, 
and the leptons and quarks in the flavour theory.  To see this, recall first 
an illuminating paper of 't~Hooft \cite{tHooft}, which pointed out that the 
standard electroweak theory, which is usually interpreted as having its local 
(flavour) $su(2)$ symmetry spontaneously broken, has a ``mathematically 
equivalent'' interpretation as a confining theory where the local $su(2)$ 
symmetry remains exact (see also \cite{banks}).  
What is broken then is only a global (often called 
the ``accidental'') symmetry, say $\widetilde{su}(2)$, hidden in the theory, 
and this global symmetry is here broken explicitly by electromagnetism (not 
spontaneously by weak hypercharge).   In this alternative interpretation 
of the electroweak theory (which we shall refer to in this paper as the 
``confinement picture'' (of 't~Hooft)), fields carrying local flavour cannot
propagate in free space but can appear only as constituents of flavour singlet
bound states confined by flavour $su(2)$.  Thus $h_W$ is the $su(2)$ singlet 
bound state confined by flavour of the fundamental scalar field $\bphi$ with 
its own conjugate $\bphi^\dagger$ in $s$-wave, $W^\pm, \gamma-Z^0$ are bound 
states of the same two constituents in $p$-wave, and leptons and quarks are 
bound states of the fundamental scalar to the fundamental fermion fields,
making them in the flavour theory the respective parallels of the $H$, $G$ 
and $F$ of the colour theory as claimed above. 

We note in passing, when the confinement picture for the flavour theory of
't~Hooft is adopted, the close parallel in the FSM between the  two 
nonabelian sectors, flavour $su(2)$ and colour $su(3)$.  Both theories are 
now confining and both are framed, for the flavour sector by {\bf (i)} above, 
and for the colour sector by construction.  In both theories, the vacuum is
degenerate, leading in each to the breaking of a global symmetry ``dual'' to 
the local gauge one, giving in the flavour sector two (up-down) flavours and in 
the colour sector three fermion generations.  And in each case, the global symmetry
is broken explicitly by electromagnetism, in the flavour sector as already 
noted above, and in the colour sector as we shall see later in Section 4.  And
yet, despite this striking parallel in formulation, the physics that emerges in 
the two sectors will be very different, as we shall see, because of some 
inherent differences between the two nonabelian symmetries, flavour $su(2)$
and colour $su(3)$.  These differences will be highlighted as points of
interest as we move along.

Let us return now to the question of detecting the $H$, $G$ and $F$.  To save 
repeating long phrases in future, it will be convenient to introduce the 
following new terms: 

\begin{itemize}

\item {\bf $B$-ons} to denote the flavour neutral bound states of flavoured 
constituents held together by flavour $su(2)$ confinement ($B_\mu$ being in our 
notation the flavour $su(2)$ gauge field).

\item {\bf $C$-ons} to denote colour neutral bound states of coloured 
constituents held together by colour $su(3)$ confinement ($C_\mu$ being in our 
notation the colour $su(3)$ gauge field).

\end{itemize}

In this terminology, the Higgs
boson $h_W$, the vector bosons\footnote{By $\gamma-Z$, we mean the component of
  the $\gamma$ and the $Z$ in flavour $su(2)$, i.e.\ $W^3$.}
$W, \gamma-Z$, 
and the quarks and leptons are all framonic $B$-ons, framonic in the sense that 
they are all obtained by binding a framon with other flavoured consitutents via 
flavour confinement.  On the other hand, $H$, $G$, and $F$ are framonic $C$-ons 
and their respective analogues, only with the roles of flavour and colour 
interchanged.  The framonic $B$-ons make up our world, that is,  the world as we 
have known it so far, which we shall call here the standard sector.  They are 
the building blocks from which, for example, baryons are obtained as higher 
level constructs via colour confinement.  Their $C$-on analogues, $H$, $G$, 
and $F$, on the other hand, have so far been hidden from us, for some reason 
yet unknown which is now our wish to find out.

To do so, we shall need first to envisage what properties these framonic 
$C$-ons are likely to possess, perhaps initially by drawing on their analogy 
in structure to particles already known to us.  For this, however, we are 
placed immediately in a dilemma.  On the one hand, as already noted, the 
framonic $C$-ons $H$, $G$, and $F$ are analogous to the framonic $B$-ons which
appear to us as point-like objects, interacting via only ``hard'' interactions 
prescribed by and derivable perturbatively from the action, the two types of
bound states being both framonic, but differing in the confining symmetry.  
On the other hand, the $H$, $G$, and $F$ are analogous also to hadrons which, 
in contrast, are bulky objects and have soft interactions, the two types being 
now both bound states via colour confinement, but differing in that the $H$, 
$G$, and $F$ each contains at least one colour framon as constituent, while 
hadrons have only quarks or antiquarks as constituents.  The question then is 
whether the $H$, $G$, and $F$ are likely to resemble more the hadrons or the 
framonic $B$-ons.

To try to answer this, let us first examine the question why the framonic $B$-ons
$h_W, W, \gamma-Z$, quarks and leptons, should be point-like while hadrons are 
fat, when both these types of particles are singlet compound states held 
together by gauge symmetry confinement.  One can, of course, ascribe the 
difference to the different confining symmetries, one being flavour $su(2)$, 
the other being colour $su(3)$, deferring to answer the question
because of our present incomplete knowledge
on how confinement comes about.  However, there is also between the two their 
structural difference, namely one type being framonic while the other not, 
which might give us a hint for their so very different properties.  Framons 
have imaginary mass, meaning in the flavour case that the $\phi^2$ terms in 
the scalar potential has a negative coefficient.  This is familiar, and is 
needed in the electroweak theory to make the vacuum degenerate and hence to 
break the flavour symmetry (local in the symmetry-breaking, global in the 
confinement picture).  In contrast, the quark constituents in hadrons are all 
possessed with a real positive mass.  Could this difference then hold the key 
to the question why hadrons have soft interactions while the framonic $B$-ons 
seem to have none? 

\begin{figure}[th]
\centering
\includegraphics[scale=0.4]{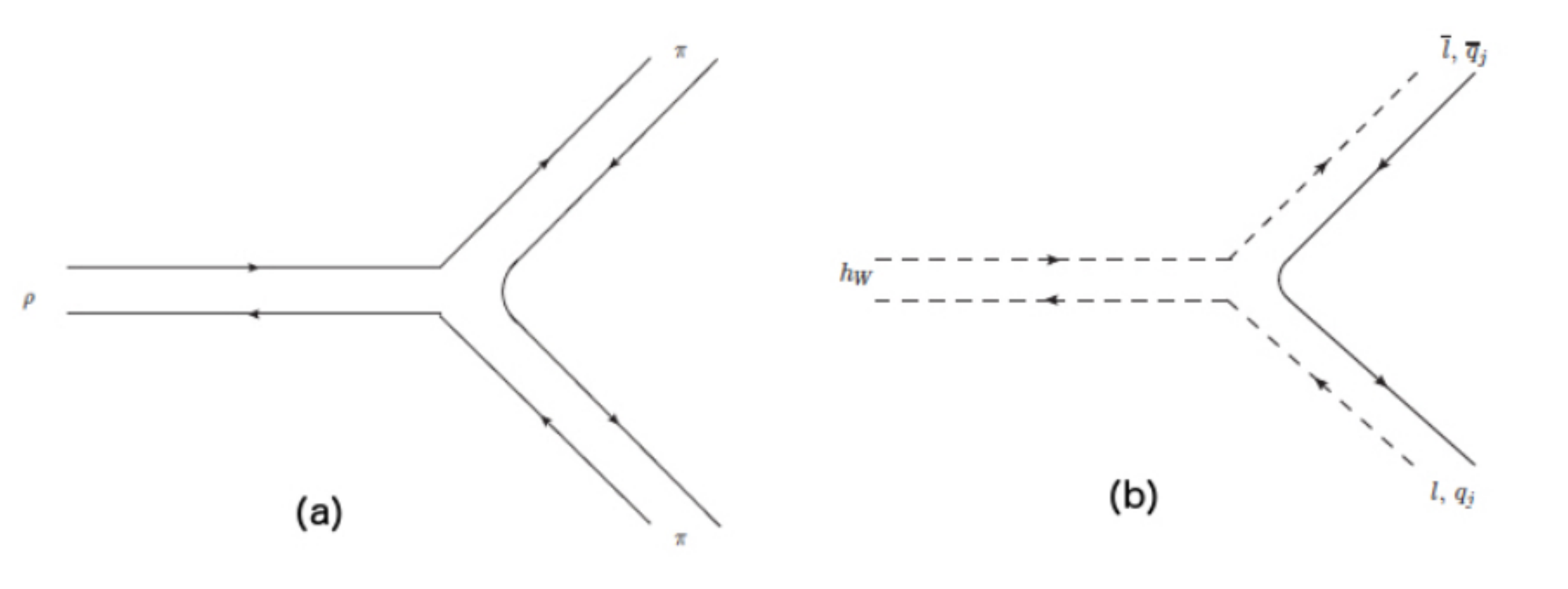}
\caption{(a) quark diagram for $\rho$-decay, (b) quark diagram for $h_W$-decay?}
\label{qdiag}
\end{figure}

Let us first recollect what little we know about soft interactions.  Soft 
interactions, being supposedly nonperturbative effects, cannot be deduced 
from theory using perturbative methods.  On the other hand, lattice methods
on which we mostly rely at present for exploring nonperturbative effects, have 
not given us much intuition on how soft interactions come about.  For that, 
we still have to rely on the phenomenology commonly done in the 1960s, as 
encapsulated in quark diagrams, which later evolved into the dual resonance 
model \cite{venez, dualres} and then to modern string theories \cite{string}.

Consider then, in the language of quark diagrams, as a typical example of soft 
interactions, the decay: $\rho \rightarrow 2 \pi$ which is represented by (a) 
of Figure \ref{qdiag}.  The quark and antiquark constituents in the $\rho$ 
recombine respectively with an antiquark and a quark of a pair emerging (say 
by quantum fluctuation) from the sea to form the 2 pions in the decay product.
This decay occurs at a very high rate which cannot be obtained in perturbative
QCD calculations.  At first sight, barring any at present unknown difference 
in dynamics between the flavour and colour theory, it would appear that a very 
similar strong decay could occur to the Higgs boson $h_W$ as indicated in 
diagram (b) of Figure \ref{qdiag} giving a lepton-antilepton or quark-antiquark 
pair.  Now, although the decay of $h_W$ into fermion-antifermion pairs does 
occur in experiment, it happens only at the standard perturbative rate obtained 
from the fundamental action, not at any unusually strong rate that the above 
analogy would lead us to expect.  Why?

The Figure \ref{qdiag}(b) differs from the Figure \ref{qdiag}(a) in that for the
$h_W$ in the former to decay softly, the flavour framon has to be separated 
from its partner so as to recombine with another to form the final state 
particles, whereas in the latter, it is only a quark-antiquark pair that has 
to be separated.  The framon differs from a quark in having an imaginary 
mass.  Now an imaginary mass 
translates to a finite life time for the framon while the quark life time (inside 
hadronic matter) will be infinite.  The possibility then arises that the 
framon, in contrast to a quark with infinite life time, may be too short-lived 
to have time finding and recombining with an alternative partner from the sea 
to form a new (flavour) singlet and emerge as a particle.  A naive intuitive 
estimate of the framon life time made in Section 9 indeed suggests that the 
framon would have practically no chance to find and recombine with a new 
partner, so that the $h_W$ would have perforce almost to eschew soft decays 
altogether. 

Generalizing the above argument to other soft interactions which all seem to
involve recombinations of a composite's constituents with particles emerging
from the sea, one would be led to suggest that they will not occur in any
of the  electroweak particles, namely $h_W, W^\pm, \gamma-Z^0$, quarks and 
leptons, these being all framonic composites in 't~Hooft's confinement
picture.  Hence,
these particles will all remain point-like and interact only weakly in contrast
to hadrons which, being nonframonic composites, can have soft interactions and 
are therefore bulky and strong. One thus seems to have found here an answer to 
the question posed above, without assuming a basic difference in dynamics
between flavour and colour.   

If this conclusion is at all acceptable, then it would seem to apply to the
framonic $C$-ons $H$, $G$ and $F$ as well.  The vacuum in the colour sector
has to be degenerate also, so as to break the generation symmetry, which means
that the coefficient of the $\bPhi^\dagger \bPhi$ term in the framon
potential (Section 3) is negative, and that the colour framon too will
have finite (short) life
time.
The same arguments as above for the framonic $B$-ons will then lead to the 
conclusion that framonic $C$-ons too will have negligible soft interactions 
and remain point-like.
  
If this is true, then only hard interactions remain for the $H$, $G$ and 
$F$, and for these interactions one can deduce a fair amount of information 
for the following fortunate reasons: 

\begin{itemize}

\item The terms in the action involving framons are strongly constrained in
form by its necessary double invariance under both the local gauge symmetry
$u(1) \times su(2) \times su(3)$ and its global dual $\tilde{u}(1) \times 
\widetilde{su}(2) \times \widetilde{su}(3)$, since physics should be independent
of the choice of both the local and the global reference frames.

\item From the action, using a method developed by 't~Hooft originally for the
electroweak theory \cite{tHooft}, one can deduce the mass matrices and couplings
at tree level of the $H$, $G$, and $F$.

\item Some of the freedom left over from the two preceding items can be tied
down further by the fit to quark and lepton data cited in \cite{tfsm}.

\end{itemize}
This programme will be carried out in some detail and reported in later sections.
From these, higher order corrections can then be calculated perturbatively.

One result of immediate interest from this programme is that it gives many
couplings among the framonic $C$-ons themselves but no direct coupling of
these particles to leptons and quarks.  Indeed, the framonic $C$-ons are found 
to couple to the light quarks and leptons which make up the present-day world
effectively only via the exchange of $\gamma$ and $Z$, and of the latter only
by virtue of some small admixture it has of framonic $C$-on.  Hence, in the 
absence of soft interactions for them as suggested above, it would seem that 
framonic $C$-ons will have difficulty communicating with our world, and will 
not be easy to produce or detect with our experiments done with matter made up 
mostly of light quarks and leptons.  If that is true, it would help to answer 
the question {\bf Q'} or {\bf Q} posed at the beginning.

But there is a surprise bonus.  This check for internal consistency of the FSM,
constructed initially just to explain the mass and mixing patterns of quarks and
leptons, has uncovered, in parallel to ours, a strange new world populated by
framonic $C$-ons.  Though seemingly quite as complex and active within itself 
as our own, this new world may be hidden from us because framonic $C$-ons have 
difficulty communicating with us and we with them.  Nevertheless, the two 
worlds came from the same roots and share the same vacuum.  We recall from 
\cite{tfsm} that it is the renormalization of the vacuum by colour framon loops 
which leads to the rotation of the quark and lepton mass matrices, and 
hence to the mixing of their up and down states and to their mass hierarchy. 
Conversely, as will be seen later, the fit in \cite{tfsm} to quark and lepton 
properties gives useful information, via the same RGE, on the mass spectrum and 
couplings of the framonic $C$-ons.

The strange new world exists in theory, but does it form part of our universe?
For this, we shall have to let our imagination loose.  Presumably, like other
coloured and flavoured constituents, framons would be present in abundance in 
the primordial soup.  As the universe cooled down and expanded, these coloured 
and flavoured objects would each struggle to find partners to neutralize their
colour or flavour so as to survive as colour or flavour neutral particles in
the confined phase.  In those primordial circumstances, the density would be
very high so that, one might suppose, even framons with their short
life times
would have no difficulty finding partners to emerge as flavour singlets (quarks,
leptons, and so on) and as colour singlets ($H$, $G$ and $F$). And these, being 
framonic, would be tenacious and hold on to their partners for good.  In the
meantime, of course, quarks in parallel would be meeting partners and combining
into baryons.  One might even argue that being binary composites, the framonic 
states, in particular the framonic $C$-ons of interest to us, would be formed 
in greater abundance than baryons which are trinary composites, since it would 
presumably be easier statistically for a framon to find one partner than a 
quark to find two partners.  The two worlds, ours of framonic $B$-ons and the 
new one of framonic $C$-ons, would then evolve separately, largely ignoring 
each other in the process.  The heavier states in our world have decayed into 
the lowest stable states, that is, the baryons and leptons.  In the other,
framonic $C$-on, world, the lowest states we have found so far are certain $H$, 
$G$ and $F$ with masses of order 10 MeV (which are a special result of the fit 
\ in Table 1), with some $F$ possibly lower still because of some see-saw 
mechanism \cite{seesaw}.   The lightest among them seem stable, with zero charge 
and little interaction with our sector.  Thus, if these conclusions are
maintained under further investigation, then those particles would be candidate 
constituents for dark matter.

However, these are early days yet, for the FSM gives a lot of information both 
in our standard sector and in the new hidden sector, which has to be checked 
through to ascertain, first, that it does not violate data already known and, 
secondly, whether some, and if so which pieces of it, can be tested against 
experiment.  Some of these points will be dealt with in this paper as they 
occur, and some in separate papers, but there are still many which will need 
further careful scrutiny, not only by us but by the community.  It is only if 
and when the FSM can manage to pass these tests will the scenario suggested 
above gain credibility, but this is a matter for the future.

Procedurally in this paper, one is dictated on by circumstances.  In venturing
into the hidden sector, where few empirical facts are known, one will have
to rely mainly on theoretical arguments.  Now in the applications we have made
so far of the FSM, there were some gaps in the formulation of the scheme which 
did not figure, and were therefore left open.  To proceed further, however,
these gaps will now have to be filled.  The first few sections of this paper
will thus be devoted to going over some old grounds with a finer comb, such as
the specifications of the framon fields and the structure of the three terms
in the action in which the framon occurs, namely the framon potential, the 
framon kinetic energy term, and the Yukawa coupling.  Moreover, we shall find 
this reappraisal highly rewarding, providing us, as it does, with both a deeper 
understanding of the facts and a clarification of the concepts than we have had 
before.  

In particular, while answering a question not posed earlier, because
not needed then,
 on the electric charges carried by the various components of the colour framon,
we came upon some new facts which have opened up for study a whole new 
phenomenology  for the FSM.  It is within the standard sector, but in an area outside 
that treated in for example  \cite{tfsm} for which the FSM was originally constructed.  This
comes about as follows.  We have already noted the close parallel between the
flavour and colour sectors in FSM, and also the analogy between the $W$-bosons 
in the one and the $G$ in the other.  Now, in the flavour theory, $W^3$ mixes
with the $u(1)$ boson to form the $\gamma$ and the $Z$.  So it is no surprise 
that in the colour theory, $G_8$  mixes with the $u(1)$ boson too.  In the
flavour sector, the charge chosen for  the Higgs (framon) field matters for keeping
the photon massless.  So, it answers to reason that in the colour theory also,
some particular choice of charges carried by the three components of the colour 
framon might do the same, as we shall show to be indeed the case in
Section 4.
Having then fixed the charges of the colour framon 
in this way, one is left with little freedom in the mixing problem.  Instead
of mixing just in the $\gamma-Z$ complex as in the electroweak theory, we have
now in FSM the mixed $\gamma-Z-G$ complex, where the matrix relating these
eigenstates of mass to the original gauge basis is given in terms of the gauge
couplings and of the vacuum expectation values of flavour and colour framon fields.  The couplings 
and the vacuum expectation value  of the Higgs field are known, leaving the problem then with only one
parameter, which we may take instead as the $G$ boson mass $m_G$.  Now such a
conclusion is phenomenologically very significant, given that the standard
mixing scheme in the electroweak theory has been tested already to very high
accuracy.  Deviations from it are easily ruled out, and any which survive 
would qualify as interesting new physics to be sought by experiment and be
tested.  Some of the resulting phenomenology on this has already been done
and reported in \cite{zmixed} which finds that the FSM has so far
survived the tests to which it has been subjected.
A summary of this can be found  in 
Section 7.3.  We consider this new 
phenonomenology an important bonus, since it provides us with the means
to test the model in a direction oblique to that for which it was
originally constructed.

Going next then into the hidden sector, one obtains from the framonic action 
the mass matrices and couplings of the $H$, $G$, and $F$.  Further, from the 
scale-dependence of these quantities deduced from the RGEs via the fit of 
\cite{tfsm}, one ends up with a fair amount of information on the mass spectra 
and interactions of the $H$, $G$, and $F$, which is what leads then, with a bit 
of guesswork, to the picture of the hidden sector outlined above.  

In the end section (Section 11), we have listed some points of interest
resulting from the present investigation, both experimental-phenomenological 
and theoretical-conceptual, at which the reader might wish to take a glance 
before getting involved with the details.

\section{The framon fields}

Being frame vectors to start with, framons carry, as do vierbeins in gravity, 
an index for the local gauge frame $r$ or $a$, as well as an index $\tilde{r}$ 
or $\tilde{a}$ for the global reference frame, and transform as representations
under both local gauge transformations and global changes of the reference 
frame.  Thus, if we denote the local gauge symmetry of SM as $G = u(1) \times 
su(2) \times su(3)$, and its corresponding (dual) global symmetry as; 
$\tilde{G} = \tilde{u}(1) \times \widetilde{su}(2) \times \widetilde{su}(3)$, 
then the framons should form together a representation of $G \times \tilde{G}$. 
The symmetries $G$ and $\tilde{G}$ being themselves product symmetries, one could 
choose for representations either the sum or the product for each pair.  In FSM, 
we choose \cite{efgt,dfsm} for the framon the representation ${\bf 1} \times 
({\bf 2} + {\bf 3})$ of $G$, this being the ``minimal representation'' in the 
sense that it requires the introduction of the smallest number of independent 
scalar fields into the theory, since counting dimensions, $1 \times 2 < 1 + 2, 
1 \times 3 < 1 + 3$, but $2 + 3 < 2 \times 3$.  But we choose for the framon 
the representation ${\bf 1} \times {\bf 2} \times {\bf 3}$ of $\tilde{G}$, 
to avoid  the flavour and colour sectors becoming completely disjoint.  With this 
choice of representations, the framon for the whole FSM breaks into two sets:
\begin{itemize}
\item {\bf (FF)} the flavour (``weak'') framon
\begin{eqnarray}
\balpha \Phi & = & \left( \alpha^{\tilde{a}} \phi_r^{\tilde{r}} \right); \ \  
   r = 1, 2; \ \ \tilde{r} = \tilde{1}, \tilde{2}; \ \ \tilde{a} = \tilde{1},
   \tilde{2}, \tilde{3}, \nonumber \\
y & = & \pm \half, \ \ \  \tilde{y} = -\third \mp \half
\label{fframon} 
\end{eqnarray}
\end{itemize}
where the columns of $\Phi$ transform as doublets of $SU(2)$ while its rows
transform as anti-doublets of $\widetilde{SU}(2)$, and:
\begin{itemize}
\item {\bf (CF)} the colour (``strong'') framon:
\begin{eqnarray}
\bbeta \bPhi & = & \left( \beta^{\tilde{r}} \phi_a^{\tilde{a}} \right); \ \  
   a = 1, 2, 3; \ \  \tilde{a} = \tilde{1}, \tilde{2}, \tilde{3}; \ \  
   \tilde{r} = \tilde{1}, \tilde{2}, \nonumber \\
y & = & -\third, \ \ + \twothirds; \ \ \ \tilde{y} = \pm \half + \third,\ \  
   \pm \half - \twothirds 
\label{cframon}
\end{eqnarray}
\end{itemize}
where the columns of $\bPhi$ transform as triplets of $SU(3)$ while its rows 
transform as anti-triplets of $\widetilde{SU}(3)$.   Each is multiplied by a 
spacetime independent factor\ \balpha\ and\ \bbeta, which can be taken as real
unit vectors without loss of generality.  Here we denote by $y$ and $\tilde{y}$ 
the chosen  $u(1)$ and $\tilde{u}(1)$ charges respectively, to be discussed a 
little further on.

For the flavour framon, the number of independent scalar fields can be reduced 
further by requiring the elements of $\Phi$ to satisfy the following 
\cite{efgt,dfsm}:
\begin{itemize}
\item {\bf (ME)} ``minimal embedding'' condition:
\begin{equation}
\phi_r^{\tilde{2}} = - \epsilon_{rs} (\phi_s^{\tilde{1}})^*. 
\label{minemb} 
\end{equation}
\end{itemize} 
We call this ``minimal embedding'' {\bf (ME}) since its implementation is akin 
to embedding\footnote{We can get a simple and clear picture of this embedding 
by identifying the elements of $SU(2)$ as unit quarternions, so that the whole 
group just sits inside $\bbr^4$ as its unit sphere.  In fact, this is a special
property of the group $SU(2)$ not shared by the other unitary groups.} the 
group $SU(2)$ in $\bbr^4$.  A similar condition, however, cannot be imposed on 
the colour framons without changing the physical dimension of some components 
with respect to the others.  The condition (\ref{minemb}) allows us to 
eliminate one of the two columns of $\Phi$, say $\bphi^{\tilde{2}}$, in terms of 
the other $\bphi^{\tilde{1}}$ (which we can call simply $\bphi$), giving then:
\begin{itemize}
\item {\bf (FF')} the flavour (``weak'') framon:
\begin{equation}
\balpha \bphi = \left( \alpha^{\tilde{a}} \phi_r \right);
    \ \ r = 1, 2; \ \  \tilde{a} = \tilde{1}, \tilde{2},
    \tilde{3},
\label{fframon'}
\end{equation}
\end{itemize}
and leaving only one independent doublet $\bphi$ to be identified with the 
Higgs field in the standard electroweak theory.

However, these minimality arguments for the choice of framon fields, whether of 
the representation or of the embedding, may be fortuitous, since they have been 
made with foreknowledge of what is needed for the standard model, and no 
physical reason is as yet known why the minimal choices are to be preferred.
Nevertheless, it seems interesting that such arguments do exist.  

We note that the condition {\bf (ME)} does not break the $\widetilde{su}(2)$ 
symmetry.  Indeed, even after the elimination of $\bphi^{\tilde{2}}$ in terms of 
$\bphi$, as in the traditional formulation, the theory still has this hidden 
$\widetilde{su}(2)$ symmetry, which is sometimes called an ``accidental'' 
symmetry (see equation (\ref{minemba})).  The only difference here is that this 
symmetry has been built into the theory as part of the framon concept, thus 
explaining in this context its origin.  To exhibit, in what follows, this 
underlying $\widetilde{su}(2)$ symmetry, it is often useful to write the 
flavour framon field in the form (\ref{fframon}), leaving the elimination by 
(\ref{minemb}) of one column in terms of the other understood to be performed 
later.

To complete the specification of framons as representations of $G \times 
\tilde{G}$, we have still to assign them $u(1)$ and $\tilde{u}(1)$ charges
$y$ and $\tilde{y}$.  For this, we shall need to specify not just the gauge 
algebra but also the gauge group\footnote{Here we would like to clarify our 
notation.  For many purposes, such as  at the level of the Lagrangian, one can 
neglect, or in fact cannot know of, any discrete identification of the group 
elements, meaning that we need, or are able, just to specify the corresponding 
Lie algebra.  However, if we want to study the charges, then we have to take 
into account discrete identifications of the group elements, meaning that we 
need to specify the Lie group itself.  To underline this point, in the first 
case we use lower case symbols, usually reserved for the Lie algebras, and in 
the second case we use upper case symbols, as is usual for the Lie  groups.}
\cite{Yang,ourbook}.  Analyses of the particle spectrum in the standard model 
give then the local gauge group as what we called in \cite{efgt} $U(3,2,1)$ 
(now renamed $U(1, 2, 3)$) which is obtained from the group $U(1) \times SU(2) 
\times SU(3)$ by identifying in it certain sextets of elements. Symbolically: 
$U(1, 2, 3) \sim \left(U(1) \times SU(2) \times SU(3)\right)/\,\bbz_6$, for which the $u(1)$ 
charges $y$ take on the following values:
\begin{eqnarray}
(1,1); \ \ \ y & = & 0 + k, \nonumber \\
(2,1); \ \ \ y & = & \half + l, \nonumber \\
(1,3); \ \ \ y & = & -\third + m, \nonumber \\
(2,3); \ \ \ y & = & \sixth +n,
\label{ycharges}
\end{eqnarray}
where the first number inside the brackets denotes the dimension of the $SU(2)$
representation, the second number that of $SU(3)$, and $k, l, m, n$ can be any 
integer, positive or negative. 

As in \cite{efgt,dfsm}, we choose to assign to the flavour framon the $U(1)$ 
charges $y = \pm \half$ with the smallest absolute values.  However, because 
of ``minimal embedding'' {\bf (ME)}, it follows that if $\bphi^{\tilde{1}}$ is 
assigned the charge $-\half$, then $\bphi^{\tilde{2}}$ must have charge $+\half$.
In other words, the global flavour $\widetilde{su}(2)$ symmetry is now broken 
along the direction $\tilde{1}$.  The breaking need not always be made along
$\tilde{1}$, which is after all only a co-ordinate choice, but can be along any 
direction specified by a vector, say  $\bgamma$, as we shall have occasion to
prefer.  In that case, introducing a further vector $\bgamma^\perp$ orthogonal to
$\bgamma$ in $\widetilde{su}(2)$ space, we rewrite {\bf (ME)} as:
\begin{equation}
\phi_r^\perp = -\epsilon_{rs}(\phi_s)^*,
\label{minemba}
\end{equation}
with $\bphi = \Phi\bgamma,\ \bphi^\perp = \Phi \bgamma^{\perp}$, which version 
is often useful for exhibiting more clearly the underlying symmetry.  We recall 
that in the confinement picture of 't~Hooft, $u(1)$ represents electromagnetism 
and $y$ the electric charge.  We can thus ascribe to the vector $\bgamma$ the 
function of specifying the direction in which electromagnetism breaks the 
global symmetry $\widetilde{su}(2)$.

The assignment of $y$ to the colour framons {\bf (CF)} is less straightforward. 
In \cite{dfsm,tfsm}, we have tentatively assigned the common charge  $-\third$ 
to all colour framons, this being the smallest in absolute value allowed them 
by (\ref{ycharges}), and we not knowing then any reason for doing otherwise.  
But that was a mistake.  It will be seen later in Section 4 that one of the three 
colour framons must be assigned  a charge $+ \tfrac{2}{3}$ to keep the photon 
massless.  Correction of this error does not change our earlier result in for 
example \cite{tfsm} which did not make use of the wrong charge assignments, 
but will lead to interesting physics in other areas, as will be spelt out in 
Section 7.3.

Lastly, we come to the $\tilde{u}(1)$ charges $\tilde{y}$.  For the factors 
$\Phi$ and $\bPhi$ in respectively {\bf (FF)} and {\bf (CF)}, just as their 
representations in the nonabelian parts of $G$ and $\tilde{G}$ are conjugates,  
so should their representations of $\tilde{u}(1)$ be the conjugate of their 
representations in $u(1)$, meaning that $\tilde{y} = - y$.  For $\Phi$ in 
{\bf (FF)}, this gives then $\tilde{y} = +\half$ for $\bphi^{\tilde{1}}$ but 
$\tilde{y} = -\half$ for $\bphi^{\tilde{2}}$.  Similarly, when we have specified 
in Section 4 separate $y$ for the 3 components of $\bPhi$ in {\bf (CF)}, each 
component will have to be assigned separately the value $\tilde{y} = - y$.
There remains, however, a question on whether the global factors $\balpha$ in 
{\bf (FF)} and $\bbeta$ in {\bf (CF)} should be assigned a $\tilde{y}$ too, and
if so, which value.

To answer this question,  it is again necessary to specify  the group 
corresponding to the global symmetry $\tilde{G} = \tilde{u}(1) \times 
\widetilde{su}(2) \times \widetilde{su}(3)$.   A reasonable choice would 
seem to be again $U(1, 2, 3)$.  One might even argue that it is necessary 
that the local and global groups be the same, as follows.  Any infinitesimal 
change of the global reference frame under the symmetry $\tilde{G}$ can be 
counteracted by the opposite change in the local frame under $G$.  The 
suggestion above that the global group is the same as the local group is 
basically just the extension of the said requirement even to finite changes.  
If so, then the list in (\ref{ycharges}) implies that the $\widetilde{su}(3)$ 
triplet $\balpha$ should be assigned a $\tilde{y} = -\third$, and the 
$\widetilde{su}(2)$ doublet $\bbeta$ a $\tilde{y} = \pm \half$.  These 
assignments are then what give the values of $\tilde{y}$ listed in 
(\ref{fframon}) and (\ref{cframon}).

Just as $u(1)$ invariance leads to charge conservation, so does $\tilde{u}(1)$
invariance lead to the conservation of the $\tilde{y}$ charge, which is in FSM
connected to baryon number and lepton number conservation.  The details of how
they are connected will be postponed to Section 7.2 after Yukawa couplings have
been considered. 

These assignments of the $y$ and $\tilde{y}$ charges complete then the
specification of the framon fields.  We note that the framons are frame vectors 
only in the internal symmetry space, and transform under internal symmetry 
operations as indicated.  They are, on the other hand, invariant under proper 
Lorentz transformations, and are therefore spacetime scalar fields.

\section{The framon self-interaction potential} 

The self-interaction potential of framons is required to be invariant under
$G \times \tilde{G}$.  Including up to quartic terms for renormalizability, 
and contracting indices in all possible ways, one can construct \cite{efgt} 
an invariant potential thus:
\begin{eqnarray}
V & =
      & - \mu'_W \tr[\Phi^\dagger \Phi] + \lambda'_W (\tr[\Phi^\dagger \Phi])^2
        + \kappa_W \tr[\Phi^\dagger \Phi \Phi^\dagger \Phi] \nonumber \\
  &   & - \mu_S \Tr[\bPhi^\dagger \bPhi] + \lambda_S (\Tr[\bPhi^\dagger \bPhi])^2
        + \kappa_S \Tr[\bPhi^\dagger \bPhi \bPhi^\dagger \bPhi] \nonumber \\
  &   & + \nu'_1 \tr[\Phi^\dagger \Phi] \Tr [\bPhi^\dagger \bPhi]
        - \nu'_2 [(\Phi \bbeta)^\dagger \cdot (\Phi \bbeta)][(\bPhi \balpha)^\dagger
          \cdot (\bPhi \balpha)].
\label{V'}
\end{eqnarray}
This is seen to be invariant under $su(2) \times su(3) \times \widetilde{su}(2) 
\times \widetilde{su}(3)$.  However, to show that this is in fact invariant 
under $u(1) \times \tilde{u}(1)$ as well, we shall have to leave until Section 
4, after we have specified the assignment of the corresponding charges to the 
various components of the colour framon. 

We note in particular the $\nu'_1, \nu'_2$ terms linking the flavour and colour 
framons, especially the $\nu'_2$ term, which breaks the global flavour symmetry 
$\widetilde{su}(2)$ explicitly via the vector $\bbeta$ coming from the colour 
framon (\ref{cframon}), and also the global colour symmetry $\widetilde{su}(3)$ 
explicitly via the vector $\balpha$ from the flavour framon (\ref{fframon}).

Eliminating one of the columns of $\Phi$ in terms of the other using 
(\ref{minemb}) according to the minimal embedding {\bf (ME)} of $SU(2)$, 
we have:
\begin{eqnarray}
V & = & - \mu_W |\bphi|^2 + \lambda_W (|\bphi|^2)^2 \nonumber \\
  &   & - \mu_S \sum_{\tilde{a}} |\bphi^{\tilde{a}}|^2
        + \lambda_S \left( \sum_{\tilde{a}} |\bphi^{\tilde{a}}|^2 \right)^2
        + \kappa_S \sum_{\tilde{a} \tilde{b}} |\bphi^{\tilde{a}*}.\bphi^{\tilde{b}}|^2 
          \nonumber \\
  &   & + \nu_1 |\bphi|^2 \sum_{\tilde{a}}|\bphi^{\tilde{a}}|^2  
        - \nu_2 |\bphi|^2 \left( \sum_{\tilde{a}} \alpha^{\tilde{a}} 
          \bphi^{\tilde{a}} \right)^\dagger 
          . \left( \sum_{\tilde{a}} \alpha^{\tilde{a}} \bphi^{\tilde{a}} \right),
\label{V}
\end{eqnarray}
with the new unprimed parameters simply given in terms of the primed parameters
in (\ref{V'}).  The $\kappa_W$ term is absorbed into the $\lambda_W$ term 
leaving just the usual Mexican hat potential for the electroweak sector.  We
note in particular, that:
\begin{itemize}
\item {\bf (ME')} the dependence of the $\nu_2$ term on $\bbeta$ drops out, 
\end{itemize}
while its dependence on $\balpha$ remains, which fact will be seen to be
important.  For this resaon, it is worthwhile making explicit how this comes 
about.  The term in (\ref{V'}) under consideration is of the form: 
\begin{equation}
[\beta^{\tilde{1}} \bphi^{\tilde{1}} + \beta^{\tilde{2}} \bphi^{\tilde{2}}]^\dagger
  \cdot[\beta^{\tilde{1}} \bphi^{\tilde{1}} + \beta^{\tilde{2}} \bphi^{\tilde{2}}],
\label{betaphi}
\end{equation}
where $\bphi$ is here a vector in $su(2)$ space and the dot denotes the inner 
product between such vectors.  Now the minimal embedding condition {\bf (ME)} 
in (\ref{minemb}), implies:
\begin{equation}
[\bphi^{\tilde{2}}]^\dagger\cdot\bphi^{\tilde{1}} = 0, \ \ \ 
|\bphi^{\tilde{1}}| = |\bphi^{\tilde{2}}| = |\bphi|^2,
\label{minembb}
\end{equation}
so the expression (\ref{betaphi}) becomes:
\begin{equation}
|\beta^{\tilde{1}}|^2|\bphi^{\tilde{1}}|^2 + |\beta^{\tilde{2}}|^2|\bphi^{\tilde{2}}|^2
  = [|\beta^{\tilde{1}}|^2 + |\beta^{\tilde{2}}|^2] |\bphi|^2 = |\bphi|^2.
\label{betaphia}
\end{equation}
Thus one sees that because of minimal embedding, the vector $\bbeta$ drops out
of the $\nu_2$ term, and indeed out of the framon potential altogether.  Hence
the vacuum obtained from the potential will not depend on $\bbeta$.

We recall that in (\ref{V'}), it was $\bbeta$ which broke explicitly the
$\widetilde{su}(2)$ symmetry.  Now, after minimal embedding, $\bbeta$ has
dropped out, but the symmetry remains broken, this time in the direction 
$\bphi^{\tilde{1}}$, or more generallly along the vector $\bgamma$.
We can say that it is the vector $\bgamma$ which prescribes the direction
of the $u(1)$ of electromagnetism that breaks the $\widetilde{su}(2)$ of the
framon potential.  However, the $\widetilde{su}(3)$ is still broken by the
vector $\balpha$, the colour equivalent of $\bbeta$ in the flavour sector.  
As we shall see, this particular lack of parallel between the flavour and colour
sectors, due directly to the special minimal embedding property of $SU(2)$,
is the source of much of the difference in physics to emerge from the FSM in
the two sectors.  

The vacuum is found by minimizing $V$, where the coefficients of the 7 terms
are by choice all positive.  In particular $\mu_W, \mu_S$ being positive means 
that the minimum is degenerate in both the flavour and colour sectors.  In the 
flavour sector, this degeneracy is the same as in the standard electroweak 
theory.  For the colour sector, it is found that any vacuum within the 
degenerate set can be cast by an appropriate choice of gauges (both local and 
global) into the following diagonal form:
\begin{equation}
\phivac \longrightarrow \zeta_S V_0 
   = \zeta_S \left( \begin{array}{ccc} Q & 0 & 0 \\
                                       0 & Q & 0 \\
                                       0 & 0 & P 
                            \end{array} \right),
\label{Phivac0}
\end{equation}
with:
\begin{eqnarray}
P & = & \sqrt{\third(1 + 2R)}, \\ 
Q & = &\sqrt{\third(1 - R)}, \\ 
R & = & \frac{\nu_2 \zeta_W^2}{2 \kappa_S \zeta_S^2}, 
\label{PQRzetaW}
\end{eqnarray}
where $\zeta_W$ and $\zeta_S$ are the vacuum expectation values of the flavour 
and colour framons respectively, and  $\balpha$, which is coupled to the 
vacuum, takes the form:
\begin{equation}
\balpha \longrightarrow \balpha_0 
   = \left( \begin{array}{c} 0 \\ 0 \\ 1 \end{array} \right).
\label{alpha0}
\end{equation}
In other words, it is the vector $\balpha$ coming from the flavour framon
(\ref{fframon}) which, we recall, broke ``explicitly'' the global colour 
symmetry of the $\nu_2$ term in $V$, that now gives a special direction 
$\tilde{3}$ to the vacuum (\ref{Phivac0}) in the chosen gauge, which, however, 
still remains degenerate in the other two (the  $\tilde{1}$ and $\tilde{2}$) 
directions.

Expanding the potential $V$ in fluctuations about the vacuum (\ref{Phivac0}),
thus: $\bPhi \rightarrow \phivac + \delta \bPhi$, with $\delta \bPhi$ 
hermitian\footnote{When $\delta \bPhi$ is antihermitian, it
  corresponds to only
a gauge transformation which is used to fix the gauge \cite{dfsm}, or, in more 
colourful language, it represents a degree of freedom which is eaten up by a 
colour gauge boson to acquire a mass.}, one obtains in this gauge the mass 
squared matrix, and the couplings to one another, of quanta we call generically 
$H$, these being, as mentioned in the introduction, the analogues in the colour 
sector of the Higgs boson $h_W$.  In 't~Hooft's confinement picture, these $H$ 
appear as bound states of a framon-antiframon pair: $\bPhi^\dagger \bPhi 
\sim \phivac^\dagger (\phivac + \delta \bPhi) = \phivac^\dagger \phivac 
+ \phivac^\dagger \delta \bPhi$, meaning therefore that an $H$ is to be labelled 
by $\zeta_S^{-1} \phivac^\dagger$ times a hermitian matrix.  As in \cite{tfsm}, we adopt as 
basis the following matrices: 
\begin{eqnarray}
V_1 & = &  \left( \begin{array}{rrr} 1 & 0 & 0 \\
                                                       0 & 0 & 0 \\
                                                       0 & 0 & 0
                                    \end{array} \right) \nonumber \\
V_2 & = &  \left( \begin{array}{rrr} 0 & 0 & 0 \\
                                                       0 & 1 & 0 \\
                                                       0 & 0 & 0
                                    \end{array} \right) \nonumber \\
V_3 & = &  \left( \begin{array}{rrr} 0 & 0 & 0 \\
                                                       0 & 0 & 0 \\
                                                       0 & 0 & 1
                                    \end{array} \right) \nonumber \\
V_4 & = & \frac{1}{\sqrt{2}} \left( \begin{array}{rrr} 0 & 1 & 0 \\
                                                       1 & 0 & 0 \\
                                                       0 & 0 & 0
                                    \end{array} \right) \nonumber \\
V_5 & = & \frac{i}{\sqrt{2}} \left( \begin{array}{rrr} 0 & -1 & 0 \\
                                                      1 & 0 & 0 \\
                                                       0 & 0 & 0
                                    \end{array} \right) \nonumber \\
V_6 & = & \frac{1}{\sqrt{(P^2+Q^2)}} \left( \begin{array}{rrr}
                                                       0 & 0 & 0 \\
                                                       0 & 0 & Q \\
                                                       0 & P & 0
                                    \end{array} \right) \nonumber \\
V_7 & = & \frac{i}{\sqrt{(P^2+Q^2)}} \left( \begin{array}{rrr} 
                                                       0 & 0 & 0 \\
                                                       0 & 0 & -Q \\
                                                       0 & P & 0
                                    \end{array} \right) \nonumber \\
V_8 & = & \frac{1}{\sqrt{(P^2+Q^2)}} \left( \begin{array}{rrr}
                                                       0 & 0 & Q \\
                                                       0 & 0 & 0 \\
                                                       P & 0 & 0
                                    \end{array} \right) \nonumber \\
V_9 & = & \frac{i}{\sqrt{(P^2+Q^2)}} \left( \begin{array}{rrr}
                                                       0 & 0 & -Q \\
                                                       0 & 0 & 0 \\
                                                       P & 0 & 0
                                    \end{array} \right) \nonumber \\.
\label{VK}
\end{eqnarray}
giving a tree-level mass (squared) matrix which is almost diagonal:
\begin{equation}
M_{H}=
\pmatrix{
4\lambda_{W}\zeta_{W}^{2} & 2\zeta_{W}\zeta_{S}(\nu_{1}-\nu_{2})
\sqrt{\frac{1+2R}{3}} &
2\sqrt{2}\zeta_{W}\zeta_{S}\nu_{1}\sqrt{\frac{1-R}{3}} & 0 \cr
\ast & 4(\kappa_{S}+\lambda_{S})\zeta_{S}^{2}\left(\frac{1+2R}{3}\right) & 
4\sqrt{2}\lambda_{S}\zeta_{S}^{2}\frac{\sqrt{(1+2R)(1-R)}}{3} & 0 \cr
\ast & \ast &
4(\kappa_{S}+2\lambda_{S})\zeta_{S}^{2}\left(\frac{1-R}{3}
\right) & 0 \cr
0 & 0 & 0 & D
}
\label{MH}
\end{equation}
where 
\begin{equation}
D=\kappa_{S}\zeta_{S}^2
\pmatrix{
4(\frac{1-R}{3})& 0 & 0 & 0 & 0 & 0 & 0 \cr
0 & 4(\frac{1-R}{3})& 0 & 0 & 0 & 0 & 0 \cr
0 & 0 & 4(\frac{1-R}{3})& 0 & 0 & 0 & 0 \cr
0 & 0 & 0 & 2(\frac{2+R}{3}) & 0 & 0 & 0 \cr
0 & 0 & 0 & 0 & 2(\frac{2+R}{3}) & 0 & 0 \cr
0 & 0 & 0 & 0 & 0 & 2(\frac{2+R}{3}) & 0 \cr
0 & 0 & 0 & 0 & 0 & 0 &2(\frac{2+R}{3})   \cr
}
\end{equation}
In (\ref{MH}), the rows and columns are labelled, for later convenience, by
respectively: $h_W, H_3, H_+ = \sqrt{\half} (H_1 + H_2), 
H_- = \sqrt{\half}(H_1 - H_2), H_4, H_5,H_6, H_7$, $H_8, H_9 $, in that 
order.  

Expanding further the potential $V$ in fluctuations about the vacuum, we get 
the tree-level couplings of these states with one another given in Appendix A.

\section{The framon kinetic energy term}

Recalling that our framon field is chosen by minimality arguments to be in the 
representation ${\bf 1 \times (2 + 3) }$ of $u(1) \times su(2) \times su(3)$, 
we can write its kinetic energy as a sum of two terms.  For the flavour framon 
we have:
\begin{equation}
{\cal A}_{\rm KE}^{\rm F} = \tr[(D_\mu \Phi)^\dagger D_\mu \Phi],
\label{KEf}
\end{equation}
with
\begin{equation}
D_\mu = \partial_\mu - i g_1 \Gamma A_\mu - \half ig_2 B_\mu, 
\label{Dmuf}
\end{equation}
and for the colour framon:
\begin{equation}
{\cal A}_{\rm KE}^{\rm C} = \Tr[(D_\mu \bPhi)^\dagger D_\mu \bPhi],
\label{KEc}
\end{equation}
with
\begin{equation}
D_\mu = \partial_\mu - i g_1 \Gamma A_\mu -\half ig_3 C_\mu. 
\label{Dmuc}
\end{equation}
In both cases, $\Gamma$ is a charge operator which specifies the $u(1)$ charge
of the various components of the framon.  This can depend only on the global 
indices $\tilde{r}$ or $\tilde{a}$, not on the local indices $r$ or $a$, since 
both local flavour and colour are confined and have to remain exact.  Thus 
$\Gamma$ can be taken as a matrix in $\widetilde{su}(2)$ for the flavour framon 
or in $\widetilde{su}(3)$ for the colour framon.\footnote{Hence, when applied 
as in (\ref{KEf}) on $\Phi$ or in (\ref{KEc}) on $\bPhi$, the rows of which are 
labelled by local indices flavour $r$ or colour $a$ and columns are labelled by 
dual flavour $\tilde{r}$ or dual colour $\tilde{a}$, $\Gamma$ as a matrix has
to operate from the right.}

Let us  first examine the expression (\ref{KEf}) for the flavour framon, 
recalling that $\bphi^{\tilde{1}}$ has charge $-\half$, and $\bphi^{\tilde{2}}$
charge $+\half$.  Eliminating then by {\bf (ME)} $\bphi^{\tilde{2}}$ in 
(\ref{KEf}) in terms of $\bphi = \bphi^{\tilde{1}}$, one has:
\begin{equation}
{\cal A}_{\rm KE}^{\rm F} = 2 [(D_\mu \bphi)^\dagger D_\mu] \bphi,
\label{KEfa}
\end{equation}
with
\begin{equation}
D_\mu = \partial_\mu + \half i g_1 A_\mu -\half ig_2 B_\mu, 
\label{Dmufa}
\end{equation}
differing from the standard electroweak theory only by a harmless factor 2 
which we shall henceforth neglect.

Proceeding from (\ref{KEfa}) to derive the masses of the vector bosons 
$ \gamma, Z^0, W^\pm $ in the usual symmetry-breaking picture is familiar.  Let 
us repeat the derivation, however, in the confinement picture of 't~Hooft, 
which we wish later to apply to the colour sector.  

In the $su(2)$ theory, there are three local gauge degrees of freedom, which 
we can use to fix the gauge by rotating, with an $SU(2)$ transformation 
$\Omega(x)$, the doublet scalar field $\bphi$ (contaning 4 parameters) to 
point, at every space-time point $x$, in the first direction and to be real, 
thus:
\begin{equation}
\bphi = \Omega \left( \begin{array}{c} \rho \\ 0 \end{array} \right)
      = \Omega \bphi_{\rm GF},
\label{Omega}
\end{equation}
with $\rho$ real.  For a theory with the usual Mexican hat potential in this 
sector, such as (\ref{V}) above, we can write:
\begin{equation}
\rho = \zeta_W + h_W,
\label{rho}
\end{equation}
where $\zeta_W$ is the vacuum expectation value of $\phi$, and $h_W$, as its 
fluctuation about the vacuum value, is the Higgs boson field.

Since to zeroth order $\rho=\zeta_W$, we can rewrite to this order the KE term 
of $\bphi$ in (\ref{KEfa}) as:
\begin{equation}
[D_\mu \bphi]^\dagger [D_\mu \bphi] = [D_\mu \Omega \bphi_{\rm GF}]^\dagger 
   [D_\mu \Omega \bphi_{\rm GF}]  = \bphi_{\rm GF}^\dagger ( D_\mu \Omega)^\dagger
   \Omega \Omega^\dagger D_\mu \Omega \bphi_{\rm GF},
\label{KEphi1}
\end{equation}
and, by introducing
\begin{equation}
\half \tilde{B}_\mu = \frac{i}{g_2} \Omega^\dagger (\partial_\mu -
\half i g_2 B_\mu) \Omega
\label{Btilde}
\end{equation}
as:
\begin{equation}
[D_\mu \bphi]^\dagger [D_\mu \bphi] = \bphi_{\rm GF}^\dagger [+\half ig_1  A_\mu 
  -\half ig_2\tilde{B}_\mu]^\dagger [ +\half ig_1  A_\mu -\half ig_2
  \tilde{B}_\mu] \bphi_{\rm GF}.
\label{KEphi2}
\end{equation}
To leading order then, this gives for the mass term, $\tilde{B}_\mu$ being 
hermitian:
\begin{equation}
(\zeta_W, 0) \quart [g_1^2 A_\mu^2 + g_2^2 \tilde{B}_\mu^2
   - 2 g_1 g_2 A_\mu \tilde{B}_\mu] \left( \begin{array}{c} \zeta_W \\ 0 
   \end{array} \right),
\label{KEphi3}
\end{equation}  
that is, $\quart \zeta_W^2$ times the 11 element of the quantity inside the square 
brackets \cite{banks}.

Next, we note that $\tilde{B}_\mu$ as defined in (\ref{Btilde}) is just:
\begin{equation}
\tilde{B}_\mu  = \Omega^\dagger  B_\mu \Omega + \frac{2i}{g_2} \Omega^\dagger 
     \partial_\mu \Omega,
\label{Btilde2}
\end{equation}
namely the gauge transform of $B$ under $\Omega$.   Then, since $\tr[G^{\mu\nu} 
G_{\mu\nu}]$ is a gauge invariant quantity, it follows that:
\begin{equation}
\tr[G^{\mu\nu} G_{\mu\nu}] = \tr[\tilde{G}^{\mu\nu}
\tilde{G}_{\mu\nu}],
\label{KEBmu}
\end{equation} 
where
\begin{equation}
\tilde{G}_{\mu\nu} = \partial_\nu \tilde{B}_\mu - \partial_\mu \tilde{B}_\nu
    +ig_2 [\tilde{B}_\mu, \tilde{B}_\nu].
\label{Gtilde}
\end{equation}
Hence, the action in terms of $\tilde{B}$ is exactly the same as the action 
obtained by the symmetry-breaking picture in terms of $B$, with a mass
squared matrix worked out from (\ref{KEphi3}) as again
\begin{equation}
 \quart \zeta_W^2\left( \begin{array}{cccc} g_2^2 & 0 & 0 & 0 \\
                           0 & g_2^2 & 0 & 0 \\
                           0 & 0 & g_2^2 & -g_1 g_2 \\
                           0 & 0 & -g_1 g_2 & g_1^2 
                           \end{array} \right),
\label{massmatew}
\end{equation}
with only the $\tilde{B}^3$ component mixing with the photon giving:
\begin{equation}
\gamma_\mu = \frac{1}{\sqrt{g_1^2 + g_2^2}} 
   [g_2 A_\mu + g_1 \tilde{B}^3_\mu],
\label{gammamu}
\end{equation}
\begin{equation}
Z_\mu = \frac{1}{\sqrt{g_1^2 + g_2^2}} 
   [- g_1 A_\mu + g_2 \tilde{B}^3_\mu],
\label{Zmu}
\end{equation}
while
\begin{equation}
W_\mu^{\pm} = \tilde{B}_\mu^{\pm}.
\label{Wmu}
\end{equation}

Although the result is the same as in the symmetry-breaking picture and even
most of the algebra used in its derivation looks familiar, one gains a very
different physical interpretation to the phenomenon.  We note first that the
massive vector bosons $W^\pm, Z$ are no longer given by the triplet of gauge 
fields $B_\mu$ of the local $su(2)$ symmery, but by the fields $\tilde{B}_\mu$
in (\ref{Btilde}).  These are $su(2)$ singlets, with all the local $su(2)$
indices originally in $D_\mu$ saturated by those in the transformation
matrices $\Omega$ and $ \Omega^{-1}$ between which it is sandwiched.  Now 
$\Omega$ is a matrix which transforms the local $su(2)$ frame to a fixed global 
frame.  Hence, the fields $\tilde{B}_\mu$ are $su(2)$ scalars, only triplets in 
a new global symmetry which we may call $\widetilde{su}(2)$.  Indeed, $\Omega$ 
as the transformation matrix between the local and global frames can also be 
taken as the vacuum expectation value of the framon matrix $\Phi$ (See Section 
2).  For this reason, the $\tilde{B}_\mu$ in (\ref{Btilde}) are 
 interpreted by 't~Hooft as a bound state of a 
$\Phi$-$\Phi^\dagger$ pair in ``$p$-wave'' (because of the $D_\mu$) formed by 
local $su(2)$ confinement.  In other words, in our adopted language here, they 
are (framonic) $B$-ons.  And they have acquired masses, as bound states usually 
do.  Now $\tilde{B}^3_\mu$ mixes with $A_\mu$ to form the $Z^0$, but this mixing 
is just the usual mixing between quantum states, and involves no mixing of the 
local gauge symmetries $u(1)$ and $su(2)$ as it is said to do in the 
symmetry-breaking picture.  It breaks the original degeneracy between 
$\tilde{B}^3$ and $\tilde{B}^\pm$, but this breaks only the global symmetry 
$\widetilde{su}(2)$, while the local $su(2)$, being confining, remains exact.

We note as an aside that, just as for the framon potential, the $\tilde{su}(2)$ 
can be broken in the $\tilde{1}$ direction as above, or along a vector 
$\bgamma$.  What really breaks the symmetry is again the $u(1)$ of 
electromagnetism.

We turn next to the kinetic energy term for the colour framons.  Since colour
is by general consensus confining and the kinetic energy term very similar to 
that in the flavour case, the above treatment in 't~Hooft's confinement picture
would seem to be tailor-made for it.  As in the flavour case, we wish first to 
gauge-fix to a convenient frame, but now with only 8 degrees of freedom in 
$su(3)$ but many more parameters in the colour framon field $\bPhi$, we can 
at best only make $\Phi$ triangular or hermitian.  In either case, we shall 
call it $\bPhi_{\rm GF}$, and write:
\begin{equation}
\bPhi = \Omega \bPhi_{\rm GF}.
\label{PhiGF}
\end{equation}
Since to zeroth order $\bPhi_{\rm GF}=\phivac$, 
we can proceed with the same manipulations as in the flavour sector.
Thus the kinetic energy is given to this order by
\begin{equation}
 \!\!\!\!\!\!\!\!\!\Tr [\bPhi_{\rm GF}^\dagger (D_\mu \Omega)^\dagger  \Omega \Omega^\dagger
            D_\mu \Omega \bPhi_{\rm GF}]
     = \Tr [\bPhi_{\rm GF}^\dagger(\!-ig_1 \Gamma A_\mu -\half i g_3 \tilde{C}_\mu)^\dagger 
       (\!-ig_1 A_\mu \Gamma -\half i g_3 \tilde{C}_\mu) \bPhi_{\rm GF}],
\label{KECt}
\end{equation}
with:
\begin{equation}
\half \tilde{C}_\mu = \frac{i}{g_3} \Omega^\dagger (\partial_\mu
-\half  ig_3 C_\mu) \Omega.
\label{Ctilde}
\end{equation}
In deriving the result in (\ref{KECt}), we have used the fact that $\Gamma$, 
being a matrix in $\widetilde{su}(3)$ space with both rows and columns labelled 
by $\tilde{a}$ indices, is not affected by sandwiching $A_\mu \Gamma$ between 
the (local) gauge fixing transformation $\Omega^\dagger \cdots \Omega$.  And 
since $A_\mu$ itself is proportional to the identity in colour, one has just:
\begin{equation}
\Omega^\dagger D_\mu \Omega \, = \, -ig_1 \Gamma A_\mu -\half ig_3 \tilde{C}_\mu.
\label{Dmun}
\end{equation}

Expanding next $\tilde{C}_\mu$ in terms of the Gell-Mann matrices, we have:
\begin{equation}
\tilde{C}_\mu \, = \,  \sum_K \tilde{C}_\mu^K \lambda_K.
\label{GK}
\end{equation}
These represent then our framonic vector $C$-ons, which we shall call $G$
generically.

To find the mass matrix of the $G$ from (\ref{KECt}), we substitute as 
usual for $\bPhi_{\rm GF}$ its vacuum expectation value, in the global 
$\widetilde{su}(3)$ gauge where it is diagonal:
\begin{equation}
\bPhi_{\rm GF} \rightarrow \frac{\zeta_S}{\sqrt{3}} \left( \begin{array}{ccc}
         \sqrt{1 - R} & 0 & 0 \\
         0 & \sqrt{1 - R} & 0 \\
         0 & 0 & \sqrt{1 + 2R} \end{array} \right) = \phivac,
\label{Phivac}
\end{equation}
and obtain the mass term as:
\begin{equation}
\Tr \left(\phivac^2 
    (-ig_1 \Gamma A_\mu -\half i g_3 \tilde{C}_\mu)^\dagger
    (-ig_1 \Gamma A_\mu -\half i g_3 \tilde{C}_\mu) \right).
\label{MG}
\end{equation}
or
\begin{equation}
\Tr \left(\phivac^2[g_1^2 A^\mu A_\mu \Gamma^\dagger \Gamma
   + \half g_1 g_2 A_\mu \sum_K \tilde{C}_\mu^K (\Gamma^\dagger \lambda_K 
       + \lambda_K \Gamma)
   + \quart \sum_K \tilde{C}_\mu^K \lambda_K \sum_J \tilde{C}_\mu^J \lambda_J] 
     \right).
\label{MGa}
\end{equation}

We now have to specify $\Gamma$ or, in other words, the charges of the framons 
labelled by the column index $\tilde{1}, \tilde{2}, \tilde{3}$ of the matrix 
$\bPhi$.  We recall that the framon charges are constrained to be those listed
in (\ref{ycharges}) for representations of $U(1, 2, 3)$, with preference for
the lowest values as being more fundamental.  Beyond this, in contrast to the flavour 
case where the condition {\bf (ME)} of (\ref{minemb}) gives a unique choice for 
the framon charges, the choice here for colour is for the moment still open.  
Now in \cite{dfsm}, it was suggested tentatively for simplicity that one gives 
to all 3 colour framon components the same minimal value $-\third$ allowed by 
(\ref{ycharges}), but it will be shown later that this leads to a mass matrix 
with no zero mode, giving thus the photon a mass, which would be physically 
unacceptable.  

However, we recall from (\ref{ycharges}) that what the gauge group $U(1,2,3)$ 
implies is only that the strong framon (colour triplet, weak $su(2)$ singlet) 
should have $u(1)$ (electric) charge $-\third + n$, $n$ being any integer, 
positive or negative.  There is no need as far as the group representation is 
concerned for all colour framons to have the same charge. 
In  fact, in the flavour sector the two framons have opposite charges 
and this  seems to have played a role in giving the mass matrix a zero 
mode, hence keeping the photon massless.  The question is thus whether we can 
devise a similar arrangement here in the colour sector.  We suggest the
following, namely that $\bphi^{\tilde{1}, \tilde{2}}$ have charge $-\third$ but that
$\bphi^{\tilde{3}}$ has charge $+\twothirds$,  both allowable as 
 $U(1,2,3)$ representations.  We have kept the 
charges the same for $\bphi^{\tilde{1}}$ and $\bphi^{\tilde{2}}$ because of the 
$\widetilde{su}(2)$ symmetry residual in our theory, and arranged the total 
charge of all three framons to be zero, in parallel to the opposite charges of 
the two framons in the electroweak theory.

Suppose we do that, what will happen?  In (\ref{KECt}), the charge operator 
$\Gamma$ now takes the form:
\begin{equation}
\Gamma = \left( \begin{array}{ccc} -\third & 0 & 0 \\
                                    0 & -\third & 0 \\
                                    0 & 0 & +\twothirds \end{array} \right),
\label{Gamma3}
\end{equation}
Evaluating now the mass matrix as per (\ref{MGa}), we have, first, from the 
$\tilde{C}^\mu \tilde{C}_\mu$ term, no nonzero crossed terms, leaving thus only 
the diagonal $(\tilde{C}^K)^2$ contributions as:
\begin{eqnarray}
\quart g_3^2 \Tr[\phivac^2 \lambda_K^2] = \tfrac{1}{6} g_3^2 \zeta_S^2 (1-R) 
    &  \ \  & {\rm for \ \  K = 1,2,3},\\
\quart g_3^2 \Tr[\phivac^2 \lambda_K^2] = \tfrac{1}{12} g_3^2 \zeta_S^2 (2+R)  
    &  \ \  & {\rm for \ \ K = 4,5,6,7},
\label{CCterm}
\end{eqnarray}
and for $K =8$
\begin{equation}
\quart g_3^2 \Tr [\phivac^2 \lambda_8^2] = \sixth g_3^2 \zeta_S^2 (1+R).
\end{equation}
Second, from the $A^\mu \tilde{C}_\mu$ term, only the $K = 8$ term gives nonzero
contribution:
\begin{equation}
\half g_1g_3 \Tr [\phivac^2 (\Gamma^\dagger \lambda_8 
   + \lambda_8 \Gamma)] = -\frac{2}{3 \sqrt{3}} g_1 g_3 \zeta_S^2 (1+R).
\label{ACtermo}
\end{equation}
Third, from the $A_\mu A_\mu$ term, we obtain:
\begin{equation}
g_1^2 \Tr[\phivac^2 \Gamma ^\dagger \Gamma]
   = \frac{2}{9} g_1^2 \zeta_S^2 (1+R).
\label{AAtermo}
\end{equation} 
The mass matrix so obtained is almost diagonal except for a mixing between  
the photon and $G_8$, with a mass submatrix of the form:
\begin{equation}
\sixth (1 + R) \zeta_S^2 \left( \begin{array}{cc} 
                             \frac{4}{3} g_1^2 & -\frac{2}{\sqrt{3}} g_1 g_3 \\
                                   -\frac{2}{\sqrt{3}} g_1 g_3 & g_3^2 
                                   \end{array} \right).
\label{Masssubmn}
\end{equation}
This matrix has a zero mode, meaning that $G_8$, though mixing with $A_\mu$,
will leave the photon massless, as we want.

We note the seemingly crucial fact that the two matrices $\Gamma$ and
$\lambda_8$ are proportional:
\begin{equation}
\Gamma= -\tfrac{1}{\sqrt{3}} \lambda_8,
\end{equation}
for one to arrive at the above result.  Given the structure of the vacuum in
(\ref{Phivac}), we believe that the choice of $\Gamma$ in (\ref{Gamma3}) is
essentially unique for the photon to remain massless although we have as yet 
no formal proof that this is so.  For example, had we taken $\Gamma$ as 
$-\third$ times the identity as we did in \cite{dfsm}, we would have found a 
mass matrix which is again diagonal except for the mixing between the photon 
and $K=8$ state as follows:
\begin{equation} \third \zeta_S^2 
\left( \begin{array}{cc}  \frac{1}{3} g_1^2& -\frac{R}{\sqrt{3}} g_1 g_3 \\
                         -\frac{R}{\sqrt{3}} g_1 g_3 & \half (1+R) g_3^2
       \end{array} \right).
\label{Masssubmo}
\end{equation}
This has no zero mode giving thus, as noted before, an unacceptable mass to the
photon. 

At this stage, it is interesting to compare the flavour and colour sectors and
note their similarities and differences.  In the electroweak  theory if we had 
used for the kinetic energy
the form (\ref{KEf}) without ({\bf ME}) instead of
(\ref{KEfa}), then the charge matrix in (\ref{Dmuf}) would be 
\begin{equation}
\Gamma = \left( \begin{array}{cc} -\half & 0 \\ 0 & \half \end{array} \right).
\label{Gamma2}
\end{equation}
The net charge is zero as for
(\ref{Gamma3}) above.  In the electroweak theory, only $\tilde{B}^3_\mu$ mixes 
with $A_\mu$, and $\Gamma$ is proportional to $\tau_3$, just as $\Gamma$ is to 
$\lambda_8$ in the colour sector.  In both cases, it is this property of 
$\Gamma$ which guarantees that the mass-mixing matrix has a zero mode leaving 
thus the photon massless.

Despite these similarities, however, there is a notable difference between the 
electroweak and colour sectors which, as we shall see, is responsible for some
marked divergences in the physics emerging for the two sectors.  First, as 
noted, it was the vector $\balpha$ coming from the weak framon which breaks 
the $\widetilde{su}(3)$ symmetry.  Given this vector, the strong vacuum 
in, say, the hermitian gauge, will automatically align itself in such a way as 
to have its long axis (for $R > 0$) pointing in the direction of $\balpha$, 
leaving the two shorter axes orthogonal to it and a residual symmetry about it. 
Now we find that the direction in which the $u(1)$ of electromagnetism is 
embedded in $\widetilde{su}(3)$ space is also given by $\balpha$, with the 
$\bphi$ pointing in the direction of $\balpha$ (again in the hermitian gauge) 
given the charge $+\twothirds$ while the two $\bphi$ orthogonal to $\balpha$ 
are given the charges $-\third$.  Now, this is quite different from what 
happens to the $\widetilde{su}(2)$ symmetry in the flavour sector which, though 
also broken by the $u(1)$ of electromagnetism via the vector $\bgamma$, this 
last, as far as is known, need not have any relation to the vector $\bbeta$ 
coming from the strong framon, since the dependence of the vacuum on $\bbeta$ 
has been eliminated by ``minimal embedding'', which is applicable to $su(2)$ 
but not to $su(3)$.

Combining the above result for the flavour and colour sectors in treating the
two kinetic energy terms together, one finds a mass matrix which is diagonal 
for all the vector states except for $A_\mu, \tilde{B}^3_\mu, \tilde{C}^8_\mu$
which mix together via the submatrix:
\begin{equation}
\left( \begin{array}{ccc}
\quart \zeta_W^2 g_1^2 + \tfrac{2}{9} (1 + R) 
   \zeta_S^2 g_1^2 & -\quart \zeta_W^2 g_1 g_2 &-\tfrac{1}{3\sqrt{3}} (1 + R) 
   \zeta_S^2 g_1 g_3   \\
   -\quart \zeta_W^2 g_1 g_2 & \quart \zeta_W^2 g_2^2 & 0 \\
   -\tfrac{1}{3\sqrt{3}} (1 + R) \zeta_S^2 g_1 g_3 & 0&\tfrac{1}{6}(1 + R) 
   \zeta_S^2 g_3^2
\end{array} \right).
\label{Masssubmfc}
\end{equation}
This has an eigenvector with zero eigenvalue giving the massless photon as:
\begin{equation}
\gamma =\left( \frac{e}{g_1} A_\mu + \frac{e}{g_2} \tilde{B}^3_\mu +
\frac{2}{\sqrt{3}} \frac{e}{g_3} \tilde{C}^8_\mu \right),
\label{photonmix}
\end{equation}
with the normalization given by the electromagnetic coupling 
\begin{equation}
\frac{1}{e^2} = \frac{1}{g_1^2} + \frac{1}{g_2^2}
+\frac{1}{\frac{3}{4} g_3^2}.
\end{equation}

Though leaving the photon massless, as is indispensable for the theory to be
viable, the result differs from the standard electroweak theory which has 
already been tested to great accuracy by experiment.  It is thus important to
check whether the deviations still remain within the limits of the present 
experimental errors and, if so, whether they can be detected as new physics
by future experiments.  An answer to these questions will require consideration
 too lengthy to be treated in this paper and so has to be delegated to another 
\cite{zmixed}, only a brief summary of which will be given in Section 7.3.   

Having now assigned $u(1)$ charges to the colour framons so as to leave the
photon massless, we are ready to turn back to answer the question left open
before on the $u(1)$ and $\tilde{u}(1)$ invariance of the framon potential
(\ref{V}).  First, we note that as for the flavour framon, $\tilde{y} = - y$
for each column of $\bPhi$.  It is then immediately clear that all terms in    
(\ref{V}) are invariant under $u(1)$ and $\tilde{u}(1)$, except perhaps for 
a hesitation over the last term with coefficient $\nu_2$.  However, even this 
last term is seen to be invariant when we recall that $\bPhi \balpha$ is just 
that component of $\bPhi$ which carries a $y$ of $+\twothirds$ and a 
$\tilde{y}$ of $-\twothirds$, so that the change in phase under either a $u(1)$
or $\tilde{u}(1)$ transformation of $\bPhi \balpha$ will cancel the
opposite change in phase of the factor $(\bPhi \balpha)^\dagger$ also present.
This is independent of what $\tilde{y}$ charges are assigned to $\balpha$ and
$\bbeta$ since these factors always occur in (\ref{V}) in conjugate pairs.
Hence, we conclude that $V$ in (\ref{V}) is indeed fully invariant under 
$G \times \tilde{G}$ as required.

We have already dealt with the tree-level mass matrix of the $G$.  Expanding 
further the expression (\ref{KECt}) gives the tree-level couplings of the $H$
to the $G$ detailed in Appendix B.

\section{Yukawa couplings of framons to fermions}

At the present stage of our understanding, the construction of Yukawa couplings 
is conceptually different from that of the framon potential and kinetic energy 
term treated above.  These latter terms involve only framons and gauge 
potentials, each of which has been assigned a geometrical significance and each 
has a specific function to discharge. Their construction requires thus merely 
an insistence that the basic invariance principles be satisfied, although at 
times this might have required some deftness to achieve.  Yukawa couplings, on 
the other hand, involve matter fermions, for which no geometrical significance 
has yet been discovered.  One does not therefore know {\it a priori} which 
fermion fields should figure in these couplings.  Input from experiment or 
other conditions is thus needed.

Let us take first as example the flavour coupling of the standard model in its 
usual formulation.  To accommodate the quarks and leptons seen in experiment, 
it is suggested that we take as fundamental  fermion fields the following:
\begin{equation}
\psi_L(\sixth, 2, 3), \ \psi_L(-\half, 2, 1),\  
\psi_R(\twothirds, 1, 3),\  \psi_R(-\third, 1, 3); \ \psi_R(-1, 1, 1), 
\ \psi_R(0, 1, 1),
\label{mfundferm}
\end{equation}
(where the first argument inside the brackets denotes the $u(1)$ charge, the 
second the dimension of the $su(2)$ representation and the third that of 
the $su(3)$ representation).  As far as the group representations are concerned,
this choice seems reasonable, these representations being indeed the simplest 
for the gauge group $U(1, 2, 3)$.  However, it is a bit of a mystery that
\begin{itemize}
\item {\bf [CH]} The flavour doublet fields are to be left-handed while the
flavour singlets are to be right-handed,
\end{itemize}
which we shall refer to as the ``chirality puzzle'', since the origin of this
requirement imposed on us by experiment is theoretically unknown.

From (\ref{mfundferm}), Yukawa couplings are constructed as:
\begin{equation}
Y_- \bar{\psi}_L^r \phi_r \psi_R^- + Y_+ \bar{\psi}_L^r \phi_r^{\perp} \psi_R^+.
\label{Yukawa}
\end{equation}
Or, to exhibit its underlying symmetry under global $\widetilde{su}(2)$, this 
can be rewritten as:
\begin{equation}
Y_- \bar{\psi}_L^r (\bphi_r \cdot\bgamma) \psi_R^-
    + Y_+ \bar{\psi}_L^r (\bphi_r \cdot\bgamma^\perp) \psi_R^+, 
\label{Yukawai} 
\end{equation} 
in terms of the vector $\bgamma$ introduced in the previous section.
 
Further, to accommodate the three generations of quarks and leptons seen in
experiment, it is postulated in the standard model that:
\begin{itemize} 
\item {\bf [GE]} There are to be three copies each of the fields in 
(\ref{mfundferm})
\end{itemize}
which we shall call the ``generation puzzle'', this being in the standard model
also theoretically unexplained.
 
Then, from these fermion fields, one constructs 6 Yukawa terms, each of the 
form (\ref{Yukawa}), and each with its own set of $(Y_-, Y_+)$, that is, 
12 independent empirical parameters altogether.  Besides, these $(Y_-, Y_+)$, 
being proportional to the quark and lepton masses, are required to take on a
hierarchical array of values, in which can be discerned:
\begin{itemize}
\item  {\bf [h1]} a hierarchy among generations of the same species, e.g.
  $m_t \gg m_c \gg m_u$,
\item  {\bf [h2]} a hierarchy between species of different charges, e.g.
$m_t \gg m_b$, $m_\tau \gg m_{\nu_3}$,
\item  {\bf [h3]} a hierarchy between quarks and leptons, e.g. $m_t \gg m_\tau, 
m_b \gg m_{\nu_3}$.
\end{itemize}
And all these properties are built into the standard model as 
empirical input. 

In the FSM, things perhaps look somewhat better.  First, the flavour framon 
{\bf (FF)} in (\ref{fframon}) carries with it a global colour vector $\balpha$, 
so that in constructing a Yukawa term with the fermion fields in (\ref{mfundferm}) 
along the lines of (\ref{Yukawa}), one would need to introduce
three copies of each to maintain $\widetilde{su}(3)$ invariance, as is demanded 
by the framon hypothesis.  One obtains then from $\psi_L(-\half, 2, 1)$ and 
$\psi_R(-1, 1, 1), \psi_R(0, 1, 1)$, the Yukawa terms for leptons
\cite{dfsm,tfsm}:
\begin{eqnarray}
\!\!\!\!\!\!\!\!{\cal A}_{\rm YF} &=& \sum_{[\tilde{a}] [b]} Y_{[b]-} \bar{\bpsi}_{[\tilde{a}]}
     \alpha^{\tilde{a}} (\Phi \bgamma) \half (1 + \gamma_5) \psi^{[b]-}
     + \sum_{[\tilde{a}] [b]} Y_{[b]+} \bar{\bpsi}_{[\tilde{a}]}
     \alpha^{\tilde{a}} (\Phi \bgamma^{\perp}) \half (1 + \gamma_5) \psi^{[b]+}
     \nonumber \\
&& {} + {\rm h.c.},
\label{Yukawaf}
\end{eqnarray}
and from $\psi_L(\sixth, 2, 3)$ and $\psi_R(\twothirds, 1, 3), 
\psi_R(-\third, 1, 3)$  similar terms for quarks.  Hence all three 
generations now appear as a natural requirement of invariance and are 
incorporated into a single Yukawa term.  Secondly, these Yukawa couplings 
give, at tree level, a rank-one mass matrix (\ref{mfact}) which is already 
putatively hierarchical, and if one believes further the result cited above 
from \cite{tfsm}, then realistic hierarchical patterns of masses for 
generations will automatically emerge as a result of (\ref{Yukawaf}), through 
the rotation of the mass matrix (\ref{mfact}) with scale under renormalization 
by framon loops and need no longer be taken as input from experiment.  In this 
sense then, the FSM appears to have solved the puzzle {\bf [GE]} why there 
should be three generations of quarks and leptons, and also the puzzle 
{\bf [h1]} why their masses should be hierarchical.  In the following Section 
6, a solution along similar lines will be suggested for the two other hierarchy 
puzzles {\bf [h2]} and {\bf (h3)}.  However, the chirality puzzle  remains 
largely unresolved.

For the flavour Yukawa coupling in the FSM, there is one further point to check 
for consistency.  The terms (\ref{Yukawaf}) is by construction invariant 
under the local gauge symmetry $u(1) \times su(2) \times su(3)$ and under the 
global symmetry $\widetilde{su}(2) \times \widetilde{su}(3)$.  But is it also
invariant under $\tilde{u}(1)$, as it is required to be?  We recall from our 
starting premises that of the fundamental fields only the framon field 
$\balpha \Phi$ has any reason to carry a global quantum number $\tilde{y}$.  
Thus, all the fundamental fermion fields $\psi$ appearing in (\ref{Yukawaf}), 
whether left-handed or right-handed, should have $\tilde{y} = 0$.  This 
statement agrees with the assertion in Section 2 that the group for the global 
symmetry $\tilde{G}$ is $U(1, 2, 3)$ so that by (\ref{ycharges}), these fermion 
fields, being singlets in both global flavour and colour, should indeed have 
$\tilde{y} = 0$.  What then in (\ref{Yukawaf}) will cancel the $\tilde{y}$ 
value $-\third \mp \half$ given in (\ref{fframon}) for flavour framon 
$\balpha \Phi$ so as to leave the Yukawa term invariant?   We notice first 
that the vectors $\bgamma, \bgamma^\perp$, though not field variables, 
are $\widetilde{su}(2)$ doublets, and so are
nevertheless required by (\ref{ycharges}) 
to have $\tilde{y} = \pm \half$, hence cancelling part of this deficit, namely 
$\mp \half$, from $\balpha \Phi$.  Secondly, we notice that appearing in 
(\ref{Yukawaf}) above is not $\balpha \Phi$ but are $\alpha^{\tilde{a}}$, the 
components of $\balpha$ in the directions $\tilde{a}$, which, being scalars 
in global colour $\widetilde{su}(3)$, should, again by (\ref{ycharges}) carry 
$\tilde{y} = 0$,  meaning that the other part of the $\tilde{y}$ charge, 
namely $\third$, supposedly carried by $\balpha \Phi$ as cited above, should 
not have appeared.  If we insist on writing (\ref{Yukawaf}) in terms of 
$\balpha \Phi$, then we should have written $\alpha^{\tilde{a}}$ as 
$(\be^{[\tilde{a}] \dagger} \cdot \balpha)$ where $\be^{[\tilde{a}]}$ are three 
orthonormal vectors associated with the three fundamental left-handed fermions 
fields $\bpsi_{[\tilde{a}]}$.  Then these $\be^\dagger$, being anti-triplets, will 
carry each an appropriate $\tilde{y} = \third$ to cancel 
off the $-\third$ from $\balpha$, leaving then both the components 
$\alpha^{\tilde{a}}$ and the whole Yukawa term (\ref{Yukawaf}) invariant under 
$\tilde{u}(1)$ as required.  Though seemingly pedantic, these considerations 
for checking the invariance of Yukawa terms under $\tilde{u}(1)$  are 
to our minds worth going through once in detail, so as to clarify later 
(Section 7.2) the relationship between $\tilde{y}$ and the lepton and baryon 
numbers.

Next, we turn to the construction of the Yukawa terms for the colour framon 
{\bf (CF)}, using (\ref{Yukawaf}) as template.  Instead of 
$\widetilde{su}(2)$ we now have 
$\widetilde{su}(3)$, which is to be broken explicitly again by 
electromagnetism.  According to the preceding section, however, 
the global nonabelian symmetry is here broken not by 
an ``external'' vector like $\bgamma$ in the flavour case, but by the vector 
$\balpha$ which comes from the flavour framon {\bf (FF)} in (\ref{fframon}).
The reason is that, as already noted, the vacuum has itself a broken symmetry 
depending on the direction of $\balpha$, which forces us to have the breaking 
by electromagnetism occurring only in that direction if we want to keep the 
photon massless.  In place of $\bgamma^\perp$ orthogonal to $\bgamma$ in the 
flavour case, let us introduce thus two corresponding (3-d) unit vectors $\bdelta$ 
and $\bdelta'$ both orthogonal to $\balpha$ and also mutually orthogonal
\footnote{This introduction in the colour Yukawa terms of three (extra)
vectors $\balpha, \bdelta, \bdelta'$, in parallel with $\bgamma,\bgamma^\perp$ 
in the flavour case, may seem somewhat arbitrary and perhaps unfounded.  The 
same may be said of the three (3d) vectors $\be^{[\bar{\alpha}]} $ introduced 
earlier.    However, this comes about from a result in (multi)linear 
algebra about tensor products, which says that there is a natural isomorphism 
(but bases-dependent) $$ \bbc^3 \otimes_\bbc H \cong H \oplus H \oplus H $$
where the right hand side represents three scalar fields in the Hilbert space 
$H$.  So we are saying that it makes good sense (as proposed above) to attach 
to each of these scalar fields a basis vector from the representation space 
of $\widetilde{su} (3)$ with its inherent linear structure, and thus make the invariance manifest and give a 
clearer mathematical formulation of the procedure.}.  The colour framon in 
the direction $\balpha$ carries the charge $+\twothirds$, but the framons in 
the directions $\bdelta, \bdelta'$ carry the charge $-\third$.

With these observations and new notations, we propose then to write the Yukawa 
term for the colour framon in the following generic form, provided we have
at our disposal the appropriate left- and right-handed fermion fields:
\begin{equation}
{\cal A}_{\rm YC} =
Z_{\delta} \bar{\bpsi}_L (\bPhi  \bdelta) \psi_R^-
  + Z_{\delta'} \bar{\psi}_L (\bPhi  \bdelta') \psi_R'^-
  + Z_{\alpha} \bar{\psi}_L (\bPhi  \balpha) \psi_R^+.
\label{Yukawaci}
\end{equation}
This is by construction invariant under the double symmetry 
$G \times \tilde{G}$, including the $\tilde{u}(1)$ factor, as can be seen 
following similar arguments for the flavour term (\ref{Yukawaf}) above.

However, this formula hides an important difference from the flavour case.
In parallel to the vector $\balpha$ in (\ref{Yukawaf}) coming 
from the flavour framon (\ref{fframon}), which played such an important 
role there in explaining the intricacies of fermion generations \cite{tfsm}, 
there ought to appear in (\ref{Yukawaci}) also the vector $\bbeta$ coming 
from the colour framon (\ref{cframon}).  In strict parallel to (\ref{Yukawaf}), 
we ought thus to have written for (\ref{Yukawaci}) something like:
\begin{eqnarray}
{\cal A}_{\rm YC} &=& \sum_{[\tilde{r}] [s]} Z_{[\tilde{s}]\delta} \bar{\bpsi}_{[\tilde{r}]L}
     \beta^{\tilde{r}} (\bPhi \bdelta) 
      \psi_R^{[\tilde{s}]-} \nonumber \\
     & + &  \sum_{[\tilde{r}] [s]} Z_{[\tilde{s}]\delta'} \bar{\bpsi}_{[\tilde{r}]L}
     \beta^{\tilde{r}} (\bPhi \bdelta') 
      \psi_R'^{[\tilde{s}]-} \nonumber \\
     & + &  \sum_{[\tilde{r}] [s]} Z_{[\tilde{s}]\alpha} \bar{\bpsi}_{[\tilde{r}]L}
     \beta^{\tilde{r}} (\bPhi \balpha) 
     \psi_R^{[\tilde{s}]+}
     \nonumber \\
     &+ & {}  {\rm h.c.}.
\label{Yukawaci1}
\end{eqnarray}
We recall, however, that in contrast to $\balpha$, which is coupled to the 
vacuum and therefore rotates with changing scale leading to all the intricacies 
of fermion generations, the vector $\bbeta$  (ultimately again because 
of the ``minimal embedding'' of (\ref{minemb}) special to $su(2)$) is decoupled 
from the vacuum and scale-independent.  Thus, for example, the expression:
\begin{equation}
\sum_{[\tilde{r}]} \bar{\bpsi}_{[\tilde{r}]L} \beta^{\tilde{r}},
\label{betadotpsi}
\end{equation}
being a constant (scale-independent)  linear combination of the $\bar{\psi}_{[\tilde{r}]L}$, represents really 
only one single field, which we could therefore just as well denote by $\bar{\psi}_L$ 
as in (\ref{Yukawaci}).    In other 
words, we could just as well remain with (\ref{Yukawaci}) above.  This means 
that there is in the colour Yukawa coupling no parallel to generations, 
a significant feature in the flavour case.

Next, we need to ask the question: for which fermions are we to construct 
such Yukawa terms?   In our previous applications of the Yukawa term in  
for example \cite{tfsm}, we did not have to answer this question, since for 
deriving the scale dependence of $\balpha$ needed by the programme  
there,  merely the generic form of the Yukawa term sufficed.  But now,  
to study the spectrum of the $F$ we shall have to do so.  
In contrast to the flavour case, where we know
from experiment  which quarks and leptons occur as the known
bound states, the same question is not easily answerable for the colour case 
since the bound state fermions $F$
are still unknown to us.  Therefore, with little other information to guide us, 
any answer, it would seem, can only be tentative, standing to be rectified 
and/or supplemented if and when further empirical information becomes available
or if and when it is understood what theoretical grounding or geometrical
significance fermions have, in parallel to those of the gauge bosons or of the
framons.  Even so, we believe that a working model for the colour Yukawa 
terms would be useful to serve as a sort of base camp for exploration and 
we propose to construct one as follows.

We start with the list (\ref{mfundferm}) of fermion fields which, from previous 
analyses, we know must be present in our theory so as to give us the quarks and 
leptons.  We need, however, to introduce three copies of these, both in the 
SM by fiat to accommodate the three generations and in the FSM to account for 
$\widetilde{su}(3)$ invariance and hence to deduce the existence of the three 
generations.  For example, we need three copies of $\psi_L(\sixth, 2, 3)$ in the
FSM to bind with the flavour framon so as to form the three generations of 
left-handed flavour-doublet quarks.  We can understand this if we regard the 
fermions in the list (\ref{mfundferm}) as fundamental quantized fields, and the 
three copies needed of them as just their quanta of excitation, since these 
quanta are automatically represented by wave functions all transforming under 
symmetry operations in the same way as the fundamental fields themselves, as we 
want our ``copies'' to do.  

This is similar in spirit to what we may call the standard scenario where the 
quark field is regarded as fundamental, when we take a $u$ quark and combine it 
with an anti-$d$ quark to form a $\pi^+$, and another $u$ quark and combine it 
with two $d$ quarks to form a neutron.  These two $u$ quarks are but two 
different quanta of excitation (identical copies) of the same ``fundamental''
quark field, which combine with two separate entities (the $\bar{d}$ and the
$dd$) respectively to form the $\pi^+$ and the neutron.  Indeed, there was a 
time when physicists would construct field theories with the compound states 
$\pi^+$ and neutron as second quantized fields, as we do now with quarks and 
leptons.  In a sense then, by accepting 't~Hooft's confinement picture one has 
gone a level deeper, by starting with the fermions (\ref{mfundferm}) as 
``fundamental fields'', while the quarks and leptons themselves appear as
compound states of the flavour framon with quanta of the fundamental fermions 
fields to give the three different generations.

Let us note that, in doing so, we make no pretence of dealing really with 
physics at a more fundamental level.  For example, we cannot tell at this stage 
which bound states should exist between the framon and the quanta of the 
fundamental fermion fields.  We are treating (\ref{mfundferm}) merely as a 
list of allowed representations to draw copies from so as to form the bound 
states we want and then to write down effective actions for them as one used to 
do in the old days for pions and nucleons before the advent of chromodynamics.  
If this does really represent a deeper level of physics, it is for the future 
to explore.

If we accept this interpretation of (\ref{mfundferm}) as fundamental fields, 
then there seems nothing in principle to stop their quanta, if these are 
coloured, from combining via colour confinement with the colour framons to form 
the $F$.  Now of the fields in (\ref{mfundferm}), three carry colour, namely:
\begin{equation}
\psi_L(\sixth, 2, 3), \ \  \psi_R(\twothirds, 1, 3), \ \ \psi_R(-\third, 1, 3). 
\label{cfundferm} 
\end{equation} 
Combining these with the colour framon $\bPhi$, what $F$-states will they give?  

First, from $\psi_L(\sixth, 2, 3)$ combined with  $\bPhi^\dagger$ we obtain the 
left-handed $F$ in the first column  in the following list (\ref{coquarks}).  
These are colour singlet bound states but are flavour doublets.  They are in 
fact the counterparts of the left-handed quarks which are flavour singlet bound 
states carrying colour, only with the roles of flavour and colour interchanged, 
and will be called co-quarks.    In order to construct a Yukawa term for each 
of these  left-handed $F$  so as to give it a (Dirac) mass, they have to be 
matched with fields of the opposite handedness, as listed in the second column, 
resulting in the mass eigenvalues in the third column, as will be shown later.  
\begin{equation}
\bPhi^\dagger \psi_L(\sixth, 2, 3) = \left( \begin{array}{c}
    \psi_L(\half, 2, 1) \\ \psi_L(\half, 2, 1) \\ \psi_L(-\half, 2, 1)
    \end{array} \right)
    : \begin{array}{c} \psi_R(\half, 2, 1) \\ \psi_R(\half, 2, 1) \\
      \psi_R(-\half, 2, 1) \end{array}
    : \begin{array}{c} Z_{Q\delta} \sqrt{\frac{1-R}{3}} \zeta_S \\ 
                       Z_{Q\delta'} \sqrt{\frac{1-R}{3}} \zeta_S  \\ 
                       Z_{Q\alpha} \sqrt{\frac{1+2R}{3}} \zeta_S \end{array}.
\label{coquarks}
\end{equation}

 Then from the other two fundamental fields on the 
list (\ref{cfundferm}), we have similarly:
\begin{equation}
\bPhi^\dagger \psi_R(\twothirds, 1, 3) = \left( \begin{array}{c}
    \psi_R(1, 1, 1) \\ \psi_R(1, 1, 1) \\ \psi_R(0, 1, 1)
    \end{array} \right)
    : \begin{array}{c} \psi_L(1, 1, 1) \\ \psi_L(1, 1, 1) \\
      \psi_L(0, 1, 1) \end{array}
    : \begin{array}{c} Z_{L1\delta} \sqrt{\frac{1-R}{3}} \zeta_S \\ 
                       Z_{L1\delta'} \sqrt{\frac{1-R}{3}} \zeta_S \\ 
                       Z_{L1\alpha} \sqrt{\frac{1+2R}{3}} \zeta_S \end{array}.
\label{coleptons1}
\end{equation}
and:
\begin{equation}
\bPhi^\dagger \psi_R(-\third, 1, 3) = \left( \begin{array}{l}
    \psi_R(0, 1, 1) \\ \psi_R(0, 1, 1) \\ \psi_R(-1, 1, 1)
    \end{array} \right)
    : \begin{array}{l} \psi_L(0, 1, 1) \\ \psi_L(0, 1, 1) \\
      \psi_L(-1, 1, 1) \end{array}
    : \begin{array}{c} Z_{L2\delta} \sqrt{\frac{1-R}{3}} \zeta_S \\ 
                       Z_{L2\delta'} \sqrt{\frac{1-R}{3}} \zeta_S \\ 
                       Z_{L2\alpha} \sqrt{\frac{1+2R}{3}} \zeta_S \end{array}, 
\label{coleptons2} 
\end{equation} 
which will be called co-leptons, with the electrically neutral members labelled also as 
co-neutrinos.

  Some of the fields listed in the second columns, specifically those coupled 
via $\bdelta$ or $\bdelta'$, carry the same quantum numbers as the charge 
conjugates of the 3 colour neutral fields in (\ref{mfundferm}), and can be 
interpreted as such.  But of those three others coupled via $\balpha$, only 
$\psi_L(0, 1, 1)$ can be taken as $\psi_R(0, 1, 1)^C$ but the other two are not 
contained in the list (\ref{mfundferm}) and have to be added as new fields 
giving the full list as:
\begin{eqnarray}
\psi_L(\sixth, 2, 3), \ \ \psi_L(-\half, 2, 1), \ \ \psi_R(\twothirds, 1, 3), 
   \nonumber \\ 
\psi_R(-\third, 1, 3), \ \ \psi_R(-1, 1, 1), \ \ \psi_R(0, 1, 1) 
   \nonumber \\ 
\psi_R(-\half, 2, 1), \ \ \psi_L(-1, 1, 1), \ \ \psi_L(0, 1, 1), 
\label{fundferm} 
\end{eqnarray} 
where the last item is repeated for convenience for later 
reference despite being the charge conjugate of $\psi_R(0, 1, 1)$ already 
listed.  This list seems to provide an interesting new take on the question of 
chirality in that the uncoloured fields in it are present initially in both 
handedness, but components of opposite handedness get pulled off in different 
directions, with half of them forming via flavour confinement the left-handed
leptons, and the other half posing as the right-handed partners of some $F$.  
However, this raises the question what if $\psi_R(-\half, 2, 1)$ should bind
with a flavour framon via flavour confinement as $\psi_L(-\half, 2, 1)$ does.
This will result in a state with the same quantum numbers as a right-handed
lepton which is not wanted.  But if it is not a right-handed lepton, then what 
is it?   To this, we shall suggest an answer later in Section 8 {\bf [g]}

With (\ref{fundferm}) as our list of fundamental fermion fields, we propose 
then to work with the colour Yukawa terms:

\begin{equation}
{\cal A}_{\rm YQ} =  \left[ 
\begin{array}{l} Z_{{\rm Q}\delta} \bar{\psi}_L(\sixth, 2, 3) \bPhi \bdelta 
                  \psi_R(\half, 2, 1) \\
              +  Z_{{\rm Q}\delta'} \bar{\psi}_L(\sixth, 2, 3) \bPhi \bdelta' 
                  \psi'_R(\half, 2, 1) \\
              +  Z_{{\rm Q}\alpha} \bar{\psi}_L(\sixth, 2, 3) \bPhi \balpha 
                  \psi_R(-\half, 2, 1) \end{array} \right] 
              + {\rm h.c.}
\label{Yukawacq}
\end{equation}
for the co-quarks, plus
\begin{equation}
{\cal A}_{\rm YL1} =  \left[ 
\begin{array}{l} Z_{{\rm L1}\delta} \bar{\psi}_R(\twothirds, 1, 3) \bPhi \bdelta 
                  \psi_L(1, 1, 1) \\
              +  Z_{{\rm L1}\delta'}\bar{\psi}_R(\twothirds, 1, 3) \bPhi \bdelta' 
                  \psi'_L(1, 1, 1) \\
              +  Z_{{\rm L1}\alpha}\bar{\psi}_R(\twothirds, 1, 3) \bPhi \balpha 
                  \psi_L(0, 1, 1) \end{array} \right] 
              + {\rm h.c.}
\label{Yukawacl1}
\end{equation}
and
\begin{equation}
{\cal A}_{\rm YL2} =  \left[ 
\begin{array}{l} Z_{{\rm L2}\delta} \bar{\psi}_R(-\third, 1, 3) \bPhi \bdelta 
                  \psi_L(0, 1, 1) \\
              +  Z_{{\rm L2}\delta'}\bar{\psi}_R(-\third, 1, 3) \bPhi \bdelta' 
                  \psi'_L(0, 1, 1) \\
              +  Z_{{\rm L2}\alpha}\bar{\psi}_R(-\third, 1, 3) \bPhi \balpha 
                  \psi_L(-1, 1, 1) \end{array} \right] 
              + {\rm h.c.}
\label{Yukawacl2}
\end{equation}
for the two sets of co-leptons.

We stress, however, that they are to be regarded as mere working hypotheses 
and do not have the same theoretical basis as the framon potential and kinetic 
energy term considered in the two preceding sections, nor the phenomenological 
justification as the flavour Yukawa couplings studied earlier in this section.

In these couplings, by substituting for the framon field $\bPhi$ its vacuum 
expectation value, one obtains the tree-level mass matrices of the $F$, which 
in the canonical gauge (\ref{Phivac0}) are already diagonal with elements 
listed in the last columns of (\ref{coquarks}), (\ref{coleptons1}) and 
(\ref{coleptons2}).  Further, by expanding about the vacuum in fluctuations 
of the framon fields, one obtains in the same gauge the tree-level coupling 
matrices of the $H_K$ to the $F$ as: 
\begin{equation}
\Gamma_K = V_K {\bf Z}  \half (1 + \gamma_5) 
           +  {\bf Z} V_K^{\dagger} \half (1 - \gamma_5),
\label{GammaK}
\end{equation}
where $V_K$ are given in (\ref{VK}), and
\begin{equation}
{\bf Z}_{\rm Q} = \left( \begin{array}{ccc} Z_{{\rm Q}\delta} & 0 & 0 \\
                                     0 & Z_{{\rm Q}\delta'} & 0 \\
                                     0 & 0 & Z_{{\rm Q}\alpha} \end{array} \right),
\label{bfZcq}
\end{equation}
with similar formulae for ${\bf Z}_{\rm L1}$ and ${\bf Z}_{\rm L2}$.  These will 
be useful in the following section.

\section{Scale-dependence from framon loops}

Apart from the terms in the action involving the framon fields detailed in the
last 3 sections, there are of course also the usual terms of the standard model
which we may call the kinetic energy terms of the gauge bosons and the fermions,
which have no explicit dependence on the framon fields.  The only question then
is how they translate in the confinement picture of 't Hooft into couplings of 
the $G$ and $F$ states.  Now in Section 4, it is shown in the parallel flavour 
case that the kinetic energy term in (\ref{KEBmu}) of the flavour gauge bosons 
$B_\mu$ is formally the same when translated into the $\tilde{B}_\mu$ states.
The same arguments will show that the kinetic energy term for the colour boson 
$C_\mu$ will also be formally the same when translated into the $\tilde{C}_\mu$ 
states representing the $G$.  This means that the couplings of the $G$ among
themselves will be the same as the colour gauge bosons among themselves.  Very
similar arguments when applied to the kinetic energy term of the fermions will
show that the couplings between the $G$ and $F$ are formally the same as 
those between the colour gauge bosons and the fundamental fermions fields.  
Though fairly straightforward, these arguments will be outlined  for 
completeness in Appendix C.
  
Together then with the tree-level couplings of the $H$ to themselves and to the 
$G$ listed in Appendices A and B, plus those to the $F$ listed at the end of
the preceding section, one can in principle proceed to evaluate the higher 
order loop corrections.  However, at this early explorative stage, we are 
obviously not yet in a position to embark on a full investigation in this 
direction.  We shall here restrict ourselves only to 1-framon loop correction 
of the fermion self-energy with only the limited aim of checking how it fits
in with the noted hierarchies in the fermion spectrum, and with the result 
obtained before in \cite{tfsm} using a Yukawa coupling which is generically 
similar but differs in detail from that suggested in the preceding section.

With hindsight, the choice of framon loop effects on the fermion self-energy
as a first example of higher order corrections is seen to be a particularly 
lucky one to study, given that the Yukawa couplings (\ref{GammaK}) are so much 
simpler than the other couplings listed in Appendix A or B, but that they 
already display that unique property of framons in carrying both local and
global indices, so that framon loops lead, with changing scales, not only to 
changing strengths to quantities as gauge boson loops do, but also to changing 
orientations in the global symmetry space, which was what we call ``rotation''
in the Introduction, from which the results of Table \ref{tfsmfit} were derived.

Fermion masses derived from Yukawa couplings have a common feature in that at
tree level the mass is given as a product of the coupling strength times the
vacuum value of the scalar (framon) field, where the latter is a property of
the vacuum and therefore independent of the fermion appearing in the coupling.
This means that the fermion mass is proportional to the coupling strength which
governs the size of the framon loop, and thus also of the renormalizing effects 
they give.  Hence, the bigger the mass, the bigger also the renormalization  
effects.

Consider first as example the flavour Yukawa terms for quarks as given in 
(\ref{Yukawaf}), which for the present purpose can be replaced simply by 
(\ref{Yukawa}) for $t$ and $b$ where we have suppressed the complexities due to generations 
given that the properties of the lower generations, according to \cite{tfsm}, 
will emerge automatically as a consequence of the rotation of $\balpha$.  Let 
us consider the scale dependence of the quark masses under renormalization by 
a framon loop, which in this case is due just to the single standard model 
Higgs boson $h_W$.  It is clear that the masses will start to run with scale 
with a speed proportional to the couplings $Y_+, Y_-$.  Given that $Y_+$ and 
$Y_-$ are proportional respectively to $m_b$ and $m_t$ at the scale we choose 
to measure these masses, it follows that $Y_+ \ll Y_-$ at that scale and that 
$Y_+/Y_-$ will decrease further as the scale increases further.  Hence, if we 
choose to write:
\begin{equation}
(Y_+, Y_-) = Y \beeta = Y (\sin \theta, \cos \theta) 
\label{eta} 
\end{equation}
then $\beeta$ will have a fixed point $(0, 1)$, or $\theta = 0$ when 
$\mu \rightarrow \infty$.  Using the language adopted in \cite{tfsm}, 
though in a different context, we may consider $\beeta$ as a vector with a 
high scale fixed point at $(0, 1)$ which rotates with scale.  

Suppose we turn the question around and choose to start with the
assertion that there is such a fixed point at $\mu=\infty$, then 
at a high finite 
scale  we expect that $\theta$ will be small, or $Y_+ \ll Y_-$, or 
$m_b \ll m_t$, obtaining this last empirical fact (Section 5 {\bf [h2]}) as a 
consequence of rotation, in much the same way as the ``leakage mechanism'' in 
\cite{tfsm} which gave the hierarchical mass spectrum for the generations, 
although the similarity is only formal, the physics being very different.  In 
this case, however, the assertion is of only conceptual but no concrete value, 
since with no other information to fix the integration constant (that
is,  the 
``initial value'') for the implied rotation equation, one cannot derive the actual value 
for $m_b/m_t$ as one would like to.

Similar arguments can be applied of course 
to the ratio $m_\tau/m_t$ for a qualitative understanding of why leptons are 
light compared with quarks (Section 5 {\bf [h3]})\footnote{They can be 
applied in principle also to the lepton mass ratio $m_{\nu_3}/m_\tau$ to complete 
our picture for the hierarchy among fermion species, but by $m_{\nu_3}$ here,
one presumably means the Dirac mass of the heaviest neutrino, which is unknown
experimentally, and not its measured physical mass which is thought to be 
affected by a see-saw mechanism.  However, in the fit of \cite{tfsm} summarized 
in Table I, a value for $m_{\nu_3} \sim 29.5$ MeV emerged, the ratio of 
which to $m_\tau$ is remarkably close to $m_b/m_t$, namely  $m_{\nu_3}/m_\tau \sim 0.166$ 
compared with $m_b/m_t \sim 0.24$.  If the running with changing scale of $m_b$
due to its renormalization by gluon loops is taken into account, the agreement 
is even closer, giving $m_b/m_t \sim 0.165$ at the $t$ mass.  It is amusing 
to note that this agreement will result if one assumes (i) that $\beeta$ is, 
for some reason, the same (or similar) for both quarks and leptons, and 
(ii) that the rotation of $\beeta$ induced by $h_W$ loop will essentially stop 
below the $h_W \sim m_t$ mass scale so that the value of $\theta$ is frozen at 
that scale.}.  Indeed, from such considerations, it would seem to follow that 
Yukawa couplings and fermion masses will in general be hierarchical if they are 
measured at scales close to the fixed point at infinity.  The only question is 
the order of the hierarchy, for which, from the above examples, it seems that 
the following rule-of-thumb from ancient folklore:
\begin{itemize}
\item {\bf [RT]} The more interactions, and hence the more self-energy it has, 
the heavier will a particle be compared with its peers
\end{itemize}
still applies (apart, of course, from the famous exception that the proton is 
lighter than the neutron, or in its modern guise, $m_u < m_d$, for which an 
explantion was already offered in \cite{tfsm}).  Thus, $m_t > m_b$ because $t$ 
has the bigger charge, and quarks are heavier than leptons because quarks have 
colour interactions while leptons do not.   

Turning next to the colour Yukawa coupling, we recall that for the working 
model suggested above in Section 5, there are altogether 9 Yukawa terms: for 
each $F$-fermion type, namely whether co-quark, co-lepton 1 or co-lepton 2, 
there are three Yukawa terms corresponding to the three columns of the colour
framon $\bPhi$ and coupled respectively via $\bdelta, \bdelta'$ and $\balpha$.  Let us 
take first just one $F$-fermion type, to be specific say the co-quark although, 
as it will be seen later, it does not really matter which.  If we were to 
introduce, in analogy to $\beeta$, some 3-vectors $\bzeta$ defined as:
\begin{equation}
(Z_{Q\delta}, Z_{Q\delta'}, Z_{Q\alpha})  \, =  \, Z_Q \bzeta_Q, 
\label{bzeta} 
\end{equation}
with $\bzeta$ again a unit vector and:
\begin{equation}
Z_Q  \, =  \, \sqrt{Z^2_{Q\alpha} + Z^2_{Q\delta} + Z^2_{Q\delta'}} 
\label{Zcq} 
\end{equation} 
and assign to $\bzeta$, in analogy to $\beeta$, a high scale fixed 
point at $(0, 0, 1)$, then at high finite scales we have 
\begin{equation}
Z_{Q\alpha} \, \gg  \, Z_{Q\delta}, Z_{Q\delta'}  
\label{bfh4}
\end{equation}
or that the couplings of the co-quarks corresponding to the three components 
will be hierarchical, in close analogy to {\bf [h2]} above for flavour.

There is a slight complication.  Compared to the flavour Yukawa coupling, the 
colour coupling differs by replacing, in the RGE, $\zeta_W$ not just by 
$\zeta_S$ but by the matrix $\phivac$ 
in (\ref{Phivac0}).  This means that the above conclusion for $Z_{Q \delta}, 
Z_{Q \delta'}, Z_{Q \alpha}$ has to be modified by some factors depending on $R$ 
and hence also on scale, which differ for the $\balpha$ and the $\bdelta, 
\bdelta'$ components.  But this will not make much difference, especially at 
high scales where it matters most, and where, if account is taken of the fit in 
\cite{tfsm}, $R \sim 0$ and the modifying factors become identical.  

Very similar considerations to the above suggest that the coupling strengths
for the $F$-fermion types would again be hierarchical, and thus by {\bf [RT]}:
\begin{equation}
Z_{\rm co-quark} \, > \, Z_{\rm co-(charged)leptons}  \,>  \,Z_{\rm co-neutrinos},
\label{bfh5}
\end{equation}
which is just the colour analogue of {\bf [h3]}.  

Account of these conclusions (\ref{bfh4}) and (\ref{bfh5}) will be taken when 
we later come to consider the mass spectrum of the $F$.

Next we turn to the question of consistency between the working model of the
preceding section and the scheme of \cite{tfsm} and the result  in Table
\ref{tfsmfit}.  The former has 9 Yukawa couplings each giving an RGE, but not
all these are of independent relevance to the problem treated there.
For instance, the three RGEs for the three $F$-fermion types which give the mass
hierarchies (\ref{bfh5}) above are seen to be equivalent as far as \cite{tfsm}
is concerned.  As noted before, the $F$ mass obtained from a Yukawa coupling
is a product, symbolically $Z_{Q \alpha} \phivac$, and so on, and the hierarchy would 
affect only the couplings $Z_{Q \alpha}$ but not  $\phivac$ which belongs to
the vacuum and should be the same whatever the $F$-fermion type it is coupled 
to.  What matters to the problem there is only the
scale-dependence of $\balpha$ induced by that of the vacuum.
In other words, the RGE for the rotation of the vacuum on which that
problem  depends would be the same irrespective of which  Yukawa term for 
the three fermion type from which it is derived. 

Next, whichever $F$ type we choose to focus on, there are still three Yukawa 
terms coupled respectively via  $\bdelta, \bdelta'$ and $\balpha$.  But only 
the last will be directly relevant for the problem in \cite{tfsm}, the results 
of which were all derived from the rotation of $\balpha$.  As explained there, 
the term coupled via $\balpha$ gives the scale-dependence of the third row of 
the rotation matrix $A$; so the terms coupled via $\bdelta$ and $\bdelta'$ will 
give the scale-dependence of the first and second row of $A$.  We recall 
however that $A$ is a rotation matrix which depends on only three parameters 
which we may take to be the Euler angles of equation (40) in \cite{tfsm}.  We 
have seen there that the equation from the term coupled via $\balpha$ already 
determines the scale-dependence of two of the Euler angles called $\theta_1$ 
and $\theta_2$ leaving only $\theta_3$ unconstrained.  It is then the 
scale-dependence of $\theta_3$ which will now result from the new equations 
implied by the terms in (\ref{Yukawacl2}) coupled via $\bdelta$ and $\bdelta'$, 
but this is in principle irrelevant for the results derived in \cite{tfsm} and
cited in Table \ref{tfsmfit}.  The only hesitation is in the very low scale 
region where one expects the heavy modes associated with $\balpha$ to decouple 
but the vectors $\bdelta$ and $\bdelta'$ continue to run, and these might give 
indirect effects of which the RGE in \cite{tfsm} has yet taken no account.

We turn our attention now on to the last remaining term coupled via $\balpha$ 
which is similar in form to the coupling studied in \cite{tfsm}.  Compared to 
equation (20) of \cite{tfsm}, this terms differs still in two respects: (i) by 
replacing $\sum_{[b]} Z_{[b]} \psi_R^{[b]}$ with $Z \psi_R$, and (ii) by replacing 
what is called $\balpha_Y$ there by $\balpha$ here.
\begin{itemize}
\item
For (i), we notice that what we call $Z \psi_R$ here is in fact just a special 
case of $\sum_{[b]} Z_{[b]} \psi_R^{[b]}$ when we take the array $(Z_{[1]}, Z_{[2]},
 Z_{[3]})$ to be $(0, 0, Z)$.  Since, in the derivation of the rotation equation
in \cite{tfsm}, $(Z_{[1]}, Z_{[2]}, Z_{[3]})$ never figures except via its norm 
$\rho_S^2 = \langle Z|Z \rangle$, the replacement above 
 will have no effect except for replacing $\rho_S$ 
there by $Z$ here.
\item
For (ii), although $\balpha_Y$ and $\balpha$ both take the value $(0, 0, 1)$
when the reference vacuum is diagonal, the former is supposed to be a fixed 
vector while the latter will be seen enventually to depend on scale.  However, 
being but a global quantity carrying no local indices, $\balpha$ does not emit 
or absorb framons, and do not thus get renormalized.  It is thus not involved 
in the renormalization calculation of \cite{tfsm} which only affect
the vacuum expectation value  of
$\bPhi$.  The vector $\balpha$ gets rotated under scale change only because it 
is coupled to the vacuum which changes direction under scale change, and so 
gets dragged along by the vacuum as the latter rotates.  For this reason, the 
rotation equation derived there for $\phivac$ remains still valid 
in form after the replacement.
\end{itemize}

We conclude therefore, that the analysis done in \cite{tfsm} would remain 
valid for the working model of Yukawa couplings suggested in the preceeding 
section, despite the apparent difference in couplings.
Indeed, from the above analysis, it would appear that almost any
other model of Yukawa couplings constructed along the lines described in
Section 5 would lead to the same result, 
since what is required from the
Yukawa couplings there is merely the rotation of $\balpha$ which 
appears in the fermion mass matrix (\ref{mfact}).  The meaning of the fitted
parameters may change depending on the choice of model, but the fitted result
would not.  And this is helpful since the present choice made in
Section 5 is only a tentative one, as warned.

\section{New physics in the standard sector}

With the last section on scale-dependence, we have completed the present round 
of theoretical scrutiny on the basic structure of the FSM, which is needed for 
its extension into the hidden sector of framonic $C$-ons: $H$, $G$, and $F$.  
Besides providing some conceptual clarifications and new insights, this 
closer study made three material changes to our earlier formulation, namely,
in order of appearance:
\begin{itemize}
\item {\bf [R1]} New assignments of the $u(1)$ (electric) charge $y$ to the 
colour framon {\bf (CF)};
\item {\bf [R2]} New assignments of the the $\tilde{u}(1)$ charge $\tilde{y}$
to the colour framon {\bf (CF)};
\item {\bf [R3]} A working model for the Yukawa terms spelling out some details 
not needed before.
\end{itemize}
Hence, before going on to study the hidden sector, which is  the stated main aim
 of this paper, we ought first to check whether, in the standard sector itself,
these changes may (i) alter the results obtained earlier, or (ii) imply new 
physics which has to be tested against experiment.  We shall deal with them
in the reverse order, leaving {\bf [R1]}, the most fundamental with the most
important consequences---and therefore the main concern of this section---to 
the last, to be considered at some length. 

\subsection{Deviations from SM predictions for some rare decays}

It was already shown in the preceding section that despite {\bf [R3]} 
the result of the fit summarized in 
Table 1 should remain valid.  
We can thus also accept some other effects deduced earlier from the rotation of $\balpha$, 
such as the deviations from the standard model in some rare Higgs decays 
\cite{Hdecay,f+mike}.  These deviations come about as follows.  The Yukawa coupling of 
the  Higgs boson $h$ to quarks and leptons in FSM is subsumed in the coupling 
of the flavour framon (\ref{fframon}) which carries as a factor the vector 
$\balpha$, the rotation of which in the resulting mass matrix for quarks and 
leptons is what leads to their hierarchical mass patterns in the FSM.  The same $
\balpha$ will thus appear in the couplings of $h$ to quark-antiquark and 
lepton-antilepton pairs, and its rotation is what in the FSM  governs the 
widths for the Higgs boson decaying into the various $q \bar{q}$ and $\ell 
\bar{\ell}$ modes.  This prediction differs from the SM where the couplings are 
given by the fermion masses.  For this reason, it was suggested in 
\cite{Hdecay} that: 
\begin{itemize}
\item the widths into lower generation fermions such as $h \rightarrow 
\bar{c} c$, $h \rightarrow \bar{s} s$ and $h \rightarrow \mu^+ \mu^-$ are all 
much suppressed compared with what the standard model would expect.  
\end{itemize}
The first two quark modes are apparently very hard to look for in LHC 
experiments because of background, but the $\mu^+ \mu^-$ mode can and has been 
searched for but has not yet been seen.  The latest bound from LHC is
given as \cite{pdg}:
\begin{equation}
\frac{\Gamma(h \rightarrow \mu^+ \mu^-)}{\rm SM \ prediction}  \, =   \, 0.1 \pm 2.5,
\label{Htomumu}
\end{equation}
with no event seen.  If the experimental error can be reduced further and still 
no such mode is seen, then it would be a result in favour of FSM.  Predictions
were made also of some flavour-violating modes such as $h \rightarrow \mu \tau$
at a low rate \cite{Hdecay,f+mike}, which might soon be experimentally accessible.

However, reservations made in \cite{Hdecay} on the tentative nature of these 
predictions persist since a systematic approach to calculating reaction 
amplitudes with a rotating mass matrix has not yet been developed.

\subsection{$\tilde{y}$ conservation versus $B$ and $L$ conservation}

It was shown in Section 5 that the Yukawa couplings constructed 
are invariant under $\tilde{u}(1)$ as required, which means that they conserve
$\tilde{y}$ as newly defined in {\bf [R2]}.  These Yukawa terms conserve 
also baryon number and lepton number separately.
It is then natural to ask, as we did, whether 
these conservation laws are connected, and if so in what way.  The conservation 
of $\tilde{y}$ comes in FSM from a gauge principle, namely $\tilde{u}(1)$ 
invariance, and so long as this symmetry is unbroken, the conservation has to 
hold.  On the other hand,  the conservation of baryon and lepton numbers are, 
as far as is known, only empirical with no generally accepted
theoretical basis \cite{yang}.
Besides, lepton number conservation is violated by the Majorana mass term for 
right-handed neutrinos, which term is wanted for neutrinoless double beta-decay 
and for the see-saw mechanism for explaining the very small physical masses of 
neutrinos.  However, even this term conserves $\tilde{y}$ since, according to 
the analysis given in Section 5, right-handed neutrinos have $\tilde{y} = 0$, 
though assigned lepton number 1 by convention.  In other words, $\tilde{u}(1)$ 
invariance or $\tilde{y}$ conservation admits the Yukawa terms in Section 5 as 
well as the Majorana mass term for right-handed neutrinos, so that, so long as 
no other terms are found which say otherwise and none are known so far, the 
FSM conserves $B$, but it conserves $L$ only up to the Majorana term, and so 
it allows both neutrinoless double beta-decay and the see-saw mechanism to 
operate.\footnote{But we were wrong before \cite{efgt} to identify $\tilde{y}$ 
with $B - L$ and to claim that we had found a gauge principle for $B - L$ 
conservation.}

\subsection{New mixing scheme in the $\gamma-Z-G$ complex giving deviations 
from the standard model}

We turn now to {\bf [R1]}, namely the assignment in (\ref{cframon}) of different
electric charges to different components of the colour framon, as necessitated 
by the imperative that the photon should remain massless.  This change is more 
fundamental since  it implies a mixing scheme for the 
vector bosons  different from that of the standard model, involving not only the 
photon and the $Z$ but also another vector boson we call $G$ (Section 4).  This will thus 
lead to departures from the standard model already at the tree level in the 
electroweak sector in which the standard model has already been tested against
experiment to great accuracy, and any sizeable departures from it would have 
already been ruled out.  An examination of these departures as a test of the 
FSM is thus urgently due.   

However, a thorough examination of this question is not yet possible at the
present stage of the FSM's development, as will be explained in the next 
paragraph.  Besides, given the vast amount of data and the sophistication with 
which they have been analysed as regards consistency with the SM mixing scheme, 
a thorough re-examination with the FSM scheme to the same breadth and precision 
is beyond our immediate capability.  We have therefore limited our analysis so 
far to only the following three very well measured quantities: 
\begin{itemize}
\item  (a) $m_Z - m_W$
\item  (b) $\Gamma(Z \rightarrow \ell \bar{\ell}$),
\item  (c) $\Gamma(Z \rightarrow q \bar{q}$).  
\end{itemize}
The details of this analysis will be reported in a separate paper \cite{zmixed},
 as they would occupy more space than can be allowed for in the present one.  
Here, we shall give as an example only an outline of the analysis for (a) 
$m_Z - m_W$, together with just a mention of the results on (b) and (c).

The comparison of the FSM to the standard model and to experiment can at 
present be done only at tree level because one is not yet in a position in the 
FSM to investigate loop corrections in general (see Section 6).  We propose 
therefore to adopt the following criterion.  Assuming that the deviations of 
loop corrections in FSM from SM to be of higher order in smallness, and that 
the SM itself is in agreement with experiment, we compare the tree level 
results of the two models, and if the difference in tree-level predictions 
for a certain quantity is less than the present experimental error in that 
quantity, we consider that the FSM prediction is also within that experimental 
error. 

Let us then look at the vector boson masses as example and see how the FSM 
differs from the standard model.  At tree level, the description of the 
$W$ boson is the same in the two models; only the neutral bosons are mixed 
differently.  Without mixing, $W$ and $Z$ would be degenerate in mass in either
model; it is the mixing which gives the shift in mass $m_Z - m_W$, and which
is what differs between the two models.  It is thus the quantity $m_{\rm shift}
= m_Z - m_W$, for which the predicted values of the two models are to be 
compared.  Let us then proceed
as follows.  From the experimental information on the mass and widths of the 
$W$ and the tree-level formula $m_W = \half g_2 \zeta_W$, we determine the 
vacuum expectation value of the Higgs scalar field as $\zeta_W \sim 246$ GeV, 
and the coupling $g_2$ of the flavour gauge field as $g_2^2 \sim 0.4271$.
These are the central values with the experimental errors yet to be folded in.
Supplying further the accurately measured value of the electron charge $e$, 
and the coupling $g_3$ as independently determined in perturbative QCD
and $Z$ hadronic decays, we
calculate the shifts of the $Z$ mass from the $W$ mass in the mixing schemes 
of the SM and of the FSM.  Then we compare the results and see whether the 
difference between the two predicted mass shifts remains within the bounds 
quoted by experiment, according to the criterion proposed in the preceeding
paragraph.  This may not be the usual way that precision tests for the 
electroweak theory are phrased, but here it makes the logic clearer since 
for the mass shifts $m_Z - m_W$ to be compared, the experimental error is 
dominated by that of $m_W$ which should therefore be folded in, 
not just that of the more precisely measure $m_Z$.

For the standard model then, from the Weinberg mixing formula:
\begin{equation}
\frac{1}{e^2} = \frac{1}{(g_1^{\rm SM})^2} + \frac{1}{g_2^2},
\label{eg1SM}
\end{equation}
and the previously settled values of $\zeta_W, g_2$ and $e$, we can calculate 
the value of $g_1^{\rm SM}$ and hence the tree-level prediction for the mass 
$m_Z^{\rm SM}$ of $Z$ as:
\begin{equation}
m_Z^{\rm SM} = \half\zeta_W  \sqrt{(g_1^{\rm SM})^2 + g_2^2} .
\label{MZSM}
\end{equation}

For the FSM, the tree-level mass of the $Z$, which we shall call $m_Z$, is to
be obtained as the lower non-zero eigenvalue of the mass matrix 
(\ref{Masssubmfc}) obtained in Section 4, or equivalently, after some algebra,
as the lower eigenvalue of the matrix:
\begin{equation}
\left( \begin{array}{cc} \ell (g_1^2 + g_2^2) &
    - \frac{1}{\sqrt{3}} \sqrt{k \ell} g_1^2 \\
    - \frac{1}{\sqrt{3}} \sqrt{k \ell} g_1^2 &
    k(\third g_1^2 + \quart g_3^2) \end{array} \right).
\label{MZG}
\end{equation}
with $\ell = \quart \zeta_W^2$ and $k = \twothirds (1 + R) \zeta_S^2$, and 
$g_1$ given by the relation:
\begin{equation}
\frac{1}{e^2} = \frac{1}{g_1^2} + \frac{1}{g_2^2}
    + \frac{4}{3} \frac{1}{g_3^2}
\label{eg1}
\end{equation}
obtained from FSM mixing \cite{zmixed}.  

The formulae for $m_Z^{\rm SM}$ and $m_Z$ look very different.  There is little 
freedom, all the couplings $e, g_2, g_3$ and $\zeta_W$ being known, with the 
only unknown parameter being $\zeta_S$, and even this, as argued in the next 
section, is loosely constrained by the fit in \cite{tfsm} to be of order TeV.  
On the other hand, the mass shift $m_Z - m_W$ is now given by the PDG \cite{pdg}
 to an impressive accuracy of about 15 MeV and is consistent with the SM 
predictions.  Whether the FSM predicted value for $m_Z - m_W$ would remain 
within experimental bounds seems thus a very stringent test for the model.

One is saved, however, by the following fortunate result:
\begin{itemize}
\item  {\bf [FR]} Provided that $\zeta_S \gg \zeta_W$, then to leading (zeroth) 
order when expanded in powers of $\zeta_W^2/\zeta_S^2$, the value of $m_Z - m_W$ 
given by FSM at tree level is identical to that of $m_Z^{SM} - m_W$ predicted
by SM also at tree level, whatever the values of $e, g_2, g_3$ and $\zeta_W$.
The deviation of the FSM from SM predictions for $m_Z$ is thus only of order  
$\zeta_W^2/\zeta_S^2$, which, however, is multiplied by the factor
$g_1^4$, that is, the deviation is of 
order $m_Z g_1^4 \zeta_W^2/\zeta_S^2$, where $g_1^4$ by (\ref{eg1}) is small 
(about $0.02$). 
\end{itemize}

That this holds can be shown by expanding the exact formula for the eigenvalue
of (\ref{MZG}) in powers of $\ell/k$, as is done in \cite{zmixed}, but it can 
be seen  already in standard first order perturbation theory, which may make 
the result more transparent, as follows. 

That $\zeta_S \gg \zeta_W$ means that in the matrix (\ref{MZG}) the mixing 
off-diagonal elements:
\begin{equation}
C = \frac{1}{\sqrt 3} g_1^2 \sqrt{k \ell} 
\label{Z} 
\end{equation} 
are small compared to the difference between the two diagonal elements:
\begin{equation}
A = \ell (g_1^2 + g_2^2), \ \
B = k(\third g_1^2 + \quart g_3^2).
\label{AB}
\end{equation}
Standard perturbation method then gives the lower of the two eigenvalues to 
first perturbative order as:
\begin{equation}
{m'}_Z^2 \sim \ell (g_1^2 + g_2^2) + \Delta, 
\label{MpZ2} 
\end{equation} 
where $\Delta$ can be expanded in powers of $\ell/k$ to give:
\begin{equation} 
\Delta = \frac{C^2}{A - B} \sim - \frac{\third g_1^4 \ell}
    {\third g_1^2 + \quart g_3^2} \left[1 + \frac{\ell}{k} \frac{g_1^2 + g_2^2}
    {\third g_1^2 + \quart g_3^2} + ... \right]
\label{Delta}
\end{equation}
Using (\ref{eg1}) and (\ref{eg1SM}), it is then easily checked that ${m'}_Z^2$ 
in (\ref{MpZ2}) to zeroth order in $\ell/k$ is the same as $(m_Z^{\rm SM})^2$ 
from (\ref{MZSM}), while the remainder has a factor $g_1^4 \ell/k$ as 
claimed in {\bf [FR]}.  

With {\bf [FR]}, the deviation of $m_Z$ from $m_Z^{\rm SM}$ of order $m_Z g_1^4 
(\zeta_W^2/\zeta_S^2)$, for $\zeta_S \sim 2$ TeV, is of order $10^{-4}$ times 
$m_Z$ and comparable to the present experimental error.  This is borne out by 
\cite{zmixed} using the exact formulae, which gives for $\zeta_S = 2$ TeV, or 
$m_G \sim 1$ TeV:   
\begin{equation}
(m_Z^{\rm SM} - m_W) - (m_Z - m_W) = 10.4\ {\rm MeV},
\end{equation}
well within the quoted experimental error of $15$ MeV.

The interesting thing is that, as shown in \cite{zmixed}, a similar scenario
obtains also in the deviations of the FSM from SM for the decay widths (b) and 
(c).  To leading order in the expansion in powers of $\zeta_W^2/\zeta_S^2$, the
tree-level decay widths calculated with the FSM mixing scheme are identical to 
those worked out from the standard model, whatever the values of the couplings, 
due again to the relations between $g_1$ and $g_1^{\rm SM}$ in (\ref{eg1}) and 
(\ref{eg1SM}).  Then to the next to leading order in the expansion, the 
deviation of FSM from SM is again only proportional to $g_1^4 (\zeta_W^2/\zeta_S^2)$
 and ensures that it remains within the present experimental bounds.  For 
example, in \cite{zmixed} one finds that, using the same parameters, the 
difference in $\Gamma (Z \longrightarrow e^+e^-)$ between the two schemes at 
tree level is only $0.03$ MeV, compared to the experimental error $0.12$ MeV. 

That {\bf [FR]} or its equivalent should hold in both the mass shift (a) and 
the decay widths (b) and (c) raises the suspicion of some deeper reason for 
the close agreement between SM and FSM that we have not yet understood.

Since the exact form of the mass mixing matrix for the FSM is given above in 
(\ref{MZG}), it is straightforward to work out, as it is done in \cite{zmixed},
a limit for $\zeta_S$ above which the predicted $m_Z - m_W$ will lie within the
present experimental bounds.  The same have been done also for the decay widths 
(b) and (c), and the conclusion is that so long as
$\zeta_S \geq 2$ TeV, then the tree-level predictions of the FSM will differ from those 
of the SM for all the 3 listed items (a), (b), and (c) only by amounts less 
than the present experimental errors, hence surviving the test we posed above. 

There are other hurdles yet to get over, of course, before one can claim the
FSM to be consistent with the extensive and very accurate data now available
in the electroweak sector, but the above examples are a good start, being
seemingly the most stringent.  It will be a massive programme to check consistency
of the FSM mixing scheme with all the data available, which we are not
yet in a
position to undertake.  In this paper, we shall adopt an optimistic view and 
tentatively assume that the programme will go through, so as to free ourselves 
for interesting explorations further afield.

Whatever the actual value of $\zeta_S$, however, there will be deviations of
the FSM from the SM.  Turning the argument around, therefore, one can regard
these deviations as new physics to be searched for in future experiments.  For
instance, the tree-level FSM prediction, worked out in \cite{zmixed}, for the 
mass shift $m_Z-m_W$ in the manner described above 
is actually smaller than that predicted by the standard model, namely
\begin{itemize}
\item The prediction of the FSM for $(m_Z - m_W) < (m_Z^{\rm SM} - m_W)$.
\end{itemize}  
In other words, had one started from the experimentally better measured value 
of the $Z$ mass and worked backwards to predict the $W$ mass, as is more 
usually done, then the FSM will give a $W$ mass somewhat larger than that 
predicted by the standard model.  It is amusing to note that
measurements \cite{atlasmw} at
LEP, Tevatron, and LHC so far actually all give central values for $m_W$ larger 
than that predicted by the standard model, though each by only $1$-$2$~$\sigma$. 
If the experimental error can, with more data, be further reduced, then it is 
not excluded that we shall soon be asking whether the deviation suggested by 
the FSM can in fact be observed.

Similar deviations of the FSM from the SM have been worked out in \cite{zmixed}
also for the partial widths of the decays (b) $Z \rightarrow \ell \bar{\ell}$ 
and (c) $Z \rightarrow q \bar{q}$, and these can again be regarded as new 
physics to be searched for in experiment.  Indeed, there being only the one 
parameter $\zeta_S$, these three deviations are correlated, as they are connected
also to the mass of the vector boson we call $G$, which, as will be discussed 
later in Section 8 {\bf [b]}, can probably be observed as an $e^+ e^-$ anomaly
in the multi-TeV range.  In short, a point that will be taken up again in the
next section and is expanded further in \cite{zmixed}, this complex of effects
holds out a promise not only of new physics to be tested in the standard sector
but also of an opening into the mysterious  world of framonic $C$-ons so 
far hidden from us.

\section{Mass spectra of $H$, $G$ and $F$}

Having survived, for the moment it seems, the most immediate tests in the 
standard sector, let us now proceed boldly to explore the new hidden sector 
populated by the framonic $C$-ons $H$, $G$ and $F$.  To navigate such uncharted 
waters, however, will require some audacity, supplementing theoretical results
sometimes with intuition or even just imagination based merely on hints from 
various bits of physics.  In case  our readers should think that we do so on 
occasion to excess, to them we proffer now our apology.
 
Let us start with the physical mass spectra of the $H$, $G$, and $F$.  In 
Sections 3, 4 and 5 the tree-level mass matrices for these states are derived 
already from the fundamental action.  These matrices, however, are 
scale-dependent according to Section 6, and so one has  to specify at
what 
scales to diagonalize them so as to evaluate the physical masses of these
physical particles.  It is a commonly accepted prescription that: 
\begin{itemize} 
\item {\bf (PM)} Physical massses should be evaluated at the mass-scales of the 
particles themselves, 
\end{itemize}
a criterion we have used in the FSM, in deriving results such as that 
in Table \ref{tfsmfit}.  That seems to have worked.  We 
aim to follow the same procedure now with the $H$, $G$ and $F$.  
 
The criterion means that the physical mass $m_x$ of a state $x$ is to be a 
solution of an equation of the form:
\begin{equation}
m_x(\mu) = \mu
\label{physmass}
\end{equation}
where $m_x(\mu)$ represents the scale-dependent eigenvalue of the mass matrix
corresponding to the state $x$.  We can distinguish three cases in the solutions
 of this equation, all of which we shall meet  in what follows:
\begin{itemize}
\item {\bf (C1)} There is a real positive solution to (\ref{physmass}) which is
unique.  In this case, the physical mass $m_x$ of the state $x$ is
unambiguously defined as that solution.
\item {\bf (C2)} There are two real positive solutions to (\ref{physmass}), in 
which case we define the physical mass $m_x$ of $x$ to be the lower of the
two solutions since the higher solution will be unstable against decay into the
lower one.
\item {\bf (C3)} There is no real postive solution to (\ref{physmass}), in 
which case  we shall interpret the state
$x$ as an inherent but covert degree of freedom in our theory which does not
materialize and manifest itself as a physical particle.
\end{itemize}

Let us take first the $H$ states, the tree-level mass squared matrix for which 
is given in (\ref{MH}).  This is seen to depend on the parameter $R$, which in 
turn is shown in Section 6 and \cite{tfsm} to depend on the scale $\mu$.  There 
may be further scale dependences coming from the other parameters on which the 
matrix also depends, but since these have not been studied, we can only ignore 
them for the moment and take account only of the scale dependence via $R$ that 
we know about.  This is also the philosophy adopted in our earlier work in
the standard sector, for example in \cite{tfsm}, which seems to have worked, 
although in that case it is the dependence on $\mu$ via the rotating vector 
$\balpha$ that matters, an interesting point of difference that we shall 
return to later.

\begin{figure}[th]
\centering
\includegraphics[scale=0.42]{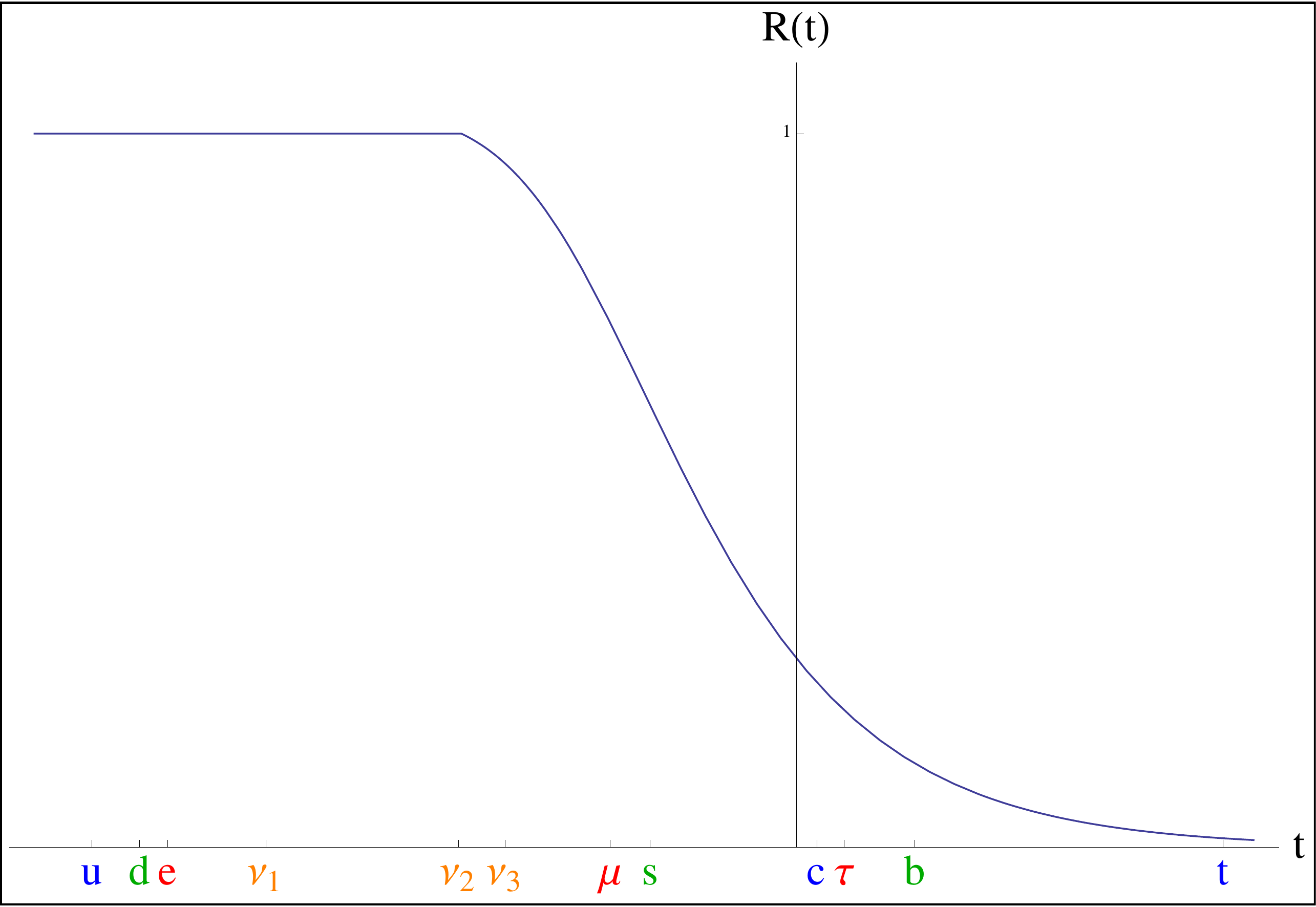}
\caption{Dependence of $R$ on scale obtained from the fit in \cite{tfsm}}
\label{Rtfsm}
\end{figure}

How then does $R$ depend on scale?  In Section 6 and \cite{tfsm}, we have 
obtained the RGEs via framon loops which are derivable from the given Yukawa 
terms.  These depend on certain parameters, and when integrated will depend on
some more integration constants, all of which are to be determined empirically 
by fitting with experiment.  At first sight, this looks difficult since these 
$H$ are still unknown:  how is one to do so?  However, one interesting 
feature---one might even be tempted to say the beauty---of the FSM scheme is 
that the hidden and standard sectors share the same vacuum, so that when the 
vacuum moves with scale, it will affect simultaneously both sectors.  This also 
means that information obtained in either sector can be used to determine how 
the vacuum moves and the conclusion will be valid for both.  Indeed, we recall 
that it is actually from the Yukawa couplings of the $F$ in the hidden 
sector that the RGEs in Section 6 and \cite{tfsm} are derived, but they are applied to 
reproduce the mass and mixing patterns of quarks and leptons in the standard 
sector.  Conversely then, now that we have determined the parameters on which 
the RGEs depend by fitting data in the standard sector, we can apply the 
result to the hidden sector to explore the mass spectra of the $H$, $G$
and $F$.  That being the case, the values of $R$ at any scale $\mu$ can just 
be read off from Figure 2 of \cite{tfsm}, reproduced here as Figure \ref{Rtfsm}
for easy reference, but of course only to the extent  that we can rely on
that result.  

As it stands, the matrix (\ref{MH}) is already almost diagonal except for the
$3 \times 3$ upper left corner block.  And apart from that labelled by
$h_W$ (the electroweak Higgs), 
the diagonal elements are of the following 3 types:
\begin{itemize}
\item (i) those with $Q = 0$ and values proportional to $1 + 2R$: 
          [$H_{\langle\tilde{3}|\tilde{3}\rangle}$];\\ \nonumber
\item (ii) those with $Q = \pm 1$ and values proportional to $\half(2
  + R)$: \\ \nonumber
      [$H_{\langle\tilde{1}|\tilde{3}\rangle}, H_{\langle\tilde{2}|\tilde{3}\rangle},
H_{\langle\tilde{3}|\tilde{1}\rangle}, 
        H_{\langle\tilde{3}|\tilde{2}\rangle}$];
\item (iii) those with $Q = 0$ and values proportional to $1 - R$: \\ \nonumber
      [$H_{\rm odd} = \tfrac{1}{\sqrt{2}}(H_{\langle\tilde{1}|\tilde{1}\rangle} 
- H_{\langle\tilde{2}|\tilde{2}\rangle}), 
        H_{\rm even} = \tfrac{1}{\sqrt{2}}(H_{\langle\tilde{1}|\tilde{1}\rangle} 
+ H_{\langle\tilde{2}|\tilde{2}\rangle}), 
        H_{\langle\tilde{1}|\tilde{2}\rangle}, H_{\langle\tilde{2}|\tilde{1}\rangle}$].  
\end{itemize}

Given that $R$ as seen in Figure \ref{Rtfsm} is about 0.02 at $\mu = m_Z$ and 
is smaller still in the TeV region that, as will be seen, interests us, we 
can largely neglect in that region the difference in $R$-dependent factors 
which distinguishes the three types (i)---(iii).  This means, first, that the 
already diagonal types (ii) and (iii) are approximately degenerate.  Secondly, 
when $R \sim 0$, the submatrix for the states labelled $H_{\langle\tilde{3}
|\tilde{3}\rangle}$ and $H_{\rm even}$ in the list above simplifies and gives as 
eigenstates:  
\begin{eqnarray}
H_{\rm high} & = & \sqrt{\tfrac{1}{3}} H_{\langle\tilde{3}|\tilde{3}\rangle} 
              + \sqrt{\tfrac{2}{3}} H_{\rm even}; \ \ \ 
              M^2 = 4(\kappa_S + 3 \lambda_S) \zeta_S^2,
\label{Hhigh}  \\
H_{\rm low}  & = & - \sqrt{\tfrac{2}{3}} H_{\langle\tilde{3}|\tilde{3}\rangle} 
              + \sqrt{\tfrac{1}{3}} H_{\rm even}; \ \ \ 
              M^2 = 4 \kappa_S \zeta_S^2,
\label{Hlow}
\end{eqnarray}
where only $H_{\rm high}$ now still mixes with $h_{\rm W}$.  The state $H_{\rm low}$ is again degenerate
with the other $H$, but $H_{\rm high}$ is singled out with a larger eigenvalue.  

To obtain the mass of the $H$ we  solve the equation 
(\ref{physmass}).  For this  we need an estimate of the vacuum
expectation value  $\zeta_S$ of 
the colour framon.  Interestingly, this same vacuum expectation value  has already been given 
a lower bound of about 2 TeV (at $\mu \sim m_Z$) (Section 7.3) when we were 
considering  $m_Z - m_W$ and $Z$-decay into quark and lepton pairs, a 
completely different area of physics to the present one.  On the other hand, 
Figure \ref{Rtfsm} gives  $R \sim 0.02$ at the $Z$ mass
scale.  If one then assumes that the dimensionless couplings $\nu_2$ and 
$\kappa_S$ appearing in $R = \nu_2 \zeta_W^2/2 \kappa_S \zeta_S^2$ both have 
values of order unity, one would obtain a value for  $\zeta_S$ of order TeV, which is not 
that far from the bound above, and converts that bound   into a 
crude, order-of-magnitude estimate for $\zeta_S$.

This means then that at $\mu \sim m_Z$ the eigenvalue $m_x(\mu)$ is of order 
TeV, that is,  larger than the scale itself for all the states listed in (i)---(iii) 
above.  As the scale increases further, the eigenvalue is expected to increase 
only logarithmically with scale and will eventually be caught up by the scale 
itself to give a solution to the equation (\ref{physmass}) in the multi-TeV 
region.  For the $H$ states of types (i) and (ii), this solution is unique 
since $m_x(\mu)$ decreases only logarithmically with decreasing scale and 
cannot catch up again with $\mu$ to give another solution, so that, by 
{\bf (C1)} above, we would obtain physical masses $m_x$ of order TeV
(or higher) for 
these states.  However, for the remaining states in (iii) with eigenvalues 
proportional to $\sqrt{1 - R}$, there will be another solution to the equation
(\ref{physmass}) as follows.  We see from Figure \ref{Rtfsm} that 
$R$ approaches $1$ rapidly at around 17 MeV so that shortly above this scale 
(whatever may be the values of the parameters $\kappa_S$ and $\zeta_S$ at this 
scale), there is bound to be another solution of the equation.  Intriguingly, 
the accuracy for the estimate of 17 MeV for this second solution is limited 
only by the credibilty of the fit in \cite{tfsm}, there being no need
to actually solve the equation (\ref{physmass}),
since what is required is that $(1 - R) \sim [17 {\rm MeV}/ 
{\rm TeV}]^2$, which is close enough to zero for us not to bother about the 
difference, especially since the curve for $R$ in Figure \ref{Rtfsm} is 
very steep there.  We conclude then by {\bf (C2)} above that the physical masses of
the states of type (iii) will be given by this second and lower solution.

These conclusions are summarized in Table \ref{Hmasses}, where we have simplified
the notation and inserted the charges as superscripts for easy reference.
\begin{table}
\center
\begin{tabular}{|l|l|l|}
\hline
Particle & State & Mass \\ 
\hline \hline
$H^0$ & mixture of $H_{\langle\tilde{3}|\tilde{3}\rangle}, H_{\rm even}$ and
            $h_{\rm W}$ 
   & $\gtrsim$ multi-TeV \\
\hline
$H^+_{\tilde{1}}$ & $H_{\langle\tilde{1}|\tilde{3}\rangle}$ & {} \\
$H^+_{\tilde{2}}$ & $H_{\langle\tilde{2}|\tilde{3}\rangle}$ & {} \\
$H^-_{\tilde{1}}$ & $H_{\langle\tilde{3}|\tilde{1}\rangle}$ & 
     $\gtrsim$ TeV \\
$H^-_{\tilde{2}}$ & $H_{\langle\tilde{3}|\tilde{2}\rangle}$ & {} \\
$H^0_{\rm low}$ & mixture of $H_{\rm even}, H_{\langle\tilde{3}|\tilde{3}\rangle}$ &{} \\
\hline
$H^0$ & $H_{\langle\tilde{1}|\tilde{2}\rangle}$ & {} \\ 
$\bar{H}^0$ & $H_{\langle\tilde{2}|\tilde{1}\rangle}$ & $\sim$ 17 MeV \\
$H^0_{\rm odd}$ & $\frac{1}{\sqrt{2}}(H_{\langle\tilde{1}|\tilde{1}\rangle}- H_{\langle\tilde{2}|\tilde{2}\rangle})$
     & {} \\
\hline \end{tabular}
\caption{Suggested spectrum of the $H$ states} 
\label{Hmasses}
\end{table}

The analysis for the $G$ states  is very similar. Apart from the mixing of $G_8$
 with $Z$ and $\gamma$, the mass matrix is already diagonal in the Gell-Mann
basis, and as seen in (\ref{CCterm}) the $G$ fall naturally into three groups,
as did the $H$, with eigenvalues proportional to, respectively: $(1 + 2R),
\half (2 + R), (1 - R)$.  Again for $\mu \sim \zeta_S$ of order TeV, $R \sim 0$ 
so that these values are all nearly degenerate, but for the last group with
eigenvalue $\propto (1 - R)$ there is a second solution for the physical mass 
at $\sim 17$ MeV near $R = 1$.  Unlike the $H$, however, the coupling 
$g_3$ here is known from  QCD, so that the spectrum depends only
on the one parameter $\zeta_S$.  Thus  once that value is known, say for 
example  from the analysis in Section 7.3, then the spectrum can actually be 
calculated.  These conclusions are summarized in Table \ref{Gmasses}, where
the equal sign in the bound denotes the mass evaluated at tree level with 
the benchmark value $\zeta_S = 2$ TeV (Section 7.3).
\begin{table}
\center
\begin{tabular}{|l|l|l|}
\hline
Particle & State & Mass \\ 
\hline \hline
$G^0$ & mixture of $G_8, Z$ and $\gamma$ & $\geq 1.1\ {\rm TeV}$ \\
\hline
$G^+$ & $\frac{1}{\sqrt{2}}[G_4 + i G_5]$ & {} \\
$G^-$ & $\frac{1}{\sqrt{2}}[G_4 - i G_5]$ & {} \\
{}    & {} &      $\geq 1.0\ {\rm TeV}$ \\
$G^{'+}$ & $\frac{1}{\sqrt{2}}[G_6 + i G_7]$ & {} \\
$G^{'-}$ & $\frac{1}{\sqrt{2}}[G_6 - i G_7]$ & {} \\
\hline
$G_1^0$ & {} & {} \\ 
$G_2^0$ & {} & $ \sim 17\ {\rm MeV}$ \\
$G_3^0$ & {} & {} \\
\hline \end{tabular} 
\caption{Suggested spectrum of the $G$ states} 
\label{Gmasses}
\end{table}

We note in Tables \ref{Hmasses} and \ref{Gmasses} the following interesting
points:

\begin{itemize}

\item {\bf [a]} 

Comparing the spectra of the $H$ and $G$ with that of the quarks and leptons, 
we find that there is a strange sort of duality between the two.  The RGE 
derived in \cite{tfsm} (see also
Section 6) gives the variations with scale of the two quantities 
$R$ and $\balpha$.  They are correlated, which is not surprising given that 
$R = \nu_2 \zeta_W^2/2 \kappa_S \zeta_S^2$ measures the relative strength of the 
$\widetilde{su}(3)$ symmetry-breaking $\nu_2$ term in the framon potential
against the $\widetilde{su}(3)$ symmetry-maintaining $\kappa_s$ term,
hence governing the amount of breaking, or in other words the direction of
$\balpha$.  The mass matrices of the quarks and leptons are independent of $R$ 
but depends on $\balpha$ whose rotation then  gives
the details of the quark and lepton spectra.  In parallel, the mass matrices 
of the $H$ and $G$ are not affected by the rotation of $\balpha$ since the
vacuum  value  of $\bPhi$ rotates covariantly with it, but they depend explicitly on $R$
and this dependence is what prescribes the spectrum.  However, despite the 
marked differences in the mechanisms by which the two spectra are generated,
and despite the fact that one concerns fermions while the other bosons, their
properties seem to echo one another.  The $H$ and $G$ each fall naturally
into three groups of decreasing masses (although the differences may not be big at
high scales where $R \sim 0$), echoing the three generations of quarks and leptons.
Of these, the lowest group has a particularly low mass because of the existence
of a second solution to the mass condition (\ref{physmass}),
 again echoing the lowest generation quarks and 
leptons which acquire their particularly low masses also by virtue of a second
solution \cite{tfsm}.  
  
\item {\bf [b]} 

Both the $H$ and $G$ spectra are such that all charged states are heavy of
order TeV while all the light states of order 17 MeV are neutral.  This is an
important property for the spectra to possess in order to remain realistically 
viable, for any charged particle of light mass is unlikely to have escaped 
detection by experiment.

\item {\bf [c]}

The heaviest states, $H^0$ in Table \ref{Hmasses} and $G^0$ in Table 
\ref{Gmasses}, are both distinguished further by being mixed states each with 
a component in the standard sector.  The mixing of the vector bosons which 
gives $G^0$ as eigenstate has already been detailed in Section 4.  For the
scalar states, the state $H_{\rm high}$ in (\ref{Hhigh}) mixes with the standard 
electroweak Higgs state $h_{\rm W}$, to give diagonal states, say, $H$ and $h$, 
where the lower mixed state $h$, which is mostly $h_{\rm W}$, is to be identified
 with the Higgs state already observed at 125 GeV, while the higher mixed state 
$H$, which is mostly $H_{\rm high}$, would be a new state yet to be observed.  

By virtue of their (small) components in the standard sector, both $H$ and
$G$ can be produced by experiments which produce the $h$ and the $Z$, and 
decay also into final states into which the $h$ and $Z$ decay.  Thus, $H$
can appear as a diphoton and the $G$ as a $\ell^+ \ell^-$ bump at LHC.  
Indeed, at one stage, a diphoton enhancement \cite{diphoton750a, diphoton750b}
was reported by ATLAS and CMS at 
a mass of around 750 GeV which could have suited $H$ although the mass is lower 
than expected, but this was in any case not confirmed by later data with higher 
statistics \cite{no750a, no750b}.  
However, looking further along these lines might well reveal the 
$H$ and $G$ in the multi-TeV range.

Of the two, $G$ looks the more promising.  Besides the lower predicted mass,
it has the virtue of depending on only the one parameter $\zeta_S$, since its 
couplings are governed by $g_3$ which is already known from perturative QCD.  
Thus, we know already its partial widths into lepton pairs and into hadrons 
\cite{zmixed}, and with more work, there is a chance that even its production 
cross section at the LHC and its total width might be estimated.  Whether
it is so is at present under investigation.

From the FSM point of view, the search for $H$ and $G$ will be extremely
worthwhile, not only for just their own sake as new particles.  By virtue of 
their being mixed states partly in the standard sector but mostly in the
hidden sector, they would serve us as valuable portals into the hidden 
sector.  Indeed, they are the only two such portals so far known to us.
For example, once produced, they would decay mostly into particles in the
hidden sector which will be all new to us, and give us a glimpse into 
that other world.

\item {\bf [d]}  

Perhaps the most striking feature of Tables \ref{Hmasses} and \ref{Gmasses} is 
the appearance of states at as low a mass as 17 MeV in a spectrum the natural 
scale of which as given by $\zeta_S$ is of order TeV.  We recall that these 
entries were made based on:
\begin{itemize}
\item (i) the $\mu$-dependence of $R$ obtained from the fit in \cite{tfsm} 
to the mass and mixing data of quarks and leptons.
\item (ii) strict adherence to the criterion ${\bf (PM)}$ that physical masses 
of particles are to be obtained from the running mass at the scale equal to the 
mass itself.
\item (iii) that when a second solution exists for the physical mass, we choose 
the lower as being more stable ${\bf (C2)}$.
\end{itemize}
Given that these are theoretically none too strong, the entries themselves are 
a rather bold assertion which can do with some phenomenological support.  

As indirect support, we can cite the example of the quark and lepton spectra 
obtained in \cite{tfsm} which was based on the same premises (i)---(iii) except 
for the replacement of (i) by the parallel $\mu$-dependence of the rotation 
of $\balpha$.  And those parallel arguments there have yielded explanations for 
the unusual properties of the lowest generation which we had previously found 
very puzzling: (i) $m_u$ of order MeV, from an input scale $m_t$ of order 100 
GeV, (ii) $m_u \sim m_d \sim m_e$ in magnitude, despite $m_c \gg m_s\gg m_\mu$,
(iii) $m_u < m_d$, despite the fact that for the higher generations, $m_t \gg 
m_b > m_\tau$ and $m_c > m_s > m_\mu$.  It is this prior experience for the quarks and 
leptons in the standard sector which gives one now some confidence to suggest 
the same interpretation ${\bf (C2)}$ of the second solution for the $H$ and 
$G$ above.

Nevertheless, some direct phenomenological support from the hidden sector 
itself will be needed for the assertion to be really creditable.  Such, of 
course, will be much harder to come by, or so we thought. One is agreeably 
surprised, however, by a recent development in an unexpected area which 
might have a bearing on this subject.  Recent observations of such phenomena 
as the $g - 2$ anomaly have led to theoretical speculations of new 
low mass particles, which in turn prompted experimental searches in this 
region.  Although most of these searches are negative, one experiment on 
excited beryllium decay \cite{Atomki}, has reported a $7 \sigma$ anomaly in 
the $e^+ e^-$ effective mass plot in the final state, which the authors 
interpret as a possible new particle with a mass of {\it coincidentally}
17 MeV, just the mass predicted for the low mass $H$ and $G$ states in 
Tables \ref{Hmasses} and \ref{Gmasses}.  The observed ratio
\begin{equation}
\frac{\Gamma(Be^{8*} \rightarrow Be^8 + X)}{\Gamma(Be^{8*} \rightarrow Be^8 
+ \gamma)}
\label{atomki}
\end{equation}
is given as $5.8 \times 10^{-6}$.  This, and the fact that it has not been
observed in other different experiments scanning this mass region, imply that
it must have rather unusual properties.  Hence, despite its not having been
independently confirmed by other experiments, the anomaly has sparked a fair 
amount of theoretical speculations \cite{yamamoto,feng,moretti}
on what this new particle could be, including
the possibility that it could be a $1^-$ state .

The interest for us is that the above considerations in FSM do predict new
particles in the hidden sector at just that mass region.  Of these particles,
that labelled $G_3$ in Table \ref{Gmasses}, in particular, can decay with a 
small width into $e^+ e^-$ via a framon loop (the framon being charged), and
can thus be a candidate for the anomaly.  We have not understood enough of FSM
dynamics as yet to ascertain whether its candidacy is indeed viable, given the
intricacy of the many bounds from other experiments which a candidate has to
satisfy.  However, if it is, and the anomaly itself should survive independent 
scrutiny, then this can be a big boost to the credibility of both, and to FSM 
as a whole.  If confirmed, then this $G_3$, like $G$ in {\bf [c]} above, will
serve as a useful window into the hidden sector.

\end{itemize}

Next, we turn to the mass spectrum of the $F$.  This depends on the Yukawa 
couplings (\ref{Yukawacq}), (\ref{Yukawacl1}), and (\ref{Yukawacl2}), 
which are, as said, but working models, and does not therefore deserve our 
confidence as much as the preceding spectra for the $H$ and $G$.

Nevertheless, to proceed with the analysis, we note first that in the lists 
(\ref{coquarks}), (\ref{coleptons1}) and (\ref{coleptons2}), those $F$ of type
 (i) which are coupled via $\bdelta$ and $\bdelta'$ will have very different 
properties from those of type (ii) coupled via $\balpha$.  Those of type (i) 
have mass matrix elements proportional to $\sqrt{1-R}$ while those of type (ii) 
have mass matrix elements proportional to $\sqrt{1+2R}$.   Further, we recall 
(Section 6) that the vector $\bzeta$ is supposed  to rotate with scale starting 
from a fixed point $(0, 0, 1)$ at infinite scale, so that at high but finite 
scales these coefficients will be hierarchical, meaning that $\zeta_\alpha \gg 
\zeta_\delta, \zeta_{\delta'}$.

Let us consider first the three  $F$ of type (ii).  They will have mass matrix 
elements of large finite values at high scales, decreasing logarithmically by 
virtue of renormalization with decreasing scales, giving thus case {\bf (C1)} 
solutions for their physical masses of order TeV in close parallel to the $H$ 
and $G$ studied above.  The six $F$ of type (i), on the other hand, will start 
at infinite scale with zero mass matrix elements because of the vanishing there 
of $\zeta_\delta$ and $\zeta_{\delta'}$.  These elements may increase as the 
scale lowers and $\bzeta$ rotates, but only slowly, if at all, since the 
overall factor $\zeta_S$ decreases.  However, the mass matrix elements have to 
vanish again as the scale nears 17 MeV because of the factor $\sqrt{1-R}$ they 
carry.  It is thus not obvious whether the matrix elements will ever match the 
scale and give solutions to the phyiscal masses.  In other words, it is unclear 
whether the $F$ of type (i) will be the case {\bf (C3)} or the case {\bf (C2)} 
as above listed.  For the charged co-quarks of (\ref{coquarks}) and charged 
co-leptons of (\ref{coleptons1}), however, it had better be the case {\bf (C3)},
 or otherwise one would have  light charged co-quarks and co-leptons with masses
 of order 17 MeV.  This would be unacceptable because any 
charged particle at this mass would be unlikely to have escaped experimental 
detection.  Thus, if this should happen, then we would have to give up the 
model of Yukawa couplings (\ref{coquarks})---(\ref{coleptons2}) altogether and
construct another.  For the two co-neutrinos of type (i) in (\ref{coleptons2}), 
however, case {\bf (C2)} solutions are not ruled out. 

Ignoring then, tentatively, the $F$ of type (i), there remain only the three 
states of type (ii) to consider: the co-quark with charge $-\half$ of 
(\ref{coquarks}), the co-lepton with charge $-1$ of (\ref{coleptons2}), and 
the co-neutrino of (\ref{coleptons1}), with coefficients $Z$ (and hence also 
masses) which, we recall, are likely to be hierarchical, presumably in that 
order according to {\bf (RT)}.  For the co-neutrinos, however, there is the 
possibility of the following added complication:

\begin{itemize} 
 
\item {\bf [e]}  
 
These states, which have zero charges, can form Majorana-type mass terms 
(Section 7.2) as right-handed neutrinos do in the standard sector.  If so, 
then they may be subject to a see-saw mechanism, which may give them lower 
(perhaps very much lower) masses than the TeV scale originally suggested.
(We recall also that by {\bf [RT]}, co-neutrinos are likely to start off in
any case with a much lower value for $Z$.)  Whether this is indeed the case, 
and what physical masses they will end up with, will be crucial to the 
question to be discussed in Section 10, namely whether $H_{\rm low}$ and 
$G_{\rm low}$ are stable and be part of the dark matter, or whether they will 
decay into co-neutrino-anti-co-neutrino pairs and thus be absent in the 
universe today.  

\end{itemize}

Taken all together then, the above arguments suggest the spectrum for $F$
listed in Table \ref{Fmasses}

\begin{itemize}
\item {\bf [f]}
The charged co-quarks in Table \ref{Fmasses} would give a step rise of size 
$\quart$ in $R = \sigma_{\rm total}/\sigma_{\rm elastic}$ in $e^+ e^-$ collision, while the 
charged leptons would give a step rise of $1$, but both only at
centre-of-mass  energies of order 
TeV, which are, however, much beyond the reach of all planned colliders at 
present.
\end{itemize}

\begin{itemize}

\item {\bf [g]}
There is, however, one question about the particle spectrum which we cannot
answer  at present, and which arises as follows.  Among the
fundamental fermion fields listed in (\ref{fundferm}) above, there is one
which stands out, namely $\psi_L(\sixth, 2, 3)$, which, unlike any other, 
carries both flavour and colour.  According to the above treatment, it combines 
with the flavour framon to form a quark, and with the colour framon to form a
co-quark, neither of which can propagate in free space, the quark being still
coloured and the co-quark, flavoured.
A quark can combine with other quarks and antiquarks via
colour confinement to form hadrons.  In parallel, we expected that co-quarks
would combine with other co-quarks and anti-co-quarks to form co-hadrons.
But cannot a quark combine with a colour anti-framon via colour confinement, 
or equivalently, a co-quark combine with a flavour anti-framon via flavour
confinement, to form a doubly framonic, simultaneous $B$-on and $C$-on?
Let us say, symbolically $\Phi^\dagger \bPhi^\dagger \psi_L(\sixth, 2, 3)$, which, for
want of anything better at present, we may call tentatively a ``lepto-$F$''.  
Conceptually, there does not seem to be any principle against ``lepto-$F$''s, 
but we are at a loss to ascertain whether they ought to exist, or to speculate
on their properties, since the tools we have been using for the other particles
do not seem to extend easily to this class of objects.  Intuitively, one would
guess that they are point-like, like framonic $B$-ons and $C$-ons, but perhaps
even more so, being doubly framonic.  They will probably be hard to form in 
the early universe, because, like baryons, they are trinary objects, but 
much smaller in size, and so they may be rather rare in nature.  But if they
exist, they may disturb  the picture we have built above
of two parallel worlds, one of framonic $B$-ons and one of framonic $C$-ons,
for, being doubly framonic they would belong to, and communicate with, both 
sectors, hence breaking the dichotomy, and playing a possibly crucial role
in the FSM framework.

Unable at present to determine whether ``lepto-$F$'' may or may not exist, we 
shall work on the hypothesis that even if they do, they will have such a high 
mass, or else interact with the rest so weakly, that the dichotomous picture 
we have been developing will still hold to a good approximation, and leave 
the problem to be sorted out, we hope, in the future.  

If the ``lepto-F'' does exist, then its right-handed component, being a bound
state of the flavour framon via flavour confinement with the right-handed $F$, 
$\psi_R(-\half, 2, 1)$, is exactly the answer to the question posed in Section 
5 at the end of the paragraph after equation (\ref{fundferm}).  That being the case, 
the chirality puzzle {\bf [CH]} simplifies with (\ref{fundferm}), since 
we require now  only the single field 
$\psi(\sixth, 2, 3)$ (but no longer $\psi(-\half, 2, 1)$) to be purely
left-handed, which might, 
in the long run, be a little easier to explain.

\end{itemize}

\begin{table}
\center
\begin{tabular}{|l|l|l||l|}
\hline
Particle & Charge & Mass & Remark \\ 
\hline \hline
co-quark &  $+ \half$ & $\gtrsim {\rm TeV}$ & {} \\
\hline
co-lepton & -1 & $\lesssim {\rm TeV}$ & {}\\
\hline
co-neutrino & 0 & $\ll {\rm TeV}$? & see-saw? \\
\hline \end{tabular} 
\caption{Suggested spectrum of the $F$ states} 
\label{Fmasses}
\end{table}

\section{Interactions of $H$, $G$ and $F$}

The situation as regards the tree-level couplings of the $H$, $G$, and $F$ as
derived from the FSM action is summarized at the beginning of Section 6.  The 
details of all these couplings are rather complicated, as are seen especially
in Appendices A and B, which would imply a lot of
complexity in the interactions of framonic $C$-ons among themselves.  They may 
be of interest in the future, but need not bother us at present.  One point to 
note, however, is that despite all these myriad interactions, our derivations 
have not revealed any couplings linking the $H$, $G$, and $F$ directly to the 
quarks and leptons with which we do our experiments.  Indeed, the only 
couplings we found which link these framonic $C$-ons to the framonic $B$-ons 
which constitute our standard sector are the few in Appendix A proportional to 
$\nu_1$ and $\nu_2$ linking the $H_K$ to the $h_W$.  

How then can the two sectors communicate?  As answers to this question, we can
identify only the following.
\begin{itemize}
\item {\bf [bc1]} A framonic $C$-on can interact with a framonic $B$-on, if
both are charged, by exchanging a photon,
\end{itemize}
the photon being mostly the $u(1)$ gauge boson $A_\mu$ and this couples to both
sides.  But they cannot interact by exchanging either a flavour or a colour 
gauge boson since a framonic $B$-on is, by definition, neutral in flavour and a 
framonic $C$-on, neutral in colour.  Furthermore, they cannot easily interact 
by exchanging one of their own members,  given the paucity of direct couplings linking the two groups; 
unless the particle exchanged happens to be a mixed state which is partly a 
$B$-on and partly a $C$-on.  Of such, we have seen above 4 examples, namely $h$ 
and $H$ which are mixtures of the usual (framonic $B$-on) Higgs $h_W$ with the 
framonic $C$-on $H_{\rm high}$, plus $Z$ and $G$, which are mixtures of $\gamma$, 
the framonic $B$-on $Z$ (of SM) and the framonic $C$-on $G_8$, as 
described in Sections 4 and 7.  These are the only examples of mixed states we
know, hence
\begin{itemize}
\item {\bf [bc2]} A framonic $C$-on can interact with a framonic $B$-on by 
exchanging the mixed vector states $Z$, $G$ or the mixed scalar states $h$, $H$.
\end{itemize} 
Further,
\begin{itemize}
\item {\bf [bc3]} In the case where one of the  interacting particles is $h_W$ or an 
$H_K$, then the interaction can also go by exchanging $h_W$ or an $H_K$ via the 
couplings proportional to $\nu_1$ or $\nu_2$ of Appendix A.  
\end{itemize}
But these are all we have found, and it does not amount to very much.  Taken at 
face value then, it would seem that, despite each sector having plenty of 
interactions within itself, the sector composed of framonic $B$-ons and the 
sector composed of framonic $C$-ons have but little communcation with each 
other.  As we shall see, this will go some way towards explaining why the 
framonic $C$-on sector containing the $H$, $G$ and $F$ should appear ``hidden'' 
to us as our sector containing quarks, leptons and us should appear ``hidden'' 
to them.

However, this is so only if we assume that the $H$, $G$ and $F$ have no 
soft interactions, as argued in the Introduction, for if there were such soft
interactions with strength much superior to the hard interactions derived 
perturbatively from the fundamental action, then they would dominate over the 
latter, making the suggestion of the sector being
``hidden'' completely irrelevant.

The centre of our attention is thus shifted to the question whether framonic
$C$-ons have soft interactions or not.  In the Introduction, we have outlined
already an argument suggesting that framonic $C$-ons have no, or at least very
little, soft interactions.  Here we shall just flesh out this argument some 
more, although we shall still be very far from clinching it.

In the Introduction, we started by asking the question why hadrons have soft 
interactions while the framonic $B$-ons (quarks, leptons and so on) seem to have 
none.  We say that framonic $B$-ons have no soft interaction because they all 
appear point-like to the furthest extent that we have probed, unlike hadrons 
which we have already seen to be bulky, that is,  about a fermi in size.  But are we sure
that the $B$-on interactions we do see, which we have ascribed to the hard
couplings derived before are not partly soft interactions in disguise?
Let us take
the decay of the standard Higgs boson $h_W \rightarrow \ell^+ \ell^-$ as an
example.  This can arise via the standard Yukawa coupling (``hard'') but it
can also arise by the  framon constituents of $h_W$ separating and recombining 
with a fundamental fermion-antifermion pair emerging from the sea as pictured 
in Figure \ref{qdiag}(b) (``soft'').  Can the two processes be distinguished from
one another?  In this case, the answer is ``yes'', and quite easily.  We recall
that the Yukawa couplings of $h_W$ to fermions depend critically on the fermion
mass.  In the SM they are simply proportional to the mass, and in the FSM they
come from the rotation of $\balpha$, but in either case, the branching
ratios satisfy
$\Gamma (h \rightarrow \tau^+ \tau^-) \gg \Gamma (h \rightarrow \mu^+ \mu^-)
\gg \Gamma (h \rightarrow e^+ e^-)$, recalling that the physical $h$
in SM is $h_W$, and in FSM it is mostly $h_W$.  
This condition is very well satisfied by
experiment, where $h \rightarrow \tau^+ \tau^-$ is relatively copious, but
$h \rightarrow \mu^+ \mu^-$ has not been seen, let alone the mode 
into $e^+ e^-$.  (See also Section 7.1 above.)  On the other hand, if the 
decay had occurred via soft interactions as pictured in Figure \ref{qdiag}(b), 
one would expect instead from soft hadron phenomenological lore that the light 
decay products be favoured over the heavy ones.  For example, we have from the
PDG tables \cite{pdg} the following:
\begin{eqnarray}
f_2(1270): \ \ r & = & 84.8/4.6 = 18.4 \\
\rho_3(1690): \ \ r & = & 23.6/1.58 = 14.9 \\
f_4(2050): \ \ r & = & 17.0/0.68 = 25.0
\label{piKratio}
\end{eqnarray}
for the ratio $r = (BR: X \rightarrow \pi \pi)/(BR: X \rightarrow K \bar{K})$.
In all three typical soft decays, the light products ($\pi \pi$) are favoured over 
the heavy ones ($K K$), that is,  directly opposite to what was seen above for $h$
decay.  It would thus seem that if there is any soft component at all in $h$
leptonic decay, it has to be very weak.

It has already been suggested in the Introduction that this difference between
hadrons being dominated by soft interactions and the framonic $B$-ons having
apparently none might be ascribed, not to possible differences in the soft
dynamics between colour $su(3)$ and flavour $su(2)$, but to the fact that hadrons
are not framonic while the framonic $B$-ons are.  It was thus argued that, for 
example, the framon and antiframon constituents of the $h$ are so short-lived 
that they will have no time to seek out and recombine each with an alternative 
partner from the sea to effect a soft decay.  If that is the case, then the 
same argument when applied to the $H$, $G$ and $F$ would suggest that they
too will have little soft interactions and remain point-like, just as framonic
$B$-ons do.   

But do we in fact know enough of the parameters of the model to make estimates 
of the framonic life time and the recombination time in soft decays so as to 
make the above assertion creditable?  An estimate of the recombination time, 
that is,  the time for a quark to find and recombine with an antiquark 
coming from the sea, can be estimated from the typical widths of hadronic 
resonances, say of order 100 MeV, corresponding to a life time of  $\sim 7 \times 
10^{-24}$ s.  The life time of the colour framon is supposedly given by 
$\sqrt{\mu_S}$, which parameter is as yet unknown but its flavour analogue 
$\sqrt{\mu_W}$ is.  Using the long known value 246 GeV for the vacuum
expectation value of the Higgs
scalar field, and the more recently known value 125 GeV of the Higgs mass, one
obtains $\sqrt{\mu_W} \sim 88\ {\rm Gev}$, corresponding to $ \sim 7.5 \times 10^{-27}$ s.  
Thus, if $\mu_S$ is anywhere near the same order as $\mu_W$, one would conclude 
that the chance of a coloured constituent from a framonic $C$-on combining with 
a quark partner from the sea within the framonic life time so as to effect a 
soft decay for the framonic $C$-on would indeed be very small. 

\begin{figure}[thb]
\centering
\includegraphics[scale=0.5]{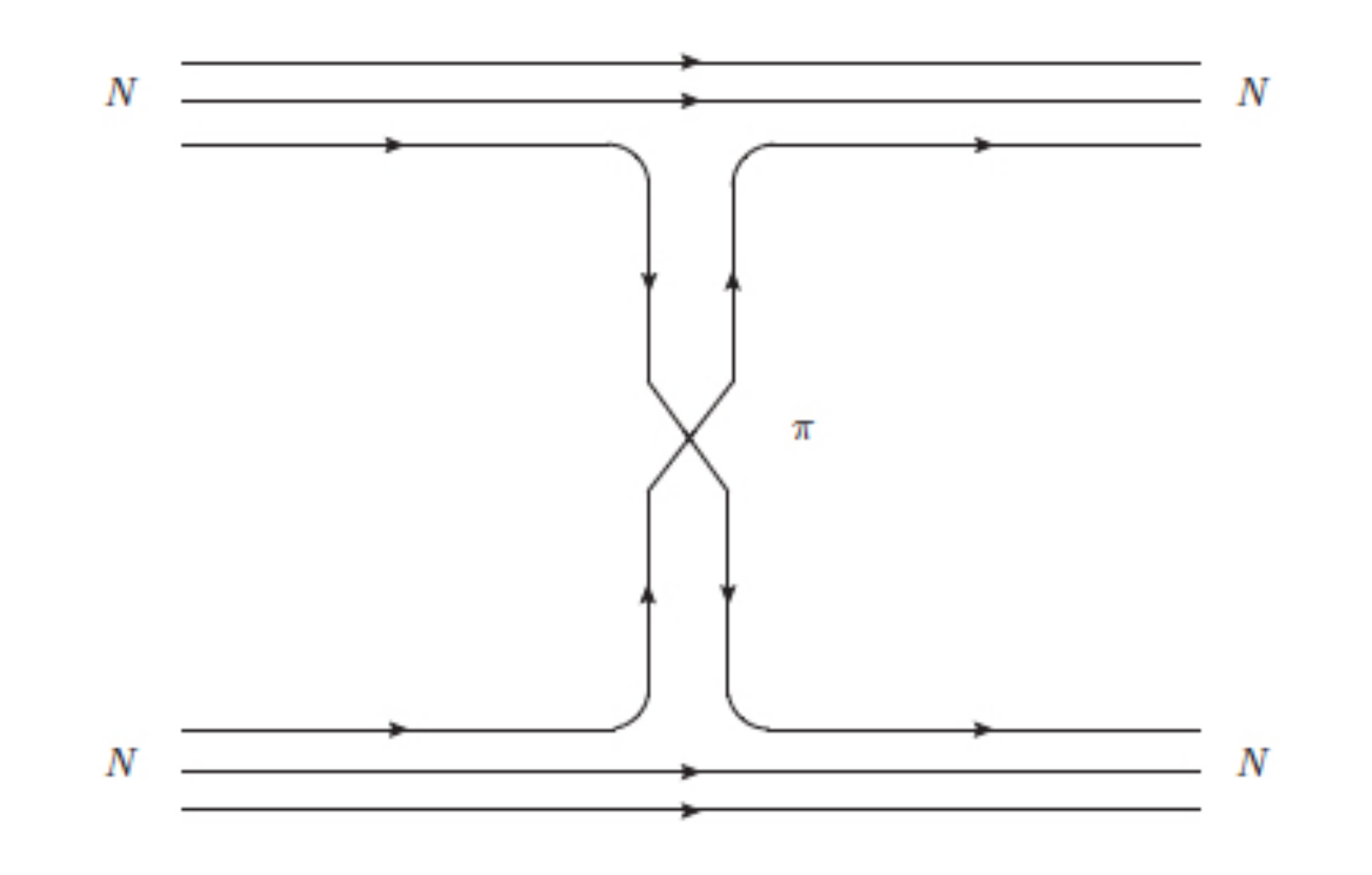}
\caption{Quark diagram of one-pion exchange giving nuclear force.}
\label{1piex}
\end{figure}

So far we have only given examples in soft decays, but similar arguments would 
apply also to suppress other soft interactions for framonic C-ons.
Let us take for example:

\begin{itemize}

\item {\bf S1} {\it No nuclear-like forces between framonic C-ons?} Nuclear 
forces between two nucleons, at least at long range, are generally thought to 
arise from one-pion exchange, which may be depicted as in Figure \ref{1piex}.  
This we interpret as a nucleon splitting off a quark which then combines with 
an antiquark from a quark-antiquark pair
coming from the sea to form a pion, while the remaining diquark 
combines with the other quark and goes off.  And then the pion gets exchanged
to the other nucleon and do the reverse.  The analogue for a nuclear-like force 
between two $H$s would be one depicted by Figure \ref{s1Hex} with the two $H$s 
exchanging another $H$.  This, one sees, will require the framon constituents 
in one $H$ to split and recombine with framons created from the sea, which we 
already claimed above to be very unlikely.  The process can go, of 
course, via the hard couplings listed in Appendices A and B, but these latter 
will not have hadronic or nuclear strength.

\begin{figure}
\centering
\includegraphics[scale=0.6]{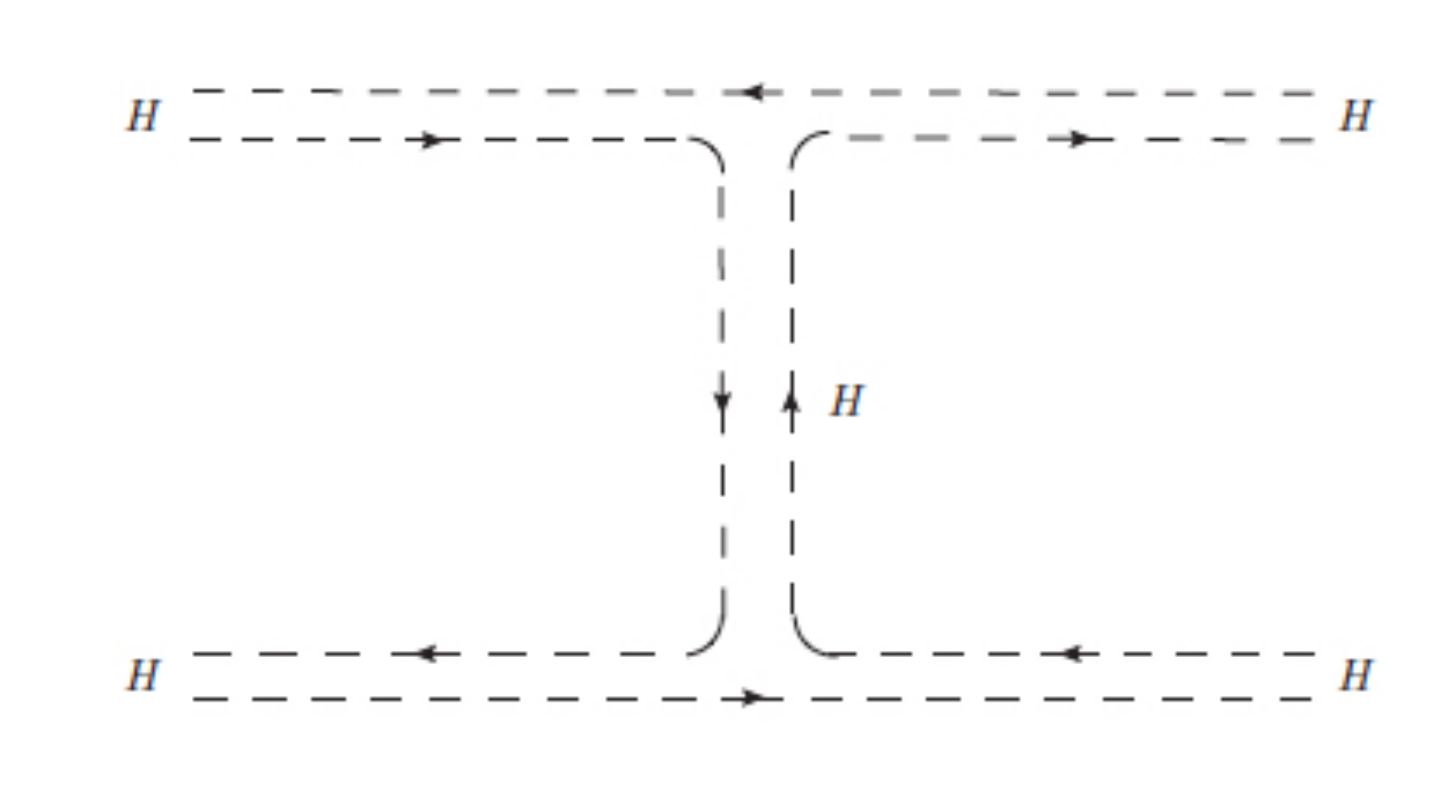}
\caption{Hypothetical quark-like diagram of one $H$-exchange giving ``soft'' 
interaction between two $H$s.}
\label{s1Hex}
\end{figure}

\item {\bf S2} {\it No framonic jets in hadron collisions?} A quark jet is 
supposedly produced in hadron collisions by hard collisions knocking off from 
the hadron a quark which then hadronizes, meaning that it combines with other 
quarks from the sea to form other hadrons which then emerge.  In a similar way, 
a framon constituent (if any) in the original hadron can be knocked
off by a hard 
collision, but our previous argument would suggest that it will not live long 
enough to find other partners from the sea to form new hadrons and to emerge as 
a jet.  Conversely, a quark knocked off by the hard collision will also find it 
hard to find a framonic partner from the sea to emerge as an $F$, the framon 
from the sea not being long-lived enough.  Hence, we argue, there will be no 
framonic jets either way.

\item {\bf S3} {\it No change of $R$ in $e^+ e^-$ collision at framon-antiframon
threshold?} The colour framon being charged, the $e^+ e^-$ can create via the
intermediate photon a framon-antiframon pair and so might seem to affect $R$.
However, the framons, being coloured, have to combine with some other coloured
object to form a colour singlet before it can emerge in the final state.  Hence,
by the same argument as in {\bf S2} above against hadronization, we claim that
it will not have time to do so.  Besides, in the parallel flavour case, the
coupling of the photon to a flavour framon-antiframon pair has also no effect
on $R$.  There, a change in $R$ occurs only at the thresholds of framonic $B$-on
pairs such as $q \bar{q}$ or $\ell \bar{\ell}$.  So also here, one would expect
changes in $R$ only at the production thresholds of the framonic $C$-ons such 
as $Q \bar{Q}$ and $L \bar{L}$ as suggested in Section 8 {\bf [e]}.

\end{itemize}

If one accepts the above suggestion that framonic $B$-ons and $C$-ons have 
no or little soft interactions, one is led to two, at first sight, rather 
astonishing conclusions.  First,

\begin{itemize}

\item {\bf [a]} 

{\it A quite revolutionarily new anwser to the old question why strong 
interactions are strong and weak interactions are weak.}  When we look at the 
actual values of the flavour and the colour gauge couplings $g_2$ and $g_3$, 
there seems to be not that much between them for size.  Why then should one 
find such disparity between the strengths of the weak and strong interactions?  
It used to be that one can give as reason that flavour is sponaneously broken 
and colour confined, but if we accept 't~Hooft's confinement picture for the 
electroweak theory, this distinction is removed.  The new answer instead would 
seem to be that the particles we know in the flavour sector, that is,  again 
the Higgs, $W^\pm, \gamma-Z^0$, quarks and leptons, are all framonic, and those 
particles we know in the colour sector, that is,  mesons and baryons, are all 
nonframonic, and in both sectors, nonframonic bound state can have soft 
interactions but framonic bound states have none, and soft interactions are 
strong but hard interactions are weak.  

\end{itemize}

Conclusion {\bf [a]} restores the parity between the flavour and colour sectors 
as far as the strengths of interactions are concerned, but begs, of course, the 
equally intriguing question why one has seen in nature so far only framonic 
$B$-ons and the nonframonic $C$-ons which are the hadrons.  To answer this question, let us 
first correct an inaccuracy in it.  When we said that hadrons are nonframonic 
as $C$-ons, we only meant that they contain no colour framons as constituent.  
Hadrons are made up of quarks and antiquarks but these, in the confinement 
picture, are themselves compound states of the flavour framon with fundamental 
fermion fields, confined by flavour.  Hadrons are thus, in this language, 
second-level constructs made by colour confinement out of the more basic and
point-like framonic $B$-ons, quarks and antiquarks.  We note that they are not 
made out of just the fundamental fermions themselves, presumably because 
hadrons are relatively large objects, and at that size-scale, the fundamental 
fermions are already confined by flavour with flavour framons to form the 
point-like flavour singlet states quarks and antiquarks.  In parallel, one 
expects framonic $C$-ons, if flavoured, to combine also via flavour confinement
to form co-hadrons which are analogues of hadrons, only with the role of 
flavour and colour interchanged.  And these, like hadrons, are in a sense only
second-level constructs made out of framonic $C$-ons and are presumably also 
fairly bulky objects.  In this language then the question asked above simplifies 
to merely why we have seen in experiments so far only framonic $B$-ons but not 
framonic $C$-ons.  And this is answerable by the second astonishing conclusion 
already mentioned, namely 

\begin{itemize}

\item {\bf [b]}

{\it That, with little interaction between framonic $B$-ons and framonic 
$C$-ons, the latter may well exist in abundance without us humans made out of 
the former being immediately aware of it.}  Whether that is indeed the case, or
in other words whether we can answer the question {\bf (Q)} or {\bf (Q')} posed
at the beginning, has to be postponed until we have considered what framonic
$C$-ons are likely to occur naturally in the world today.  But if it is, then 
we would end up with the picture given in the Introduction of two worlds which 
are not communicating much with each other, although each is as complex and 
interactive within itself as the other.  We choose to call ours (the framonic 
$B$-on world) the standard sector, but the other (the framonic $C$-on world) 
the hidden sector, only because we ourselves happen to be composed of framonic 
$B$-ons.  This lack of communication between the two sectors is not absolute.  
There are the interactions {\bf [bc1]}---{\bf [bc3]} already listed which can be 
used by us to probe the hidden sector.  Besides, the suppression of soft 
interactions in framonic $B$-ons and $C$-ons as argued is not due to any 
conservation law or selection rule, only to lack of probability. This means 
that these soft interactions may still occur though generally at a very low 
rate, that is, except in unusual circumstances as could exist in, for example, 
the early universe.  

\end{itemize}

\section{A speculative survey of the hidden sector}

Let us try next to imagine what the world would look like if there were indeed
such a ``hidden'' sector in the particle spectrum.  This would be instructive, 
we think, even though, given the flimsy knowledge we have at present, any 
picture our imagination can produce is bound to be inaccurate and incomplete.

In the very early universe when it was still very hot and dense and when 
presumably both flavour and colour were deconfined, the fundamental particles 
including framons of both types would be swimming around free.  When the 
universe began to expand and cool, however, these particles would be looking 
for appropriate partners to form flavour and colour neutral objects so as to 
survive into the next epoch when both flavour and colour would be confined.  
The first objects to form, it seems, would be the tightly bound (in the sense 
of appearing point-like at scales familiar to us) framonic $B$-ons and $C$-ons, 
and later the more loosely bound hadrons and co-hadrons, these being bound 
states by respectively colour and flavour confinement of the coloured framonic 
$B$-ons (quarks) and flavoured framonic $C$-ons (co-quarks).  It is 
interesting---perhaps even sobering---to note that baryons, which are destined 
to become the staple of our standard sector, would appear to be late-comers and
of a rather rare occurence in this formation, requiring as they would the 
concurrence of three objects of the appropriate combination of colours.  It
would be rare compared with almost all the other particles, namely the framonic 
$B$-ons (for example leptons), $C$-ons (for example  $H$, $G$ and $F$) and 
even the co-baryons, which require the concurrence of only two objects of 
the appropriate flavour or colour. This observation may go towards an eventual 
understanding of why dark matter makes up the bulk of our material universe.

As soon as formed, some of the composite particles would start to decay. In the 
standard sector, the scenario is familiar.  All the unstable particles would 
have decayed away by our epoch, leaving only the stable few: the photon, the 
neutrinos, the electron, the proton, and on some occasions the neutron too if
it happens by luck to be bound up in a stable nucleus.  

What would happen in the hidden sector?  There, besides the already familiar 
$H$, $G$ and $F$, we should take account of the co-hadrons too. These are 
of two general types, co-mesons made up of a co-quark and an anti-co-quark, 
thus $Q_r\bar{Q}_r$, and co-baryons made up of two co-quarks $\epsilon_{rs} 
Q_rQ_s$.  We note that unlike ordinary baryons, the co-baryons are bosonic, 
not fermionic.  Like ordinary hadrons, however, co-hadrons are probably bulky 
objects, liable to soft interactions, and likely to exist in many excited 
(resonance) states.  The excited states, being short-lived, would have decayed
quickly via soft interactions into the ground state, so that by now we have
only these latter to contend with.  We do not know whether the $H$, $G$ and
$F$ exist also in excited states.  For their counterparts in the flavour 
sector, that is,  the framonic $B$-ons, no excited states are known, but this 
may only mean that they are too high in mass to have been detected.  In any 
case, even if excited states for the framonic $C$-ons exist, they can all easily
decay into the ground states by emitting $H_{\rm light}$ and $G_{\rm  light}$, namely
those with masses $\sim 17$ MeV, and into $F_{\rm light} \bar{F}_{\rm light}$ 
pairs via many of the couplings listed in Appendices A, B and Section 5, 
leaving again only their ground states to be considered.  Let us then go 
through the remaining ground state particles in the hidden sector and see 
what one might expect to happen to them.

Let us start with the heaviest states $H^0$ and $G^0$ in Tables \ref{Hmasses} 
and \ref{Gmasses}.  Both of these can easily decay into $H_{\rm light}$ and
$G_{\rm light}$ via the couplings listed in Appendices A and B, and also into 
ordinary particles via their mixing respectively with $h_W$ and $\gamma-Z$, 
so that by now no such particles would remain.  Next, consider the charged $H$
and $G$ which, we recall, are still of order TeV in mass.  Those of opposite 
charges would have attracted one another and annihilated into $H_{\rm light}$, 
$G_{\rm light}$, and even into ordinary light particles, but some might have 
survived.  These survivors, however, could not decay further into $H_{\rm light}$ 
or $G_{\rm light}$ for these latter are neutral.  Like their $B$-on counterparts, 
they are more likely, if the masses involved permit, to decay into 
fermion-antifermion pairs, for example into a light co-neutrino plus a heavy 
charged anti-co-lepton, or vice versa.  If this decay is permitted, then again 
no heavy charged $H$ and $G$ would remain.  Of the remaining light states in 
Table \ref{Hmasses}, $H_{\rm odd}$, can decay further into $e^+ e^-$ by mixing 
with $h_W$ via a framon loop and disappear, and so would $G_3$, one of the 
$G_{\rm light}$, by coupling to  $\gamma$ via a loop.  The rest would be stable 
unless co-neutrinos can acquire a mass lower than 8 MeV via a see-saw 
mechanism, as envisaged in Section 8 {\bf [e]}, in which case none of the $H$ 
and $G$ would survive.

We recall that, partly by choice, all charged $F$ are heavy (that is, with 
mases of order TeV) while all light $F$ (that is,  with masses $\leq 17$
MeV) are neutral, so that unless there are other couplings than those we have
derived from the action above, there is no way for the heavy charged $F$ to
decay.  As pointed out in Section 5, co-quarks and co-leptons have no 
equivalent to the generations of quarks and leptons, so that there are no 
lighter charged states for them to decay into via normal hard
couplings derivable from the action.  However, given that the suppression of 
soft interactions suggested in Section 9 is not meant to be absolute, it is
possible that some residue of soft interaction would allow the charged co-quark 
and co-lepton to decay into ordinary charged mesons (e.g. $\pi^\pm$) plus 
co-neutrinos.  We think that is likely, although we have not been able to work 
it through.  If not, then some heavy co-quarks and co-leptons would remain, 
forming co-protons and perhaps even co-hydrogen atoms.  We do not know whether 
this latter scenario has any chance at all of being viable.

However, independently of what happens to the charged co-quark and/or co-lepton,
the light $H$, light $G$ and light co-neutrino will be there in abundance, both 
from formation in the early universe, and as decay products from the heavier 
states as described.  And these, being neutral, cannot interact with our 
standard sector via the exchange of a photon {\bf [bc1]}.  They can, however,
interact with particles in the standard sector via {\bf [bc2]} by exchanging a
$Z$ as neutrinos do, except that the interaction rate will be suppressed 
compared with that for neutrinos by the smallness of the component of $Z$ in 
$G^8$ which is what allows the $Z$ to couple into the framonic $C$-on sector.  
Now this mixing element has already been calculated in \cite{zmixed} and given 
there (in equations (28), (60)) a value of around 0.16, for $\zeta_S$ at its 
smallest permitted value of about 2 TeV.  This means that the light $H$, light 
$G$ and light co-neutrinos are all estimated to interact with ordinary matter
via $Z$-exchange with cross sections of no more than a few percent of that for 
neutrinos.  They can interact with the standard sector also by exchanging the 
other mixed vector state $G$, but the cross section will be even smaller, 
because $G$ has both a higher mass and a smaller component in $Z$.  Further, 
the light $H$, light $G$ and light co-neutrinos can interact with the standard 
sector via {\bf [bc2]} also by exchanging an $h$ or an $H$, of which the mixing 
is less known though similar.  But these interactions can be safely neglected 
compared to $Z$-exchange, since they all involve the $h_W$-coupling to light 
quarks and leptons, which are all that we have naturally occurring at present 
in the standard sector.  The same can be said as well for the interactions of 
the light $H$ via {\bf [bc3]} by exchanging $h_W$.  Hence, having gone through 
the already limited list in Section 9 above and found no interactions with our 
sector of significance, we suggest that the light $H$, the light $G$ and the 
light co-neutrinos may all qualify as candidates for dark matter.  They
are light, of mass $\leq 17$ MeV, but very numerous compared with baryons as 
already noted, and could thus make up an appreciable component of the missing 
dark matter, although without further investigation one cannot
ascertain whether they could make up the bulk of it.
But, being light, they would have escaped detection by current 
dark matter experiments  which gives stringent bounds on 
heavy dark matter particles but almost no constraint on dark matter particles 
in the mass range of our concern \cite{darkmatter}.

This scenario suggested by the spectra for $H$, $G$, and $F$ listed in Tables 
\ref{Hmasses}---\ref{Fmasses} would seem to give a tentative answer also to 
the question {\bf Q} or {\bf Q'} posed at the beginning.  The heavy states are 
supposedly unstable, so that none would occur naturally.  Their high masses 
and meagre interactions with the standard sector mean that they would not 
have been produced easily by our experiments to-date.  On the other hand, the 
stable, low mass states which remain are neutral and barely interacting with
us, and so the whole sector could so far have escaped our notice altogether 
(except, of course, through their gravitational effect).  However, if the 
estimate of $\zeta_S \sim {\rm TeV}$ is anything to go by, one may be starting 
soon to catch glimpses of this new sector populated by framonic $C$-ons which 
has hitherto been hidden from our view.

\section{Concluding summary}

The material dealt with in this paper, being well beyond the original physics
for which the FSM was first constructed, and still mostly at an exploratory 
stage, does not admit yet of a proper conclusion, in lieu of which, therefore, 
only a summary will be given of what seems so far to have emerged.

\subsection{Points of experimental and phenomenological interest}

\noindent {\bf Reproduction of standard model results}

\vspace{.2cm}

We start by listing what the FSM has done previously \cite{tfsm} in reproducing 
standard model results, near qualitatively with many fewer parameters.
\begin{itemize}
\item {\bf [i]} It gives the standard model Higgs boson as part of the flavour
framon.
\item {\bf [ii]} It gives three generations (each) of quarks and leptons, with
generations arising as the global dual of the local colour symmetry.
\item {\bf [iii]} It reproduces the hierarchical mass spectrum for the 3 
generations of quarks and leptons.
\item {\bf [iv]} It reproduces the CKM mixing matrix for quarks including the
Kobayashi-Maskawa CP-violating phase. 
\item {\bf [v]} It reproduces the 3 mixing angles in neutrino oscillations, 
that is,  the PMNS matrix for leptons except a possible CP-violating phase.
\end{itemize}
Of these, {\bf [iii]}---{\bf [v]} come about as consequences of the quark
and lepton mass matrices rotating with scale, which, as shown in Section 6,
is itself a conseqence of renormalization under framon loops.  As a result, 
the number of empirical paramaters is much reduced compared with the standard 
model (17 SM parameters replaced by 7 for FSM). 
 
\vspace{.3cm}

\noindent {\bf Deviations from the SM in the standard sector}

\vspace{.2cm}

When probed deeper, the FSM reveals deviations (not always small) from the 
standard model, among which are (Section 7.1) the following, 
\begin{itemize}
\item {\bf [vi]} Departures from SM in rare decays, such as the suppresson of
$h \rightarrow \mu^+ \mu^-$ and the likely occurrence of some flavour-violating 
modes such as  $h \rightarrow \tau \bar{\mu}$.
\end{itemize}
which were suggested some time ago \cite{Hdecay,f+mike} but may soon be detectable.
These deviations, like {\bf [iii]}---{\bf [v]} above, are consequences of the
scale dependence derived in Section 6.  But there are deviations from SM which
are more fundamental (Section 7.3):
\begin{itemize}
\item {\bf [vii]} Departures from SM in the electroweak mixing scheme:
  for example,
$m_Z - m_W$, $\Gamma(Z \rightarrow \ell^+ \ell^-)$, $\Gamma(Z \rightarrow 
q^+ q^-)$, all slightly different from that predicted by SM.
\end{itemize}
For the vacuum expectation value of the colour framon $\zeta_S \geq 2$  TeV, 
these deviations are all within the present experimental errors \cite{zmixed} 
but they should show up when experimental accuracy is improved.

\vspace{.3cm}

\noindent {\bf Glimpses of the hidden sector}

\vspace{.2cm}

Framonic $C$-ons in the hidden sector are supposed to communicate little with 
our standard sector except via the photon coupled to charges on either side, 
and via the mixing of their $G$ with our $Z$ and the photon (Section 4), 
and the mixing of their $H_{\rm high}$ and $H_{\rm even}$ with the standard Higgs 
$h_W$ (Sections 3 and 8).  But through these few chinks, one can deduce the 
following:
\begin{itemize}
\item {\bf [viii]} Possibly at LHC, a $\ell^+ \ell^-$ bump at the invariant mass of 
$G$ and a diphoton bump at the invariant mass of the $H$, again both
in the TeV range (Section 8 {\bf [b]}). 
\item {\bf [ix]} Atomki-like anomalies \cite{Atomki} at around 17 MeV (Section 8
 {\bf [d]}).
\item {\bf [x]} Step increases in $R(e^+ e^-)$ at production thresholds of 
$Q \bar{Q}$ (co-quark, anti-co-quark), $L \bar{L}$ (co-lepton, anti-co-lepton) 
(Section 8 {\bf [f]}), and a peak in $R(e^+ e^-)$ at the $G$ mass (Section 8 
{\bf [c]}, \cite{zmixed}), all in the TeV range, unfortunately beyond that of
any collider being planned.
\end{itemize}

\vspace{.3cm}

\noindent {\bf Within the hidden sector itself}

Most heavy framonic $C$-ons would have decayed away by our epoch leaving only 
the lowest as dark matter candidates.

\vspace{.2cm}
\begin{itemize}
\item {\bf [xi]} These dark matter candidates are light, of masses $\sim 17$ 
Mev, and can be both bosonic and fermionic if no see-saw mechanism occurs for 
co-neutrinos, but can be of masses $< 8$ MeV and all fermionic, if there is 
see-saw for co-neutrinos.  They are estimated to have cross sections with 
ordinary matter of the order of a few percent of neutrino cross sections.
\end{itemize}
Though light, these dark candidates can occur in abundance and make up an 
appreciable portion of the missing dark matter, though not necessarily
the bulk of it (Section 10).

\subsection{Points of conceptual and theoretical interest}

\hspace*{\parindent}{\bf [a] Parallel between the flavour and colour sectors}

\vspace{.2cm}

Perhaps the most striking feature of the FSM as here formulated is the close
parallel in structure between the flavour and colour sectors.  One is used to
thinking of the flavour sector as describing what are called weak interactions,
and of the colour sector as describing what are called strong interactions. 
The physics one sees in weak and strong interactions are very different, and 
one would have expected the theories governing these two types of interactions 
to be very different too.  This is true in the usual SM formulation where the 
the flavour $su(2)$ symmetry is spontaneously broken, while the colour $su(3)$ 
symmetry is confining and exact.  However, the confinement picture of 't~Hooft 
for the electroweak theory adopted in this paper changes all that, for one can
now regard the flavour theory as confining also, in parallel with the colour 
theory, which parallel is further accentuated in the FSM, where both theories 
are framed as well.  It might therefore appear mysterious how so very different 
physics in the two sectors can emerge.

Some of these differences are traced to the difference in basic properties 
between the flavour $SU(2)$ and colour $SU(3)$ groups.  As pointed out in 
Section 2, the number of independent framons in the flavour symmetry can be 
reduced by imposing the condition (\ref{minemb}) while a similar reduction 
is not available for the colour framon without changing the physical dimension 
of some of its components.  And this difference is traced directly to the 
special property of the $SU(2)$ group being embeddable in $\bbr^4$ (Footnote 3).

Now the difference just noted between flavour and colour is the source in the 
FSM of some great differences in the physics outcome for the two sectors.  In 
Section 3, it is this reduction in the number of flavour framons which makes the 
factor $\bbeta$ coming from the colour framon {\bf (CF)} in  (\ref{cframon}) 
disappear from the framon potential (\ref{V}), while the corresponding factor 
$\balpha$ coming from the flavour framon {\bf (FF)} in (\ref{fframon}) remains. 
As a result, the FSM vacuum depends on $\balpha$ but not on $\bbeta$, and makes
the vector $\balpha$ rotate with changing scales, while $\bbeta$ is unchanged. 
It is this rotation of $\balpha$ which leads in \cite{tfsm} to three generations of 
quarks and leptons with hierarchical masses, to CKM mixing for quarks, and to 
neutrino oscillations for leptons.  But, because of the above, there is no 
equivalent to all these in the $F$ spectrum.  

These observations, however, still do not explain the following:

\vspace{.3cm}

{\bf [b] Why flavour interactions appear weak while colour interactions strong}

\vspace{.2cm}

By this, what we really mean is why hadrons which occur in the colour theory 
interact strongly while particles occurring in the flavour theory, such as the 
leptons, the vector and Higgs bosons do not.  It is not that the hadrons do 
not have interactions similar to those of the ``weak'' particles. They do, and 
these their ``hard interactions'' are the subject of study of perturbative QCD. 
The real difference is that apart from these hard interactions, hadrons have in 
addition strong ``soft interactions'' which cannot be derived perturbatively 
from the fundamental Lagrangian.  Instead of ascribing this difference to the
possible difference in dynamics between the flavour and colour symmetries, as
one may be tempted to, the FSM suggests that the difference stems instead
from the difference in structure between the ``weak'' particles and hadrons, namely 
that the former are framonic while the latter are not, as set out in Section 9.
And the framon, being short-lived---the argument goes---will find it hard to 
separate from its partners in framonic states to recombine with alternative 
partners from the sea, as is needed to effect a soft interaction.  Admittedly, 
the arguments presented there are no better grounded theoretically than the 
previous assumption of different dynamics operating for the flavour and colour 
symmetries.  But we think that the hypothesis is worth entertaining in that it 
restores the parallel observed above between the flavour and colour sectors, 
while opening a door to a possible hidden sector rich in structure for us to 
explore.

As to the question whether soft interactions for framonic composites, be they
$B$-ons or $C$-ons, are suppressed or not, it is not easily settled definitvely since it
requires an understanding of nonperturbative effects which we do not have.
The only tool one has at present for probing non-perturbative physics is by 
lattice calculations, and we very much hope that experts in this field might
consider throwing some light on to the matter.

\vspace{.3cm}

{\bf [c] Strong and weak CP-violations identified}

\vspace{.2cm}

The observation in Sectiion 9 {\bf [b]} separates physics according to the FSM 
into what we call the standard and the hidden sectors.  In the standard sector, 
FSM has reproduced the results listed in {\bf [i]}---{\bf [v]} in the preceding 
subsection.  This does not end with just numbers but make conceptual changes as 
well.  For example, geometrical significance has been given to the Higgs
boson as a framon, and to generations as the dual  to colour; the kaleidoscopic 
mass and mixing patterns of quarks and leptons are not just accidents of nature
but some at least have a dynamical origin.  But these have already been much 
discussed in our earlier work and need not be repeated.  We recall only one 
point, namely that FSM offers a solution of the strong CP problem without axions
but links it instead via the mass matrix rotation to the Kobayashi-Maskawa 
phase in the CKM matrix.   This we think worth emphasizing again, now that the 
axion is proving elusive to experimental searches.    

Traditionally, in the standard model, CP-violation for quarks can come from two 
sources: 
\begin{itemize}
\item (i) the Kobayashi-Maskawa phase admissible in the $3 \times 3$ quark 
mixing matrix,
\item (ii) the CP-violating theta-angle term in the strong Lagrangian allowed 
by gauge invariance \cite{weinberg}.
\end{itemize}
That the phase (i) is admissible is just a property of a $3 \times 3$ unitary
matrix, but the standard model gives no physical reason for its presence nor 
an estimate for its size.  That the theta-angle term (ii) is allowed by gauge 
invariance is a bit of an embarrassment for the standard model, given that the 
natural choice of order unity for this theta-angle would give CP-violations 
for strong interactions many orders larger than acceptable to experiment.  The 
FSM scheme interestingly identifies the two problems, exploiting the existence of 
zero eigenvalues in a rank-one quark mass matrix and using a well-known 
loop-hole in the problem in this case to transform away the theta-angle term,
 avoiding thus gross CP-violation in strong interactions, while making use of 
rotation in the mass matrix to transmit the effect into the CKM matrix to 
predict a KM phase of the right magnitude \cite{strongcp}.  It has, in a sense, killed two 
birds with a single stone, solving, on the one hand, the strong CP problem
(without axions) and on the other the weak CP problem by supplying a {\it 
raison d'etre} and even an estimate for the KM phase.

\vspace{.3cm}   

{\bf [d] New scale at order 10 MeV}

\vspace{.2cm}

In the fit to data summarized in Table \ref{tfsmfit}, one obtains a trajectory
for $\balpha$ which passes through $\theta = 0$ at a scale of around 17 MeV,
at which point its normal curvature changes sign, while the following ratio of 
parameters $R = \nu_2 \zeta_W^2/2 \kappa_S \zeta_S^2$
goes to the value 1, and stays there when the scale lowers further.  The actual 
value of $\mu$ where this happens, given in \cite{tfsm} as 17 MeV, depends on 
details but the outcome that there is such a point does not seem to.  From 
this, we recall, one has obtained the following quite significant consequences:
\begin{itemize}
\item that $m_u < m_d$,  on which crucial 
empirical fact depends the stability of the proton and hence our own existence,
\item that there are possibly Atomki-like anomalies as listed in {\bf [x]} of 
the preceeding subsection,
\item that there are likely to be dark particles at or below that mass, as 
listed in {\bf (xi)} of the preceeding subsection.
\end{itemize}

Superficially, $R \rightarrow 1$ means that the vacuum value of the colour 
framon in its canonical gauge (\ref{Phivac}) vanishes in 2 directions.  Or if,
as suggested in equation (90) of \cite{dfsm}, in analogy to the vierbeins in 
gravity, this is interpreted as the ``inverse square root'' of the metric in 
$\widetilde{su}(3)$ space, thus:
\begin{equation}
(g_{\tilde{a}\tilde{b}}) = 3 \zeta_S^{-2} \left( \begin{array}{ccc}
                      (1 - R)^{-1} & 0 & 0 \\
                      0 & (1 - R)^{-1} & 0 \\
                      0 & 0 & (1 +2R)^{-1} \end{array} \right)
\label{metricinsu3t}
\end{equation}
then it would mean that generation space has collapsed at that scale from 3 
into only 2 dimensions.  We have not yet succeeded in undertanding what this
really means, nor theoretically how and why the trajectory for $\balpha$ should 
have such a behaviour, nor yet why there should be such an additional scale in 
the problem.  But it is clearly something worth understanding in 
future\footnote{It is perhaps interesting to note in passing that in his big theory 
under constructiion which has adopted as one ingredient our scheme of the 
rotating mass matrix (though not the details of the FSM here) \cite{Bjwesite} 
Bjorken has suggested a new scale at around the same order (7 MeV for him vs 
17 MeV for us, but close enough at this juncture) which he ascribes to the 
Zeldovitch effect in gravity.  However, we have not understood enough as yet 
either to concur or to disagree with his suggestion.}.

\subsection{Remark}

In the above summaries of Sections 11.1, 11.2, we might have been guilty of
emphasizing the positives while ignoring possible negatives.  If we have done 
so, however, it was due merely to the common human weakness of optimism with 
no intention to deceive.  The fact is that, having embarked on an exploration 
of a domain both vast and unfamiliar, much more so than we at first suspected, 
we quickly found our understanding and knowledge under strain, and barely equal 
to the task.  Guided only by some distant spots of light to direct our search, 
we have just reported what we found, but have not had yet the sagacity or the
courage, nor even the time, to dig into dark corners for things we might have 
missed.  We have no doubt that there are many mysteries in the FSM that we 
have not yet understood.  In summarizing now the results we find interesting, 
when the investigation is still far from complete, we do so in the hope that 
abler minds than ours might be tempted to turn their power on to the problem 
to give it the illumination it needs.  

\vspace*{3mm}

\begin{flushleft}
{\bf Appendix A. Tree-level couplings of $H_K$ and $h_W$ among themselves}
\end{flushleft}

In the framon potential (\ref{V}) we shall omit the purely electroweak
part (the first two terms), since they are identical to the standard
model, and shall consider the strong framon terms and the terms linking
the strong and weak framons. 

In order to obtain the tree-level interactions between the fields we expand the framon fields around their vacuum values:
\begin{eqnarray*}
\bphi &=& \zeta_W + h_W \nonumber \\
\bPhi &=& \phivac + \delta \bPhi
\label{vav+fluc}
\end{eqnarray*}
with $\balpha = (0,0,1)^T$ and $\delta \bPhi = \sum_{K=\pm,3}^8 V_K H_K$
\footnote{For ease of writing, we put $H_+=\frac{1}{\sqrt{2}} (H_1+H_2),\,  H_-=\frac{1}{\sqrt{2}} (H_1-H_2)$.}.

In the subsequent expansion linear terms in the fields vanish after
considering the minimum condition of the potential and 
we shall omit the quadratic terms that give the mass matrix of the
framon fields already worked 
out in the text (Section 3). In the following we shall consider the
tree-level interactions of the 
framons among themselves, that is,  cubic and quadratic terms in the
fields. In a self-explanatory 
notation we distinguish the following terms:
\[
V^{int.} \, = \, V_S \, + \, V_{SW}
\]
Each term in turn is decomposed according to their couplings into:
\begin{eqnarray*}
V_S &=& \lambda_S V_S^{(\lambda_S)}+ \kappa_S V_S^{(\kappa_S)}
\nonumber\\
V_{SW} &=& \nu_1 V_{SW}^{(\nu_1)} - \nu_2 V_{SW}^{(\nu_2)}
\end{eqnarray*}
where
\begin{eqnarray*}
V_S^{(\lambda_S)}&=&  (\Tr[\delta \bPhi^\dagger \, \delta \bPhi])^2  \, + \, 2 
\Tr[\delta \bPhi^\dagger \, \phivac \, + \,\phivac^\dagger  \delta \bPhi]   \Tr[\delta \bPhi^\dagger \, \delta \bPhi] \nonumber\\
V_S^{(\kappa_S)} &=&  \Tr[\delta \bPhi^\dagger \, \delta \bPhi \, \delta \bPhi^\dagger \, \delta \bPhi]  \, + \, 2 
\Tr[(\delta \bPhi^\dagger \, \phivac \, + \,\phivac^\dagger  \delta \bPhi )\delta \bPhi^\dagger \, \delta \bPhi]
\nonumber\\ 
 V_{SW}^{(\nu_1)} &=& 2 \zeta_W h_W \Tr[\delta \bPhi^\dagger \, \delta \bPhi] \, + \, h_W^2 
 \Tr[\delta \bPhi^\dagger \, \phivac \, + \,\phivac^\dagger  \delta \bPhi] 
\nonumber\\
 & +&  h_W^2 Tr[ \delta \bPhi^\dagger \, \delta \bPhi]
\nonumber\\
V_{SW}^{(\nu_2)} &=&  2 h_W \zeta_W [(\delta \bPhi \balpha)^\dagger \cdot (\delta \bPhi \balpha)] \, + \,
 h_W^2 [(\delta \bPhi \balpha)^\dagger \cdot (\delta \bPhi \balpha)]
\nonumber\\
& + &
h_W^2  [(\delta \bPhi \balpha)^\dagger \cdot (\phivac \balpha)+ (\phivac \balpha)^\dagger \cdot(\delta \bPhi \balpha)] 
\label{terms}
\end{eqnarray*}

After the expansion, the interaction terms in the potential are the following.

$1.$ Cubic terms proportional to $\lambda_S$ coming from
$V_S^{(\lambda_S)}$, containing
the combination of the fields $H_I H_J^2$.
\[
V_S^{(\lambda_S)} \, = \,
4 \sqrt{2}  \zeta_S 
\left[
Q  H_+ + \frac{P}{\sqrt{2}} H_3 \right]
\left(\sum_{K=\pm,3}^8 H_K^2 \right) 
\]

$2.$ Cubic terms proportional to $\kappa_S$ coming from
$V_S^{(\kappa_S)}$,
containing the combination of the fields $H_I H_J H_K$.
\begin{eqnarray*}
&&
V_S^{(\kappa_S)} \, = \,
2 \sqrt{2} \zeta_S 
\Bigg\{
Q H_+^3 + \sqrt{2} P  H_3^3 + 3Q H_+ \left(H_-^2 + H_4^2 + H_5^2\right) 
\\
&& 
+ Q 
\frac{2P^2+Q^2}{P^2+Q^2} 
\Big[
H_+\left(H_6^2 + H_7^2 + H_8^2 + H_9^2\right) + H_-\left(-H_6^2 - H_7^2 + H_8^2 + H_9^2\right)
\\
&& 
+
 2 H_4 \left(H_6H_8 + H_7H_9 \right)  + 2 H_5 \left(H_6H_9 - H_7H_8\right) 
\Big]
\\
&& 
+
\left.
\sqrt{2} P 
\frac{2P^2+Q^2}{P^2+Q^2} H_3 \left(H_6^2 + H_7^2 + H_8^2 + H_9^2\right)
\right\}
\end{eqnarray*} \\

$3.$ Cubic terms proportional to $\nu_1$ coming from $
V_{SW}^{(\nu_1)}$, mixing strong and weak framon fields, 
containing the combination of the fields $h_W^2 H_K$ and $h_W H_K^2$.
 \[
V_{SW}^{(\nu_1)} \, = \,
\zeta_S h_W^2 2 \sqrt{2} \left( Q  H_+ +  \frac{P}{\sqrt{2}} H_3 \right) + 2 \zeta_W h_W 
\left(\sum_{K=\pm,3}^8 H_K^2 \right) 
\]

$4.$ Cubic terms proportional to $\nu_2$ coming from $ V_{SW}^{(\nu_2)}$, mixing strong and weak framon fields:
\[
V_{SW}^{(\nu_2)} \, = \,  2 \zeta_S 
P h_W^2 H_3 + 2 \zeta_W h_W
\left[ H_3^2 + \frac{Q^2}{P^2+Q^2} \left( H_6^2 + H_7^2 + H_8^2 + H_9^2 \right)
\right]
\nonumber
\]

$5.$ Quartic terms proportional to $\lambda_S$ coming from
$V_S^{(\lambda_S)}$, 
containiing the combination of the fields $H_I^2 H_J^2$.
\[
V_S^{(\lambda_S)}  \, = \, \lambda_S \left(\sum_{K=\pm,3}^8 H_K^2 \right)^2
\]

$6.$ Quartic terms proportional to $\kappa_S$ coming from
$V_S^{(\kappa_S)}$, 
containing the combination of the fields $H_I H_J H_K H_L$.

\begin{eqnarray*}
&& \!\!\!\! \!\!\!\!
V_{SW}^{(\kappa_S)} \, = \,
\frac{1}{2} \left[H_+^4 + H_-^4 + \left( H_4^4 + H_5^4 \right)^2\right] + H_3^4
\\
&&  \!\!\!\! \!\!\!\!
+ 
H_+^2 
\Big[ 3 \left(H_-^2+H_4^2+H_5^2 \right) + H_6^2 + H_7^2 + H_8^2 + H_9^2
\Big] 
\\
&&  \!\!\!\! \!\!\!\!
+2H_3^2 
\left(H_6^2 + H_7^2 + H_8^2 + H_9^2\right)
\\
&&  \!\!\!\! \!\!\!\!
+
H_-^2 \left(H_4^2+H_5^2+H_6^2 + H_7^2 + H_8^2 + H_9^2\right)
+2H_3^2 
\left(H_6^2 + H_7^2 + H_8^2 + H_9^2\right)
\\
&& \!\!\!\! \!\!\!\!
+
 2H_+H_-
\left(-H_6^2 - H_7^2 + H_8^2 + H_9^2\right) 
+
 \left( H_4^2+H_5^2 \right) \left (H_6^2 + H_7^2 + H_8^2 + H_9^2 \right)
\\
&& \!\!\!\! \!\!\!\!
+
4 H_+ \left(H_4H_6H_8 +H_4H_7H_9 + H_5H_6H_9-H_5H_7H_8\right)
\\
&&  \!\!\!\! \!\!\!\!
+
\frac{2\sqrt{2}PQ}{P^2+Q^2}
\Big[ H_-H_3\left(-H_6^2 - H_7^2 + H_8^2 + H_9^2\right)+H_+H_3 
\left(H_6^2 + H_7^2 
\right.
\Big.
\\
&&  \!\!\!\! \!\!\!\!
\Big. \left.
+ H_8^2 + H_9^2\right)
+2 H_3 \left(H_4H_6H_8 +H_4H_7H_9 + H_5H_6H_9-H_5H_7H_8\right)
\Big]
\\
&&  \!\!\!\! \!\!\!\!
+
\frac{P^4+Q^4}{(P^2+Q^2)^2} 
\left( H_6^2 +H_7^2 +H_8^2 +H_9^2 \right)^2
\end{eqnarray*}

$7.$ Quartic terms proportional to $\nu_1$ coming from
$V_{SW}^{(\nu_1)}$, 
containing the combination of the fields $h_W^2 H_K^2$.

 \[
V_{SW}^{(\nu_1)} \, = \, h_W^2 \left(\sum_{K=\pm,3}^8 H_K^2 \right) 
\nonumber 
 \]

$8.$ Quartic terms proportional to $\nu_2$ coming from
$V_{SW}^{(\nu_2)}$, 
containing the combination of the fields $h_W^2 H_K^2$.
 \[
V_{SW}^{(\nu_2)} \, = \, h_W^2 \, \left\{ H^2_3 +  \frac{Q^2}{P^2+Q^2} \,\Big( H_6^2 + H_7^2 + H_8^2 +H_9^2 \Big) \right\}
\nonumber
 \]

\vspace*{3mm}

\begin{flushleft}
{\bf Appendix B. Tree-level couplings of the $H$ with the $\tilde{C}$
  (or $G$)}
\end{flushleft}

The component of the kinetic energy corresponding to the colour framon
is given in (\ref{KEc}). 
After expanding the framon fields about their vacuum values as it has
been done in Appendix A the 
result can be decomposed according to the field content in
 \[
{\cal A}_{\rm KE}^{\rm C} =  K^{(2)} +  K^{(3)} +  K^{(4)}
 \]
The first term ($K^{(2)}$) corresponds to the mass matrix of the
$\tilde{C}^i_\mu$ and $A_\mu$ 
fields that has been worked out in the text (Sections 4 and 8). The
remaining terms give the 
tree-level interactions between the $H_K$ with the $\tilde{C}^\alpha_\mu$ and $A_\mu$ and will be given in the following.

We start with the term cubic in the fields and write:
\begin{eqnarray*}
 K^{(3)} & = & g_3^2 \, \zeta _S \, K^{(3)}_1 +g_1^2 \, \zeta _S \, K^{(3)}_2 +2 g_3 g_1 \, \zeta _S \, K^{(3)}_3 
\end{eqnarray*}
with each term given by:
\begin{eqnarray*}
 K^{(3)}_1 & = &   \,  \frac{1}{ \zeta _S} \, \sum_{\alpha, \beta} \, \tilde{C}^\alpha \, \tilde{C}^\beta  \, \,
   {\mbox Tr} \,\left( \,\lambda_\alpha  \lambda_\beta \left[\phivac \,  \, \delta \bPhi^\dagger  \, +\,\delta\bPhi \,  \phivac^\dagger \right]  \right)
\nonumber \\
K^{(3)}_2 & = &  \, \frac{1}{ \zeta _S}A_\mu^2 \,
 {\mbox Tr} \,  \left( \Gamma \left[ \phivac^\dagger \, \delta\bPhi  +  \, \delta\bPhi^\dagger \phivac \right] \Gamma   \right)
\nonumber \\
K^{(3)}_3 & = &\, \frac{1}{2ƒ \zeta _S}A_\mu \, \sum_{\alpha} \,\tilde{C}^\alpha\,  \,    
{\mbox Tr} \,  \left( \Gamma \, \delta\bPhi^\dagger  \lambda_\alpha \phivac +   \Gamma \phivac^\dagger \lambda_\alpha  \, \delta\bPhi \, \right)
 \label{kinematic3} 
\end{eqnarray*}

Working out the traces of the matrix fields we get the following parts.
\medskip

$1.$ Terms proportional to $g_3^2 \, \zeta _S$ coming from
$K^{(3)}_1$,
containing the combinations of fields $\tilde{C}_{\mu}^\alpha \,\tilde{C}_{\mu}^\beta \,H_K$.

\begin{eqnarray*}
K^{(3)}_1 & = &
2 \left\{ \tilde{C}_{\mu}^4 \tilde{C}_{\mu}^4 +\tilde{C}_{\mu}^5 \tilde{C}_{\mu}^5+
\frac{1}{3} \tilde{C}_{\mu}^8 \tilde{C}_{\mu}^8 \right\}
\left\{ P H_3 +\frac{Q}{\sqrt{2}} (H_+ + H_-) \right\}
\Bigg.
\nonumber \\
+ &&
\!\!\!\!\!\!\!\!\!\!\!\!
2 \left\{ \tilde{C}_{\mu}^6 \tilde{C}_{\mu}^6 +\tilde{C}_{\mu}^7 \tilde{C}_{\mu}^7 +
\frac{1}{3} \tilde{C}_{\mu}^8 \tilde{C}_{\mu}^8 \right\}
\left\{ P H_3  +\frac{Q}{\sqrt{2}} (H_+ - H_-) \right\}+
\frac{4 P}{3} \tilde{C}_{\mu}^8 \tilde{C}_{\mu}^8
\nonumber \\
+ &&
\!\!\!\!\!\!\!\!\!\!\!\!
\frac{4\,P\,Q}{\sqrt{P ^2+Q^2}} \,
\Bigg\{
\Big(\tilde{C}_{\mu}^1 \tilde{C}_{\mu}^4  + \tilde{C}_{\mu}^2 \tilde{C}_{\mu}^5 - \tilde{C}_{\mu}^3 \tilde{C}_{\mu}^6 - \frac{1}{\sqrt{3}} \tilde{C}_{\mu}^6 \tilde{C}_{\mu}^8 \Big) H_6
\Bigg.
\nonumber \\
+ &&
\!\!\!\!\!\!\!\!\!\!\!\!
\Big(-\tilde{C}_{\mu}^2 \tilde{C}_{\mu}^4  + \tilde{C}_{\mu}^1 \tilde{C}_{\mu}^5 - \tilde{C}_{\mu}^3 \tilde{C}_{\mu}^7 - \frac{1}{\sqrt{3}} \tilde{C}_{\mu}^7 \tilde{C}_{\mu}^8 \Big) H_7
\nonumber \\
+ &&
\!\!\!\!\!\!\!\!\!\!\!\!
 \Big(\tilde{C}_{\mu}^3 \tilde{C}_{\mu}^4  + \tilde{C}_{\mu}^1 \tilde{C}_{\mu}^6 - \tilde{C}_{\mu}^2 \tilde{C}_{\mu}^7 - \frac{1}{\sqrt{3}} \tilde{C}_{\mu}^4 \tilde{C}_{\mu}^8 \Big) H_8
\nonumber \\
+ &&
\!\!\!\!\!\!\!\!\!!\!\!
 \Big(\tilde{C}_{\mu}^3 \tilde{C}_{\mu}^5  + \tilde{C}_{\mu}^2 \tilde{C}_{\mu}^6 + \tilde{C}_{\mu}^1 \tilde{C}_{\mu}^7 - \frac{1}{\sqrt{3}} \tilde{C}_{\mu}^5 \tilde{C}_{\mu}^8 \Big) H_9
\nonumber \\
+ &&
\!\!\!\!\!\!\!\!\!\!\!\!
2 \sqrt{2} Q 
\Bigg\{
\Big(
\tilde{C}_{\mu}^1\, \tilde{C}_{\mu} ^1  + \tilde{C}_{\mu}^2 \tilde{C}_{\mu}^2 + \tilde{C}_{\mu}^3 \tilde{C}_{\mu}^3 
\Big)  H_+
+ 
\frac{2}{\sqrt{3}} \tilde{C}_{\mu}^3 \tilde{C}_{\mu}^8  H_-
\Bigg.
\nonumber \\
+ &&
\!\!\!\!\!\!\!\!\!\!\!\!
\Big(
\tilde{C}_{\mu}^4 \, \tilde{C}_{\mu}^6  + \tilde{C}_{\mu}^5 \tilde{C}_{\mu}^7 +  \frac{2}{\sqrt{3}} \tilde{C}_{\mu}^1 \tilde{C}_{\mu}^8 
\Big)  H_4
+
\Big(
\tilde{C}_{\mu}^5 \, \tilde{C}_{\mu}^6  - \tilde{C}_{\mu}^4 \tilde{C}_{\mu}^7 +  \frac{2}{\sqrt{3}} \tilde{C}_{\mu}^2 \tilde{C}_{\mu}^8 
\Big)  H_5
\Bigg\}
\nonumber \\
\end{eqnarray*}

$2.$ Terms proportional to $g_1^2 \, \zeta _S$ coming from
$K^{(3)}_2$,
containing the combinations of fields $A _{\mu}^2\, H_K$.
 \[
K^{(3)}_2 \, = \, \frac{2}{9} \, A_{\mu}^2 \left( 4 \, \,P  \, H_3 + Q \sqrt{2} H_+  \right)
 \]

$3.$ Terms proportional to $g_1 \, g_3 \, \zeta _S$ coming from
$K^{(3)}_3$,
containing the combinations of fields $A _{\mu} \, \tilde{C}_{\mu}^\alpha \,H_K$.

\begin{eqnarray*}
&&
K^{(3)}_3 \, = \, \frac{2}{3} \,
 A_{\mu} \, \Bigg\{ 
 \sqrt{2} Q \, 
\left(
-\tilde{C}_{\mu}^1\, H_4 - \tilde{C}_{\mu}^2 \, H_5 - \sqrt{2} \, \tilde{C}_{\mu}^3 \, H_- 
\right) 
 \Bigg.
 \nonumber \\
&& 
\left.
+ \,
\frac{P\,Q }{\sqrt{\,P ^2+\, Q^2}}    
 \left(
  \, \tilde{C}_{\mu}^4 \, H_8 \, +  \,\tilde{C}_{\mu}^5 \, H_9 \, +  \, \tilde{C}_{\mu}^6 \, H_6 \, +  \,\tilde{C}_{\mu}^7 \, H_7   \,
 \right)
 \right.
 \nonumber \\
&& 
\Bigg. 
+ \frac{1} {\sqrt{3}}  \, \tilde{C}_{\mu}^8  \left(-4 \,P  \,  H_3  - Q \sqrt{2} H_+  \right)
 \Bigg\}
 \nonumber \\
\end{eqnarray*}

Next we write the quartic terms:

\begin{eqnarray*}
K^{(4)} & = & K^{(4)}_1 +2 g_3 \, K^{(4)}_2 +2 g_1 \,K^{(4)}_3 +g_3^2 \, K^{(4)}_4 +g_1^2 \,K^{(4)}_5 +2 g_3 g_1 \,K^{(4)}_6
\nonumber \\
\end{eqnarray*}

where:
\begin{eqnarray*}
 K^{(4)}_1 & = & \,{\mbox Tr} \, 
\left(   
\left[\partial_\mu \,\delta \bPhi \right]^\dagger  \, \left[ \partial_\mu \,\delta \bPhi   \right]  \right) \,
 \nonumber \\ 
 K^{(4)}_2 & = & \frac{-i}{2} \, \sum_{\alpha=1}^8 \, \tilde{C}^\alpha_\mu \, {\mbox Tr} \,  \left(    
\left[ \partial_\mu \,\delta \bPhi \right]^\dagger  \, \lambda_\alpha\, \delta \bPhi \, -  \,\delta \bPhi^\dagger   \lambda_\alpha \left[ \partial_\mu \,\delta \bPhi \right] \right)
 \nonumber \\ 
 K^{(4)}_3 & = & \frac{-i}{2} \, A_\mu \, {\mbox Tr} \,  \left(    
\left[ \partial_\mu \,\delta \bPhi \right]^\dagger  \, \delta \bPhi \, \Gamma  \, - \, \Gamma \,  \delta \bPhi^\dagger  \left[ \partial_\mu \,\delta \bPhi \right] \right)
\nonumber \\
 K^{(4)}_4 & = & \sum_{\alpha, \beta=1}^8 \, \tilde{C}^\alpha_\mu\,\tilde{C}\beta_\mu\,  {\mbox Tr} \, \left(\, \delta \bPhi^\dagger \lambda_\alpha  \lambda_\beta \,\delta \bPhi  \right)
\nonumber \\
K^{(4)}_5 & = & A_\mu^2 \,   {\mbox Tr} \,  \left( \Gamma\, \delta \bPhi^\dagger \,\delta \bPhi  \Gamma   \right) 
 \nonumber \\
 K^{(4)}_6 & = & \frac{1}{2} A_\mu \, \sum_{\alpha=1}^8 \,\tilde{C}^\alpha_\mu \,    {\mbox Tr} \, \left( \,\delta \bPhi^\dagger \lambda_\alpha \Gamma  \,\delta \bPhi \, + \,\Gamma \,\delta \bPhi^\dagger \lambda_\alpha  \delta \bPhi \right)
 \label{kinematic4} 
\end{eqnarray*}

Working out the traces of the matrix fields we get the following terms.
\medskip
 
$1.$ Terms from $K^{(4)}_1$, containing the combinations of fields $\partial _{\mu}  H_K^2 $.

 \[
K^{(4)}_1 = \partial_{\mu} H_+^2 +\partial_{\mu}H_-^2 +\partial _{\mu}  H_3^2+\partial _{\mu}  H_4^2+\partial _{\mu}  H_5^2+\partial _{\mu}  H_6^2+\partial _{\mu}  H_7^2+\partial _{\mu}  H_8^2+\partial _{\mu}  H_9^2
\nonumber \\
 \]

$2.$ Terms proportional to $g_3$ coming from $K^{(4)}_2$,
containing the combinations of fields $\tilde{C}_{\mu}^\alpha \, H_I \,  \partial_{\mu}  H_J$.

\begin{eqnarray*}
&&
\!\!\!\!\!\!\! \!\!\!  
K^{(4)}_2 \, = \,
\nonumber \\
&&
\!\!\!\!\!\!
 = \tilde{C}_{\mu}^1 
\Big\{
H_5  \partial_{\mu}  H_- - H_- \partial_{\mu}  H_5 +
\frac{Q^2}{P^2+Q^2}
\Big[H_6 \partial_{\mu}  H_9 - H_9   \partial_{\mu} H_6  +
H_8  \partial_{\mu} H_7 -H_7 \partial_{\mu}  H_8 
\Big] 
\Big\}  
\nonumber \\
&&
\!\!\!\!
+
\tilde{C}_{\mu}^2 
\Big\{ H_-\partial _{\mu}  H_4 - H_4 \partial_{\mu}  H_- 
  + \frac{Q^2}{P^2+Q^2} 
\Big[ H_8  \partial_{\mu}  H_6 -  H_6  \partial_{\mu}  H_8 + H_9   \partial_{\mu}  H_7  - H_7   \partial_{\mu}  H_9 
\Big]
\Big\}  
\nonumber \\
&&
\!\!\!\!
+
\tilde{C}_{\mu}^3 
\Big\{
H_4 \partial_{\mu}  H_5 - H_5 \partial_{\mu}  H_4 +
\frac{Q^2}{P^2+Q^2}  
\Big[H_6  \partial_{\mu}  H_7- H_7  \partial_{\mu}  H_6  +   H_8  \partial_{\mu}  H_9 -  H_9 \partial_{\mu}  H_8 \Big]
\Big\} 
\nonumber \\
&&
\!\!\!\!
+
\frac{1}{\sqrt{P^2+Q^2}}  \tilde{C}_{\mu}^4 
\Big\{
\frac{P}{\sqrt{2}}
\Big[  
H_9 \partial_{\mu}H_+ - H_+ \partial_{\mu}  H_9 + H_9 \partial_{\mu}H_-  -
H_- \partial_{\mu} H_9 + H_7 \partial_{\mu}  H_4   
\Big.
\nonumber \\
&&
\!\!\!\!
\Big.
- H_4 \partial_{\mu}  H_7 +
H_6 \partial_{\mu}  H_5  - H_5 \partial_{\mu}  H_6  
\Big] +
Q  \Big[  H_3  \partial_{\mu}  H_9 -H_9 \partial_{\mu}  H_3 \Big]
\Big\}
\nonumber \\
&&
\!\!\!\!
+
 \frac{1}{\sqrt{2(P^2+Q^2)}}  \tilde{C}_{\mu}^5 
\Big\{
\frac{P}{\sqrt{2}}
\Big[
H_+ \partial_{\mu} H_8- H_8 \partial_{\mu} H_+ + H_- \partial_{\mu} H_8 - H_8 \partial_{\mu}  H_- - H_6 \partial_{\mu}  H_4 
\Big.
\nonumber \\
&& 
\!\!\!\!
\Big.
+ H_4 \partial_{\mu}  H_6 + H_7 \partial_{\mu}  H_5 - H_5 \partial_{\mu}  H_7
\Big] 
+ Q 
\Big[
H_8 \partial_{\mu}  H_3 - H_3 \partial_{\mu}  H_8 \Big]
\Big\}  
\nonumber \\
&&
\!\!\!\!
+
 \frac{1}{\sqrt{2(P^2+Q^2)}} \tilde{C}_{\mu}^6  
\Big\{
\frac{P}{\sqrt{2}}
\Big[
H_7 \partial_{\mu}H_+ -H_+ \partial_{\mu}  H_7 - H_7 \partial_{\mu} H_- + 
H_- \partial_{\mu}  H_7 +  H_9  \partial_{\mu}  H_4 
\Big.
\nonumber \\
&&
\!\!\!\!
\Big.
-H_4 \partial_{\mu}  H_9  -  
H_8 \partial_{\mu}  H_5 + H_5 \partial_{\mu}  H_8 
\Big]
+ Q \, 
\Big[ - H_7 \partial_{\mu}  H_3 + H_3 \partial_{\mu}  H_7 \Big]
\Big\}  
\nonumber \\
&&
\!\!\!\!
+
\frac{1}{\sqrt{2(P^2+Q^2)}}  \tilde{C}_{\mu}^7  
\Big\{
\frac{P}{\sqrt{2}}  
\Big[ 
H_+ \partial_{\mu}  H_6- H_6 \partial_{\mu} H_+  + H_6 \partial_{\mu} H_- - H_- \partial_{\mu}  H_6 -
H_8 \partial_{\mu} H_4 
\Big.
\nonumber \\
&&
\!\!\!\!
\Big.
+ H_4 \partial_{\mu}  H_8 -
H_9 \partial_{\mu}  H_5 + H_5 \partial_{\mu}  H_9  
\Big]
+   Q \Big[ H_6 \,  \partial_{\mu}  H_3 - \,H_3\partial _{\mu}  H_6 \Big]
\Big\}  
\nonumber \\
&&
\!\!\!\!
+
\frac{2 P^2+Q^2}{\sqrt{3}(P^2+Q^2)}  \, \tilde{C}_{\mu}^8 \,
\Big\{
H_6 \partial_{\mu}  H_7 -H_7 \partial_{\mu}  H_6-  H_9 \partial_{\mu}  H_8+H_8 \partial_{\mu}  H_9
\Big\}
\nonumber \\
\end{eqnarray*}

$3.$ Terms proportional to $g_1$ coming from $K^{(4)}_3$,
containing the combinations of fields $A_\mu \, H_I \,  \partial_{\mu}  H_J$.

\[
K^{(4)}_3 \,= \, \frac{1}{3 \left(P^2+Q^2\right)}  \, A_{\mu}  
\Big\{
- H_7  \partial_{\mu}  H_6  + H_6   \partial_{\mu}  H_7- H_9 \partial_{\mu}  H_8  + H_8  \partial_{\mu} H_9 
\Big\}
\]

$4.$ Terms proportional to $g_3^2$ coming from $K^{(4)}_4$,t
containing the combinations of fields  $\tilde{C}_{\mu}^\alpha \, \tilde{C}_{\mu}^\beta \, H_I \,  H_J$.

\begin{eqnarray*}
&& 
\!\!\!
K^{(4)}_4 = 
\Big\{
\tilde{C}_{\mu}^1 \,\tilde{C}_{\mu}^1 + \tilde{C}_{\mu}^2 \, \tilde{C}_{\mu}^2 + 
\tilde{C}_{\mu}^3 \, \tilde{C}_{\mu}^3 \Big\} 
\nonumber \\
&&
\Big\{
H_+^2+H_-^2+H_4^2+\, H_5^2 +
\frac{Q^2}{P^2+Q^2}
\Big[ H_6^2 + H_7^2 + H_8^2 + H_9^2  \Big] 
\Big\}
\nonumber \\
&&
+\, \frac{2}{\sqrt{P^2+Q^2}} \,
\Big\{
\tilde{C}_{\mu}^1 \, \tilde{C}_{\mu}^4 + \tilde{C}_{\mu}^2 \, \tilde{C}_{\mu}^5 - 
\tilde{C}_{\mu}^3 \, \tilde{C}_{\mu}^6 - \frac{1}{\sqrt{3}} \tilde{C}_{\mu}^6 \, \tilde{C}_{\mu}^8
\Big\}
\nonumber \\
&&
\Big\{
\frac{P}{\sqrt{2}} 
\Big[ H_+ \, H_6 - H_- H_6 + H_4 H_8 + H_5 H_9  \Big] 
+ Q \, H_3  H_6 
\Big\}
\nonumber \\
&& 
+\, \frac{2}{\sqrt{P^2+Q^2}} \, 
\Big\{
\tilde{C}_{\mu}^1 \, \tilde{C}_{\mu}^5 - \tilde{C}_{\mu}^2 \, \tilde{C}_{\mu}^4 - 
\tilde{C}_{\mu}^3 \, \tilde{C}_{\mu}^7 - \frac{1}{\sqrt{3}} \tilde{C}_{\mu}^7 \, \tilde{C}_{\mu}^8
\Big\}
\nonumber \\
&&
\Big\{
\frac{P}{\sqrt{2}} 
\Big[ H_+ H_7 - H_- H_7 + H_4 H_9 - H_5 H_8 \Big] 
+ Q \, H_3 H_7 
\Big\}
\nonumber \\
&& 
+\, \frac{2}{\sqrt{P^2+Q^2}} \,
\Big\{
\tilde{C}_{\mu}^1 \,\tilde{C}_{\mu}^6 - \tilde{C}_{\mu}^2 \, \tilde{C}_{\mu}^7 +
\tilde{C}_{\mu}^3 \,\tilde{C}_{\mu}^4 - \frac{1}{\sqrt{3}} \tilde{C}_{\mu}^4\,\tilde{C}_{\mu}^8
\Big\}
\nonumber\\
&&
\Big\{
\frac{P}{\sqrt{2}} 
\Big[H_+ H_8 + H_- H_8 + H_4 H_6 - H_5 H_7 \Big] 
+ Q \, H_3 H_8 \, 
\Big\}
\nonumber \\
&& 
+\, \frac{2}{\sqrt{P^2+Q^2}} \,
\Big\{
\tilde{C}_{\mu}^1 \, \tilde{C}_{\mu}^7 + \tilde{C}_{\mu}^2 \, \tilde{C}_{\mu}^6 +
\tilde{C}_{\mu}^3 \, \tilde{C}_{\mu}^5 - \frac{1}{\sqrt{3}} \tilde{C}_{\mu}^5 \, \tilde{C}_{\mu}^8
\Big\}
\nonumber\\
&&
\Big\{
\frac{P}{\sqrt{2}} 
\Big[H_+ H_9 + H_- H_9 + H_4 H_7 + H_5 H_6 \Big]
+ Q\, H_3 H_9 \, 
\Big\}
\nonumber \\
&& 
+\Big\{
\frac{4}{\sqrt{3}} \tilde{C}_{\mu}^1 \, \tilde{C}_{\mu}^8 + 2\tilde{C}_{\mu}^4\,\tilde{C}_{\mu}^6 +
2 \tilde{C}_{\mu}^5 \, \tilde{C}_{\mu}^7
\Big\} 
\,
\Big\{
H_+ H_4 + \frac{Q^2}{P^2+Q^2} \, 
\Big[H_6 H_8 + H_7 H_9 \Big]
\Big\}
\nonumber\\
&& 
+
\Big\{
 \frac{4}{\sqrt{3}} \tilde{C}_{\mu}^2 \, \tilde{C}_{\mu}^8 - 2 \tilde{C}_{\mu}^4\,\tilde{C}_{\mu}^7 + 
2 \tilde{C}_{\mu}^5 \, \tilde{C}_{\mu}^6 
\Big\}
\,
\Big\{
H_+ H_5 + 
\frac{Q^2}{P^2+Q^2}
\Big[ H_6 H_9 - H_7 H_8 \Big]
\Big\}  
\nonumber \\
&& 
+ \frac{2}{\sqrt{3}} \, 
\tilde{C}_{\mu}^3 \, \tilde{C}_{\mu}^8
\Big\{
2 H_+ H_- +
\frac{Q^2}{P^2+Q^2} \, 
\Big[- H_6^2 - H_7^2 + H_8^2 + H_9^2 \Big]
\Big\} 
\nonumber\\
&& 
+
\Big\{\tilde{C}_{\mu}^4\,\tilde{C}_{\mu}^4 + \tilde{C}_{\mu}^5 \, \tilde{C}_{\mu}^5\Big\} 
\nonumber \\
&&
\Big\{
\frac{(H_+ + H_-)^2}{2} + H_3^2 + \frac{\, H_4 \,^2}{2} + \frac{\, H_5^2}{2} +
\frac{P^2}{P^2+Q^2}
\Big[H_6^2 + H_7^2 \Big] + H_8^2 + H_9^2 
\Big\}
\nonumber \\
&& 
+ 
\Big\{\tilde{C}_{\mu}^6 \, \tilde{C}_{\mu}^6  + \tilde{C}_{\mu}^7 \, \tilde{C}_{\mu}^7 \Big\}
\nonumber \\
&&
\Big\{
\frac{(H_+ - H_-)^2}{2} + H_3^2 + \frac{H_4^2}{2} + \frac{H_5^2}{2} +
H_6^2 + H_7^2 + \frac{P^2}{P^2+Q^2} \,
\Big[ H_8^2 + H_9^2 \Big]
\Big\}
\nonumber \\
&& 
+
\frac{1}{3}
\tilde{C}_{\mu}^8 \, \tilde{C}_{\mu}^8
\Big\{
H_+^2 + H_-^2 + 4 H_3^2 + H_4^2 + H_5^2 +
\frac{4 P^2 +Q^2}{P^2+Q^2}
\Big[H_6^2 + H_7^2 + H_8^2 + H_9^2 \Big]
\Big\}
\nonumber
\end{eqnarray*}

$5.$ Terms proportional to $g_1^2$ coming from $K^{(4)}_5$,
containiing the combinations of fields  $A_\mu^2 \, H_I \,  H_J $.

\[
K^{(4)}_5 \, = \,
A_{\mu}^2 \, \frac{1}{9}
\Big\{
H_+^2 + H_-^2 + 4 H_3^2 + H_4^2+ H_5^2  +
\frac{P^2+4 Q^2}{P^2+Q^2}
\Big[H_6^2 + H_7^2 + H_8^2 + H_9^2 \Big]
\Big\}
\]

$6.$ Terms proportional to $g_1 \, g_3$ coming from $K^{(4)}_6$,
containing the combinations of fields  $A_\mu \, \tilde{C}_{\mu}^\alpha  \, H_I \,  H_J$.

\begin{eqnarray*}
&& \!\!\!\!\!
K^{(4)}_6 =  A_{\mu}\tilde{C}_{\mu}^1 \, \frac{2}{3} \,
\Big\{
- H_+  H_4  +
\frac{2 Q^2}{\left(P^2+Q^2\right)} 
\Big[ H_6 H_8 + H_7 H_9 \Big]
\Big\}
\nonumber \\
&+& \!\!\!\!\!
A_{\mu}\tilde{C}_{\mu}^2 \, \frac{2}{3} \,
\Big\{
- H_+ H_5 +
\frac{2 Q^2}{P^2+Q^2} 
\Big[ H_6  H_9 - H_7  H_8 \Big]
\Big\}
\nonumber \\
&+& \!\!\!\!\!
A_{\mu}\tilde{C}_{\mu}^3 \, \frac{2}{3} \,
\Big\{
-H_+ H_- +
\frac{Q^2}{P^2+Q^2}
\Big[H_8^2 + H_9^2 - H_6^2 - H_7^2 \Big]
\Big\}
\nonumber \\
&+& \!\!\!\!\!
A_{\mu}\tilde{C}_{\mu}^4 \,\frac{2}{3 \sqrt{P^2+Q^2}}
\Big\{
\frac{P}{\sqrt{2}} \Big[ - H_+ H_8 - H_- H_8 - H_4 H_6 + H_5 H_7 \Big]
+ 2 Q  H_3 H_8
\Big\}
\nonumber \\
&+&  \!\!\!\!\!
A_{\mu}\tilde{C}_{\mu}^5 \, \frac{2}{3 \sqrt{P^2+Q^2}}
\Big\{
\frac{P}{\sqrt{2}} \Big[- H_+ H_9 - H_- H_9 - H_5 H_6 - H_4 H_7 \Big] 
+ 2 Q H_3 H_9   
\Big\}
 \nonumber \\
&+&  \!\!\!\!\!
A_{\mu}\tilde{C}_{\mu}^6 \, \frac{2}{3 \sqrt{P^2+Q^2}}
\Big\{
\frac{P}{\sqrt{2}} \Big[- H_+ H_6 + H_- H_6 - H_4 H_8 - H_5 H_9 \Big] 
+ 2 Q H_3 H_6   
\Big\}
 \nonumber \\
&+& \!\!\!\!\! 
A_{\mu}\tilde{C}_{\mu}^7 \,\frac{2}{3 \sqrt{P^2+Q^2}}
\Big\{
\frac{P}{\sqrt{2}} \Big[- H_+ H_7 + H_- H_7 + H_5 H_8 - H_4 H_9 \Big]
+ 2 Q H_3 H_7
\Big\}
 \nonumber \\
&+& \!\!\!\!\!
A_{\mu}\tilde{C}_{\mu}^8  \,\frac{1}{3 \sqrt{3}} 
\Big\{
 - H_+^2 - H_-^2  - 4 H_3^2 - H_4^2 -  H_5^2 + 2 H_6^2 + 2 H_7^2 + 2 H_8^2 + 2 H_9^2
\Big\}
\nonumber \\ 
\end{eqnarray*}

\vspace*{3mm}

{\bf Appendix C.  Tree-level couplings of the $F$ with the $\tilde{C}_\mu$
(or $G$)}

\vspace{.3cm}

The kinetic energy term of a fundamental fermion field $\psi$, say a flavour doublet
and colour triplet, can be written as:
\[
\bar{\psi} D_\mu \psi
   = \bar{\psi}(\partial_\mu - ig_1 \Gamma A_\mu - i \half g_2 B_\mu
                - i \half g_3 C_\mu) \psi,
\]
where $\Gamma$ is the charge operator operating on whatever follows.
Using the operator $\Omega$ introduced in (\ref{PhiGF}) to fix the gauge so that
the colour framon field $\bPhi$ becomes hermitian as in (\ref{Phivac}), we can
rewrite the above as:

\begin{eqnarray*}
\bar{\psi} D_\mu \psi
  & = & \bar{\psi} \Omega \Omega^{-1} (\partial_\mu - ig_1 \Gamma A_\mu 
        - i \half g_2 B_\mu - i \half g_3 C_\mu) \Omega \Omega^{-1} \psi, \nonumber\\
  & = & \bar{\chi} (\Omega^{-1} \partial_\mu \Omega - ig_1 \Gamma A_\mu
        - i \half g_2 B_\mu - i \half g_3 \Omega^{-1} C_\mu \Omega) \chi,
\end{eqnarray*}
where $\chi = \Omega^{-1} \psi$, being a bound state of the fundamental fermion
field $\psi$ with $\bPhi^\dagger$ (approximated by $\Omega^{-1}$ as explained in
Section 4) represents in our present language an $F$ field.  We note also that
since $\Omega$ acts on only the colour sector, it leaves $A_\mu$ and $B_\mu$ in
the preceding formula unchanged.  Using then (\ref{Ctilde}) of Section 4, the
above formula can be rewritten again as:
\[
\bar{\psi} D_\mu \psi
   = \bar{\chi}(\partial_\mu - ig_1 \Gamma A_\mu - i \half g_2 B_\mu
                - i \half g_3 \tilde{C}_\mu) \chi,
\]
where $\tilde{C}_\mu$ now represents our $G$.  This formula is the same in form 
as the first equation, except for the replacement of $\psi$ by $\chi$ and $C_\mu$
by $\tilde{C}_\mu$.  From this we deduce that the $F$ do couple to the 
$G$ in the same way that the fundamental fields $\psi$ couple to the colour 
gauge fields $C_\mu$, as claimed in Sections 6 and 9.

We note that the above derivation is essentially just a paraphrase of that
given by 't Hooft \cite{tHooft} and Banks and Rabinovici \cite{banks} 
for the quarks and leptons in the confinement picture of the electroweak 
theory, only with flavour and colour interchanged.  Indeed, had we taken 
$\Omega$ for colour above as  the $\Omega$ for 
flavour in (\ref{Omega}) of Section 4, we would reproduce their result, 
namely that quarks and leptons, as (in our language) framonic $B$-ons, have 
interactions with $W^\pm, \gamma-Z$ (also as framonic $B$-ons) the same as 
the fundamental fermions $\psi$ have with the original gauge fields.


\begin{thebibliography}{99}

\bibitem{tfsm} Jos\'e Bordes, Chan Hong-Mo and Tsou Sheung Tsun, Int. J. Mod. 
   Phys. A30 (2015) 1550051; arXiv:1410.8022.

\bibitem{EKS} See e.g. A. Trautman, ``Einstein--Cartan Theory'', in
    {\it Encyclopaedia of Mathematical Physics}, ed. J.  P. Fran\c{c}oise, G. L.
    Naber and Tsou Sheung Tsun, Elsevier (Oxford) 2006.

\bibitem{tHooft} G. 't~Hooft, Acta Phys.\ Austr., Suppl.\ 22,
    531  (1980).

\bibitem{banks} T. Banks and E. Rabinovici, Nucl.\ Phys.\ B160
  (1979) 347.

\bibitem{venez} G. Veneziano, Nuovo Cimento 57A (1968) 190.

\bibitem{dualres} J. E. Paton and Chan Hong-Mo, Nucl. Phys. B10 (1969) 516.

\bibitem{string} There are many papers and books on string theory; see
  e.g. M. B. Green, J. H. Schwarz and E. Witten, {\em Superstring theory},
  Cambridge University Press (1987). 

\bibitem{seesaw} M.  Gell-Mann, P. Ramond and R. Slansky (1979), 
in Supergravity, edited by P. van Nieuwenhuizen and D. Z. Freedman, North Holland 
Publ. Co.;
P. Minkowski, Phys. Lett. B 67 (1977) 421;
R. N. Mohapatra and G. Senjanovic, Phys. Rev. Lett. 44 (1980) 912.

\bibitem{zmixed} Jos\'e Bordes, Chan Hong-Mo and Tsou Sheung Tsun, ``The
  $Z$ Boson in the Framed Standard Model''. arXiv:1806.08271.

\bibitem{efgt} Chan Hong-Mo and Tsou Sheung Tsun, Int. J. Mod. Phys. A27 (2012)
   1230002; arXiv:1111.3832.

\bibitem{dfsm} Michael J. Baker, Jos\'e Bordes, Chan Hong-Mo and Tsou Sheung Tsun,
   Int. J. Mod. Phys. A27 (2012) 1250087; arXiv:1111.5591.

\bibitem{Yang} C. N. Yang, Phys.\ Rev.\ D1 (1979) 2360.

\bibitem{ourbook} See e.g. Chan Hong-Mo and Tsou Sheung Tsun,
    {\it Some Elementary Gauge Theory Concepts}, World Scientific, Singapore, 1993.

\bibitem{Hdecay} Jos\'e Bordes, Chan Hong-Mo and Tsou Sheung Tsun,
  Eur. Phys. J. C65; 537-542, 2010; arXiv:0908.1750.

\bibitem{f+mike} Michael J. Baker and Tsou Sheung Tsun,
Eur. Phys. J. C70; 1009, 2010, DOI 10.1140/epjc/s100052-010-1506-0;
  arXiv:1005.2676.

\bibitem{pdg} M . Tanabashi et al. (Particle Data Group), Phys. Rev. D98, 030001 (2018). Also at PDG website: http://durpdg.dur.ac.uk/lbl/ 

\bibitem{yang} T. D. Lee and C. N. Yang, Phys.\ Rev.\ 98 (1955) 1501.

\bibitem{atlasmw} M. Aaboud et al.
(ATLAS Collaboration). Eur. Phys. J. C (2018) 78:110. arXiv:1701.07240.
Also at https://phys.org/news/2016-12-atlas-mass-lhc.html

\bibitem{diphoton750a} M. Aaboud et al.
(ATLAS Collaboration) (September 2016).
``Search for resonances in diphoton events at √s = 13 TeV with the ATLAS
detector''.
Journal of High Energy Physics. 2016 (9): 1.

\bibitem{diphoton750b} V. Khachatryan et al.
(CMS Collaboration) (28 July 2016).
``Search for resonant production of high-mass photon pairs in proton-proton
collisions at √s = 8 and 13 TeV''.
Phys. Rev. Lett. 117 (5): 051802; arXiv:1606.04093 

\bibitem{no750a} M. Aaboud et al.
(ATLAS Collaboration)
``Search for new phenomena in high-mass diphoton final states using 37 fb−1
of proton--proton collisions collected at s√=13 TeV with the ATLAS detector''.
Phys. Lett. B775 (2017) 105-125.

\bibitem{no750b} V. Khachatryan et al.
(CMS Collaboration)
``Search for high-mass diphoton resonances in proton–proton collisions at 13
TeV and combination with 8 TeV search''.
Phys. Lett. B767 (2017) 147-170; arXiv:1707.04147

\bibitem{Atomki} A. J. Krasznahorkay et al., Phys. Rev. Lett. 116, 042501
(2016).\\ arXiv:1504.01527[nucl-ex].

\bibitem{yamamoto} Teppei Kitahara and Yasuhiro Yamamoto, Phys.\ Rev.\
  D95, 015008 (2017); arXiv:1609.01605.

\bibitem{feng} Jonathan L. Feng et al., Phys.\ Rev.\ D95, 035017
  (2017);\\ arXiv:1608.03591.

\bibitem{moretti} Luigi Delle Rose, Shaaban Khalil, Stefano Moretti,
  Phys.\ Rev.\ D96, 115024 (2017); arXiv:1704.03436.

\bibitem{darkmatter} M. Drees and G. Gerbier, review article in
  \cite{pdg}, and references therein.

\bibitem{weinberg} See for example: S. Weinberg, {\it The Quantum 
   Theory of Fields II} (Cambridge University Press, New York, 1996).

\bibitem{strongcp} Jos\'e Bordes,  Chan Hong-Mo and Tsou Sheung Tsun,
  Int. J. Mod. Phys.  A24 (2009) 101-112; arXiv:0707.3358.


\bibitem{Bjwesite} James Bjorken, private communication; see also the
  website\\ bjphysicsnotes.com

\end{thebibliography}
\end{document}